\documentclass[traditabstract]{aa} 

\usepackage{graphicx}
\usepackage{amsmath}	
\usepackage{amssymb}	
\usepackage{txfonts}
\usepackage[]{hyperref}
\hypersetup{colorlinks = true, linkcolor = blue, urlcolor=blue, citecolor=blue} 
\usepackage[normalem]{ulem}
\usepackage{tabularx}
\usepackage{rotating, booktabs}
\usepackage{inputenc}
\usepackage{geometry}
\usepackage{changepage}
\usepackage{color}
\usepackage{multirow}

\def\msun{\hbox{M$_\odot$}}

\newcommand{\angstrom}{\mbox{\normalfont\AA}}

\begin{document} 
\title{A Virgo Environmental Survey Tracing Ionised Gas Emission. 
   (VESTIGE)}

\subtitle{XVIII. Reconstructing the star formation history of early-type galaxies through the combination of their UV and H$\alpha$ emission}

\author{S. Martocchia\inst{1, 2}, 
          A. Boselli\inst{1,3}, 
          C. Maraston\inst{4},
          D. Thomas\inst{4},
          M. Boquien\inst{5}, 
          Y. Roehlly\inst{1},
          M. Fossati\inst{6,7},
          L.-M. Seill\'e\inst{1},
          P. Amram\inst{1},
          S. Boissier\inst{1},
          V. Buat\inst{1}, 
          P. C{\^o}t{\'e}\inst{8},
          J-C. Cuillandre\inst{9},
          L. Ferrarese\inst{8}, 
          S. Gwyn\inst{8},
          J. Hutchings\inst{8},
          Junais\inst{10,11},
          C. R. Morgan\inst{12,13},
          J. Postma\inst{14}, 
          T. E. Woods\inst{15},
          J. Roediger\inst{8},
          A. Subramaniam\inst{16},
          M. Sun\inst{17} and
          H.-X. Zhang\inst{18}}

\institute{Aix Marseille Universit{\'e}, CNRS, CNES, LAM, Marseille, France
        \email{silvia.martocchia@lam.fr}
    \and
        Astronomisches Rechen-Institut, Zentrum f\"ur Astronomie der Universit\"at Heidelberg, M\"onchhofstraße 12-14, D-69120 Heidelberg, Germany
    \and 
        INAF - Osservatorio Astronomico di Cagliari, via della Scienza 5, 09047 Selargius, Italy
    \and
        Institute of Cosmology, University of Portsmouth, Burnaby Road, Portsmouth, PO1 3FX, UK
    \and 
        Universit{\'e} C{\^o}te d'Azur, Observatoire de la C{\^o}te d'Azur, CNRS, Laboratoire Lagrange, 06000, Nice France
    \and
        Universit\`a di Milano-Bicocca, Piazza della scienza 3, 20100 Milano, Italy
    \and 
        INAF - Osservatorio Astronomico di Brera, Via Brera 28, 20159 Milano, Italy
    \and
        National Research Council of Canada, Herzberg Astronomy and Astrophysics Research Centre, Victoria, BC V9E 2E7, Canada    
    \and 
         AIM, CEA, CNRS, Universit\'e Paris-Saclay, Universit\'e de Paris, 91191 Gif-sur-Yvette, France
    \and
         Instituto de Astrof\'{i}sica de Canarias, V\'{i}a L\'{a}ctea S/N, E-38205 La Laguna, Spain
    \and
        Departamento de Astrof\'{i}sica, Universidad de La Laguna, E-38206 La Laguna, Spain
    \and 
        Waterloo Centre for Astrophysics, University of Waterloo, Waterloo, ON N2L 3G1, Canada
    \and 
        Department of Physics and Astronomy, University of Waterloo, Waterloo, ON N2L 3G1, Canada
    \and
        Dept. of Physics and Astronomy, University of Calgary, Calgary, AB T2N 1N4, Canada
    \and 
        Department of Physics and Astronomy, University of Manitoba, Winnipeg, Manitoba R3T 2N2, Canada
    \and
        Indian Institute of Astrophysics, Bangalore, India
    \and
        Department of Physics \& Astronomy, University of Alabama in Huntsville, 3001 Sparkman Drive, Huntsville, AL 35899, USA
    \and Department of Astronomy, University of Science and Technology of China, Hefei, Anhui 230026, People’s Republic of China
            }

   \date{Received XX; accepted YY}

 
  \abstract
   {We reconstruct the star formation histories of 7 massive ($M_{\star}\gtrsim 10^{10}$\msun) early-type galaxies (ETGs) in the Virgo cluster by analysing their spatially resolved stellar population (SP) properties including their ultraviolet (UV) and H$\alpha$ emission.
   As part of the Virgo Environmental Survey Tracing Ionised Gas Emission (VESTIGE), we used H$\alpha$ images to select ETGs that show no signs of ongoing star formation. We combined VESTIGE with images from Astrosat/UVIT, GALEX and CFHT/MegaCam from the Next Generation Virgo Cluster Survey (NGVS) to analyse radial spectral energy distributions (SEDs) from the far-UV to the near-infrared. The UV emission in these galaxies is likely due to old, low-mass stars in post main sequence (MS) phases, the so-called UV upturn. 
   We fit the radial SEDs with novel SP models that include an old, hot stellar component of post-MS stars with various temperatures and energetics (fuels). This way, we explore the main stellar parameters responsible for UV upturn stars irregardless of their evolutionary path. We make these models publicly available through the SED fitting code \texttt{CIGALE}.
   Standard models are not able to reproduce the galaxies' central FUV emission (SMA/$R_{\rm eff}\lesssim1$), while the new models well characterise it through post-MS stars with temperatures T$\gtrsim$25000~K. All galaxies are old (mass-weighted ages $\gtrsim10$~Gyr) and the most massive M49 and M87 are supersolar ($Z\simeq2Z_{\odot}$) within their inner regions (SMA/$R_{\rm eff}\lesssim0.2$). Overall, we found flat age gradients ($\nabla$Log(Age)$\sim -0.04 - 0$ dex) and shallow metallicity gradients ($\nabla$Log(Z)$<-0.2$ dex), except for M87 ($\nabla$Log($Z_{\rm M87}$)$\simeq-0.45$ dex).  
   Our results show that these ETGs formed with timescales $\tau\lesssim1500~$Myr, having assembled between $\sim40-90\%$ of their stellar mass at $z\sim5$. This is consistent with recent JWST observations of quiescent massive galaxies at high$-z$, which are likely the ancestors of the largest ETGs in the nearby Universe. The derived flat/shallow stellar gradients indicate that major mergers might have contributed to the formation and evolution of these galaxies.}


   \authorrunning{S. Martocchia}
   \titlerunning{SFH of Virgo ETGs through their UV and H$\alpha$ emission}
   \maketitle
%

\section{Introduction}
\label{sec:intro}

Historically, galaxies that are named early-type (ETGs, \citealt{hubble26}) are red and spheroidal; they lack gas and spiral arms and are pressure$-$supported. Understanding the formation of ETGs, which appear smooth and well-behaved today, poses a significant challenge in galaxy mass assembly and is closely tied to the evolution of the Universe as outlined by the $\Lambda$CDM model.
According to the $\Lambda$CDM paradigm, structures of dark matter haloes are expected to assemble in a hierarchical way, where larger haloes form through the mergers of smaller ones born at earlier epochs (e.g., \citealt{delucia04, springel05, laporte13}).
Within this evolutionary picture, ETGs are thought to form in a two-phase scenario (e.g., \citealt{delucia06, oser10}). First, a phase of dissipative collapse at $z\gtrsim2$ produces \textit{in-situ} star formation, which is rapidly quenched. This is followed by an \textit{accretion} phase, where ex-situ stars are accreted from mostly dry, minor mergers with smaller satellite galaxies at $z\lesssim3$. Therefore, in agreement with the $\Lambda$CDM view, stars in ETGs are formed very early on, but the galaxy's mass growth is assembled at later stages.

The launch of the \textit{James Webb Space Telescope} (JWST) has triggered a renewed interest in these galaxies. Indeed, recent JWST results revealed that high-redshift red and quiescent galaxies are already very massive ($>10^{10}$\msun) at $z>3$ (e.g., \citealt{antwi-danso23, baggen23, carnall23, valentino23, labbe23, boyett24, nanayakkara24, glazebrook24, carnall24}), and that they formed their stars and quenched very rapidly, up to $\sim$200 Myr after formation (e.g., \citealt{degraaff24}). Consequently, these observations challenge both current galaxy formation theories and the $\Lambda$CDM cosmological model (e.g., \citealt{menci22, boylan23}), unless some extreme efficiencies in converting baryon into stars are invoked (e.g., \citealt{degraaff24, carnall24}).

This tension is not only present at high redshift, but it is reflected in studies of ETGs in the local Universe. Indeed, local ETGs, whose properties can be studied at high spatial resolution and sensitivity, represent the direct descendants of high redshift galaxies. Therefore, reconstructing their formation history serves as a crucial testbed to constrain cosmological models.

ETGs in the nearby Universe are observed to be the oldest, most metal-rich and the most abundant in [$\alpha$/Fe] ratio (e.g., \citealt{worthey92, trager00a, thomas05, clemens06, thomas10, greene13, parikh19}). This is indicative of rapid star formation timescales (e.g., \citealt{worthey92, trager00a,gavazzi02, thomas05, spolaor10,worthey14}), consistent with observations in the early Universe. Recent results from the ATLAS$^{3D}$ survey of nearby ETGs \citep{cappellari11} revealed unexpected properties for this class of galaxies. The majority of ETGs in ATLAS$^{3D}$ were found to be fast rotators (FR), while only a small fraction of the most massive galaxies are pressure-supported (or slow rotators, SR, \citealt{emsellem11}); in addition, $\sim$40\% of ETGs host gas in both molecular and ionised phases (e.g. \citealt{serra12,rampazzo13,young14}), indicating that some rejuvenation in star formation (SF) had to occur.
All this evidence suggests that several unique formation pathways are likely necessary to explain the broad and complex nature of ETGs (e.g., \citealt{naab14}). Besides, it is still unclear what is the exact contribution of the physical processes that are responsible for quenching the SF of ETGs in early phases (e.g., \citealt{man18} and references therein), as well as the role of possible variations in their stellar initial mass function (e.g., \citealt{bastian10, conroy_vandokkum, tortora13, martinnavarro15, yan21}). 
As a consequence, a self-consistent explanation on how present-day ETGs form and evolve remains elusive (e.g., \citealt{naab17, lagos22}).

In this work, we aim at placing new constraints on how massive local ETGs form by analysing their resolved ultraviolet (UV) properties. As mentioned earlier, there is observational evidence of residual SF in a number of ETGs (e.g., \citealt{kaviraj07, yi11, vazdekis16, rampazzo17,gavazzi18, sr22}), where the UV emission is due to young stellar populations. The UV light in this case is dominated by massive A stars with ages $\sim100$ Myr (e.g. \citealt{lequeux88,boselli09}). However, there are ETGs that are genuinely old systems that also show a UV excess in their spectra, at $\lambda<2500$\AA, the so-called phenomenon of the UV upturn (e.g., \citealt{code79, oconnell99}). This UV emission is boosted by old, low-mass stars in late evolutionary stages, although its exact origin is still debated (e.g., post-asymptotic giant branch stars, blue horizontal branch or interacting binary systems, \citealt{greggio90, bressan94, mt00,brown00, maraston05, han07, yi08, han10, hernandez13} and references therein).
For this reason, the investigation of the UV domain is imperative to reconstruct the star formation history (SFH) of ETGs (see, e.g., \citealt{brown00, gildepaz07, boselli14}); furthermore, it represents a powerful diagnostic of low mass stars evolution and old stellar populations (e.g., \citealt{burstein88, brown04, boselli05, yi08}). The presence of the upturn regardless of the local galaxy density \citep{boissier18,ali19, phillips20} suggests that the UV upturn is an intrinsic property of ETGs, which forms at early times and cannot be interpreted in terms of residual SF.

While the UV upturn has been previously investigated in ETG samples at different redshifts (e.g., \citealt{burstein88, peletier90, ohl98, brown00, boselli05, ree07, yi08, ali18b, ali18,  lecras16, boissier18, lonoce20, dantas20, werle20, depropris22, akhil24}), the goal of this paper is to characterise the UV properties of local ETGs in spatially resolved stellar populations. In this context, we aim to reconstruct the SFH of nearby ETGs as a function of radius, combining multiwavelength data from the far-UV (FUV) to optical ranges (including H$\alpha$), with unprecedented angular resolution and sensitivity. Another main novelty of this work is the computation of state-of-the-art stellar population models, starting from those by \cite{mt00} and \cite{lecras16}, that are specifically tailored to reproduce and describe the properties of the UV upturn in ETGs across cosmic history. The models are implemented in the publicly available spectral energy distribution (SED) fitting code \texttt{CIGALE} \citep{boquien19}, thus we will enable the determination of galaxy properties across diverse ranges of ETG samples.

\begin{table*}
\centering
\caption{Properties of the target galaxies.}
{\tiny 
\begin{tabular}{ccccccccc}
\hline
\noalign{\smallskip}
Name & Units &  NGC~4262 & NGC~4374/M84 & NGC4406/M86 & NGC~4417 & NGC~4442 & NGC4472/M49 & NGC4486/M87\\
\hline
\noalign{\smallskip}
VCC Name$^{(a)}$ & & VCC355 & VCC763 & VCC881 & VCC944 & VCC1062 & VCC1226 & VCC1316\\
Virgo Substructure $^{(b)}$ & & A & A & A & B & B & B & A\\
           & &  &    &   &   &   & (central) & (central)\\
Morph. Type$^{(a)}$ & & SB02/03 & E1 & E3/S013 & S017 & SB016 & E2/S012 & E0\\
Nuclear Act.$^{(c)}$ & & passive & LINER;Sy2 & 
passive & passive & passive & LINER & 
NLRG;Sy\\
$cz$$^{(a)}$ & (km/s) & 1359 & 1017 & -224 & 828 & 547 & 981 & 1284\\
d$^{(d)}$ & (Mpc) & 17 & 17 & 17 & 23 & 23 & 17 & 17 \\ 
$R_{\rm eff}(i)$$^{(e)}$ & (kpc) & 0.62 & 6.62 & 13.04 & 1.63 & 1.92 & 9.58 & 11.36\\ 
Log($M_{\star}$)$^{(f)}$ & (Log(\msun)) & 9.88 &
11.18 & 11.22 & 10.65 & 10.67 & 11.59 & 11.36\\
$\sigma_{c}^{(g)}$ & (km/s) & $195$ & $288$ & $217$ & $152$ & $194$ & $288$ & $314$\\
Fast/Slow Rotator$^{(h)}$ & & FR? & SR & SR & FR & FR & SR & SR\\
\hline
\end{tabular} 
References: $^{(a)}$ \cite{binggeli85}. $^{(b)}$: \cite{boselli10}.  $^{(c)}$: \cite{gavazzi13}. $^{(d)}$: \cite{cortese12}. $^{(e)}$: calculated from NGVS $i$ band. $^{(f)}$: \cite{cortese12}. $^{(g)}$: central ($R_{\rm eff}/8$) velocity dispersion from \cite{cappellari13}. $^{(h)}$: \cite{emsellem11}.}
\label{tab:data}
\end{table*}

In this pilot work, we report the SED analysis of seven massive ($M_{\star}>10^{9.8} M_{\odot}$) ETGs (E+S0) in the Virgo cluster. Virgo, our closest galaxy cluster only $\sim$17~Mpc away \citep{gavazzi99,mei07,cantiello18,cantiello24}, represents an excellent laboratory to study the origin of ETGs. Its proximity enables a detailed radial analysis of the different galactic components at an unprecedented sensitivity and angular resolution (1$''$ on the sky corresponds to $\sim$80 pc at 17~Mpc), reinforced by the availability of several multiwavelength surveys covering the entire extent of the cluster. 
Additionally, both elliptical and lenticular galaxies are mainly found in high density environments, with the former being more numerous in the core of rich galaxy clusters (e.g. \citealt{dressler80, whitmore93}). Hence, the formation of ETGs is strongly shaped by the environment in which they reside (e.g. \citealt{thomas10, labarbera12, boselli14, pasquali19}).

As we are interested in the underlying stellar populations, we selected galaxies with no evidence of recent SF, thanks to the Virgo Environmental Survey Tracing Ionised Gas Emission (VESTIGE), a blind and deep narrow H$\alpha$+[NII] imaging survey of the Virgo cluster with the MegaCam at the Canada-France-Hawaii Telescope (CFHT,  \citealt{boselli18}). In addition, we use imaging data in the FUV, thanks to the exquisite sensitivity and angular resolution of the UltraViolet Imaging Telescope (UVIT) on board the AstroSat satellite \citep{agrawal06}. We combine this with near-UV (NUV) images from the Galaxy Evolution Explorer (GALEX), optical images from the Next Generation Virgo Cluster Survey (NGVS, \citealt{ferrarese12}, MegaCam at CFHT), and integral field unit (IFU) spectroscopy data with the Multi Unit Spectroscopic Explorer (MUSE) at the Very Large Telescope (VLT), whenever available.


This paper is structured as follows: Sect. \ref{sec:sample} describes the sample, while Sect. \ref{sec:data} reports information on the data used. In Sect. \ref{sec:analysis} we show the analysis of the imaging and spectroscopic data. Sect. \ref{sec:sed} presents the new stellar population models used to determine the properties of the target galaxies and describes the SED fitting analysis. We report on the results in Sect. \ref{sec:res}, while we discuss in Sect. \ref{sec:disc}. Finally, we summarise and conclude in Sect. \ref{sec:concl}.

\section{The Sample}
\label{sec:sample}

The sample is composed of seven massive ($M_{\star}>10^{9.8} M_{\odot}$) ETGs in the Virgo cluster. These galaxies have been selected among the most massive and extended ETGs of the cluster with available AstroSat/UVIT data in the FUV band (see Sect. \ref{subsec:uvit}). They do not show any clear star formation activity in the form of clumps or knots, thus suggesting that their UV emission is dominated by old stellar populations\footnote{Although one galaxy - NGC~4262 - shows HII regions in its outskirts, see Sect. \ref{subsec:ha_analy} and Appendix \ref{app:notes_gals} for more details.}. We refer to Sect. \ref{subsec:ha_analy} for more details about this. 

Table \ref{tab:data} reports the main properties of the target galaxies. They are all well-known ETG systems, some of which already studied in detail, also in the UV spectral domain (e.g., \citealt{burstein88, peletier90, brown95, ohl98, mt00, bettoni10, boselli22}). As defined, the sample is not complete in any sense. 
It consists of four ellipticals (E) and three lenticulars (S0).  The E galaxies in the sample are the well- known M84, M86, M49 and M87, while the S0 sample is composed of NGC~4262, NGC~4417 and NGC~4442. These galaxies belong to different substructures within the Virgo cluster (see Table \ref{tab:data}). Indeed, Virgo is composed of several subgroups of galaxies with different mean recessional velocities and positions within the cluster \citep{boselli14}. M87 and M49 are the central galaxies of two different substructures, called A and B, respectively \citep{gavazzi99}.

\section{Observations and Data Reduction}
\label{sec:data}

The analysis presented in this work is based on a set of multifrequency data, some of which was gathered during untargeted surveys of the whole Virgo cluster region.

\subsection{VESTIGE H$\alpha$ imaging}
\label{subsec:vestige}

In order to check for contamination from young stars or recent SF in the target galaxies, we used narrow-band H$\alpha$ images from the VESTIGE survey \citep{boselli18}. The data are obtained from the narrow band filter MP9603 ($\lambda_c = 6590\AA$; $\Delta\lambda=106$\AA) installed on the MegaCam at the CFHT. At the redshift of these galaxies, the filter includes the H$\alpha$ line together with the two [NII] lines at $\lambda=6548, 6583\AA$. Hereafter, we refer to the filter simply as H$\alpha$, unless otherwise stated. Galaxies were observed with a typical $\sim$2 hours of integration time in the narrow band H$\alpha$ filter and $\sim$12 min in the broad-band $r$ MegaCam filter ($\lambda_c =6369\AA$).  
An extensive description about the observations and data reduction is reported in \cite{boselli18, boselli22}. We refer the interested reader to these papers for more details.
The reached sensitivity of the survey at full depth is $\Sigma({\rm H}\alpha)\simeq 2 \times 10^{-18}$ erg s$^{-1}$ cm$^{-2}$ arcsec$^{-2}$ (1$\sigma$ after smoothing to 3$''$) for the extended sources, while it is $f({\rm H}\alpha)\simeq 4 \times 10^{-17}$ erg s$^{-1}$ cm$^{-2}$ (5$\sigma$) for point sources.

\begin{table*}
\centering
\caption{Main information on the archival and proprietary AstroSat/UVIT BaF2 images for the target galaxies.}
\begin{tabular}{ccccc}
\hline
\noalign{\smallskip}
Name & Obs. Date & Obs. ID & Exp. time & PI\\
(units) &           &      &     (s)       &   \\
\hline
\noalign{\smallskip}
NGC~4262 & 2021-06-03 & A10\_071T44\_9000004434 & 10755 & A. Boselli\\
NGC~4374/M84 & 2021-06-03 & A10\_071T48\_9000004478 & 7577 & A. Boselli\\
NGC~4406/M86 & 2020-05-06 & A08\_003T13\_9000003646 & 7074 & J. Hutchings\\
NGC~4417 & 2021-06-07 & A10\_071T18\_9000004448 & 10941 & A. Boselli \\
NGC~4442 & 2021-06-24 & A10\_071T54\_9000004482 & 7544 & A. Boselli \\
NGC~4472/M49 & 2017-04-22 & A03\_030T01\_9000001 & 6949 & P. Shastri \\
NGC~4486/M87 & 2018-04-21 & A04\_115T02\_9000002 & 14062 & C. Varsha\\
\hline
\end{tabular}
\label{tab:uvit_tab}
\end{table*}

The observations in the $r$ filter are necessary to subtract the stellar continuum of the target galaxies. This step is particularly critical for ETGs where the emission of ionised gas is marginal compared to the emission of stars. As extensively described in \cite{boselli19,boselli22}, the stellar continuum was estimated by combining the $r$ band image with a $g-r$ colour map, where the $g-$band is taken from the NGVS $g$ band images (see Sect. \ref{subsec:ngvs}). We refer the interested reader to these papers for more details about the continuum subtraction procedure. The continuum$-$subtracted H$\alpha$ images of the target galaxies are reported in Figure \ref{fig:ha}. As it is visible in almost each galaxy, a possible spurious emission in the continuum$-$subtracted image might be present. This emission is not real, but due to the way the continuum$-$subtracted image is created. The H$\alpha$ emission would look clumpy and/or filamentary rather than smooth as in NGC~4417, NGC~4442 or NGC~4472. Indeed, since the contribution of the stellar continuum is derived using the $r$ and $g-r$ images, artefacts might be artificially created in those regions where
the emission is strongly peaked and colour gradients are present, which typically happens in the core of bright ellipticals \citep{boselli19}. Here, possible differences in the seeing of the three images (H$\alpha$, $r$, $g$) and minimal misalignment may produce major effects in the resulting continuum$-$subtracted images (see also the intense emission visible in the galactic nuclei in Fig. \ref{fig:ha}). Any possible emission in these particular regions must be confirmed with spectroscopic data (see Sect. \ref{subsec:ha_analy} for more details about the analysis of the ionised gas emission).

\subsection{Optical imaging - NGVS}
\label{subsec:ngvs}

We used optical images from the Next Generation Virgo cluster Survey (NGVS, \citealt{ferrarese12}), which observed the Virgo cluster with the MegaCam instrument onboard the CFHT. The images are available in the following bands: $u$ ($\lambda_c = 3799$\AA), $g$ ($\lambda_c = 4846.4$\AA), $i$ ($\lambda_c = 7543$\AA), $z$ ($\lambda_c = 8823$\AA). They have an angular resolution (FWHM) $\leq 1.1''$ and a pixel$-$scale of $0.187''$ \citep{ferrarese12}. For more detailed information on the data reduction, sky subtraction and exposure times, we refer the interested reader to \cite{ferrarese12}. In this work, we used images that are a merger between long and short exposures. The pixels of long exposure images are replaced in the centre of galaxies with short exposure pixels, to avoid saturation.

\subsection{AstroSat/UVIT imaging}
\label{subsec:uvit}

AstroSat/UVIT data for the target galaxies are either archival or proprietary (from proposals A08\_003 and A10\_071). Information on the FUV data used are reported in Table \ref{tab:uvit_tab}. We used images in the FUV filter BaF2, with $\lambda_c = 1541$\AA, whose reached angular resolution is $\sim1.5''$ \citep{tandon20}.
The reduction was performed following the prescriptions reported in \cite{tandon20} and the UVIT dedicated webpages, with a zeropoint $Z_p=17.778$ mag\footnote{See \url{https://uvit.iiap.res.in/Instrument/Filters}}. Figure \ref{fig:uvit} shows the images of the target galaxies in the FUV BaF2 filter with surface brightness contours superimposed in white. The FUV images span a sensitivity down to $22-23$ AB mag depending on the galaxy and considered regions (see Table \ref{tab:uvit_tab} for the exposure times). 

\begin{figure*}
\centering
\includegraphics[scale=0.18]{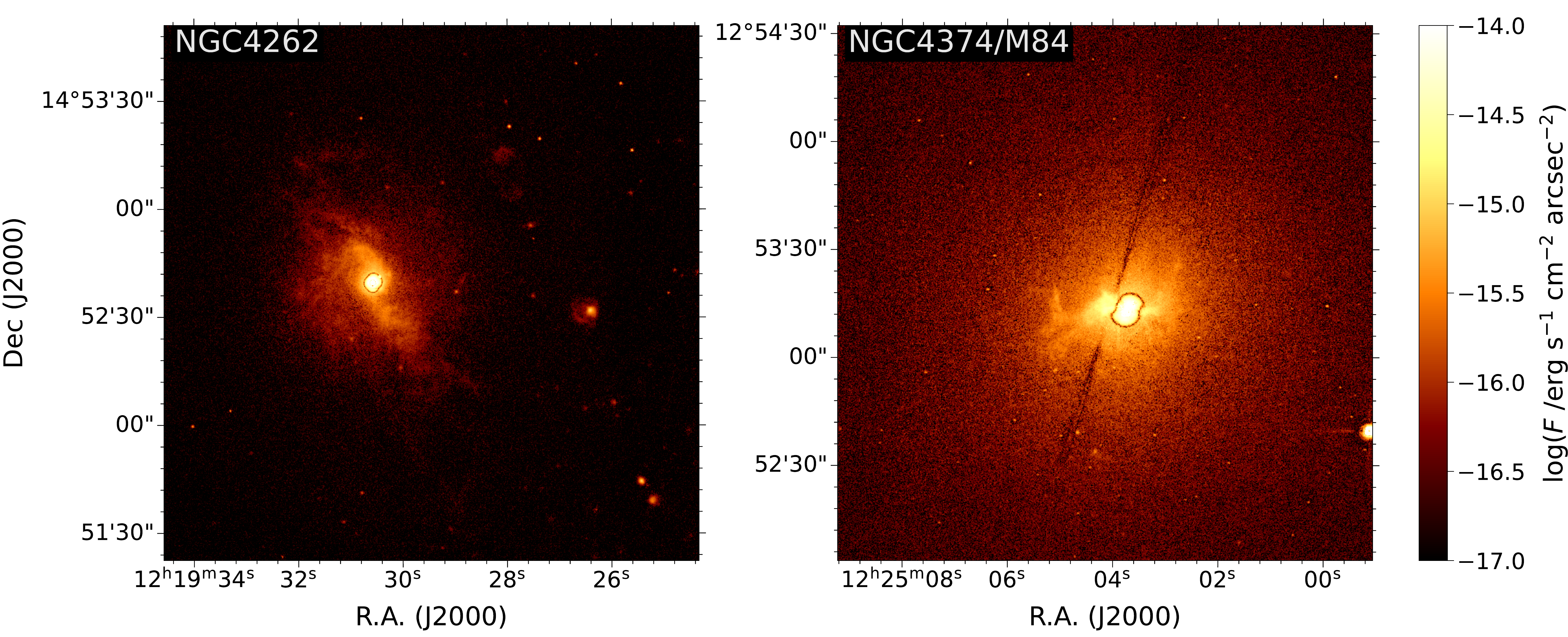}
\centering
\includegraphics[scale=0.18]{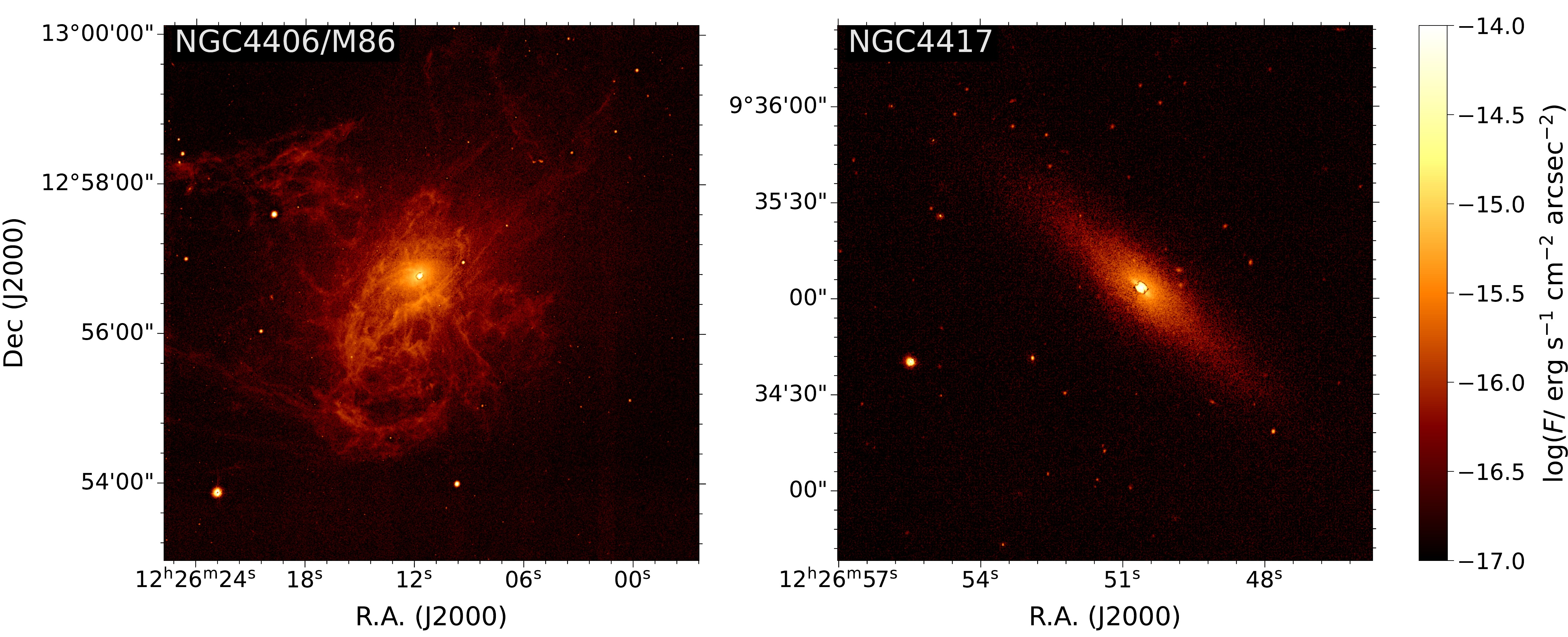}
\centering
\includegraphics[scale=0.18]{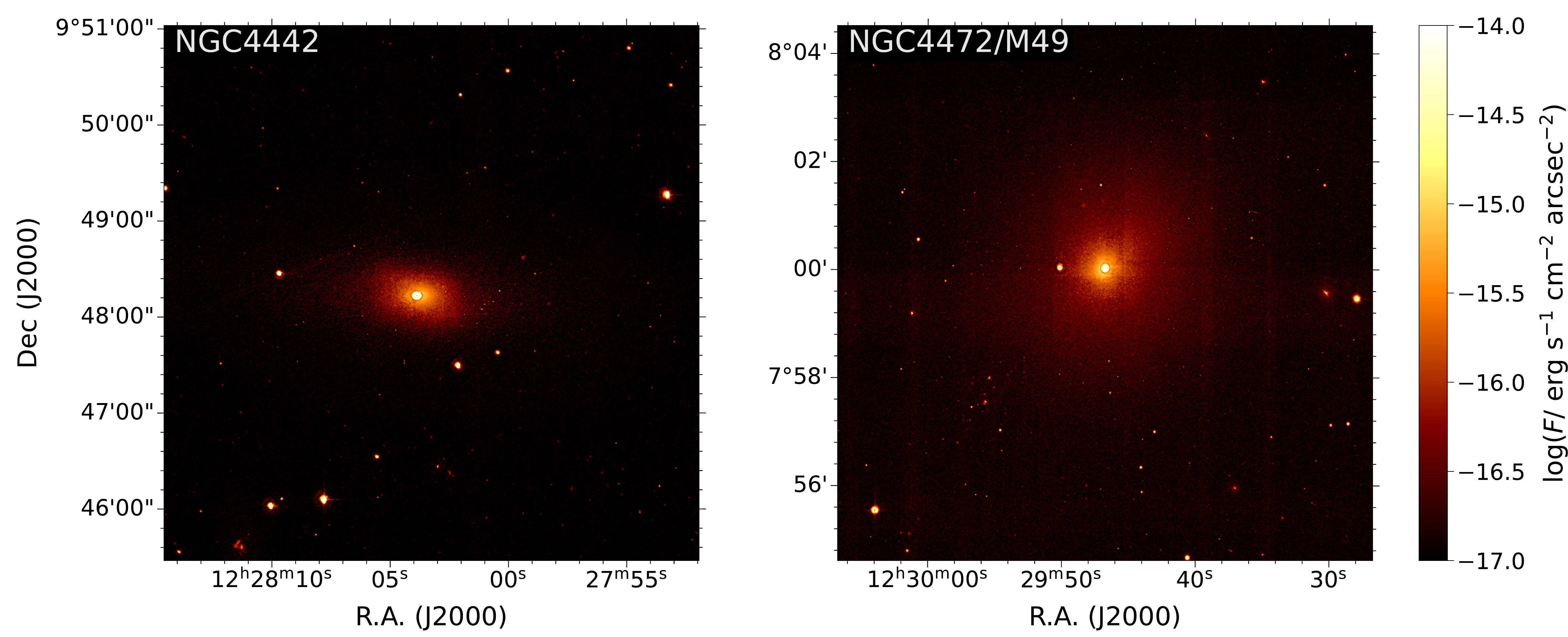}
\includegraphics[scale=0.19]{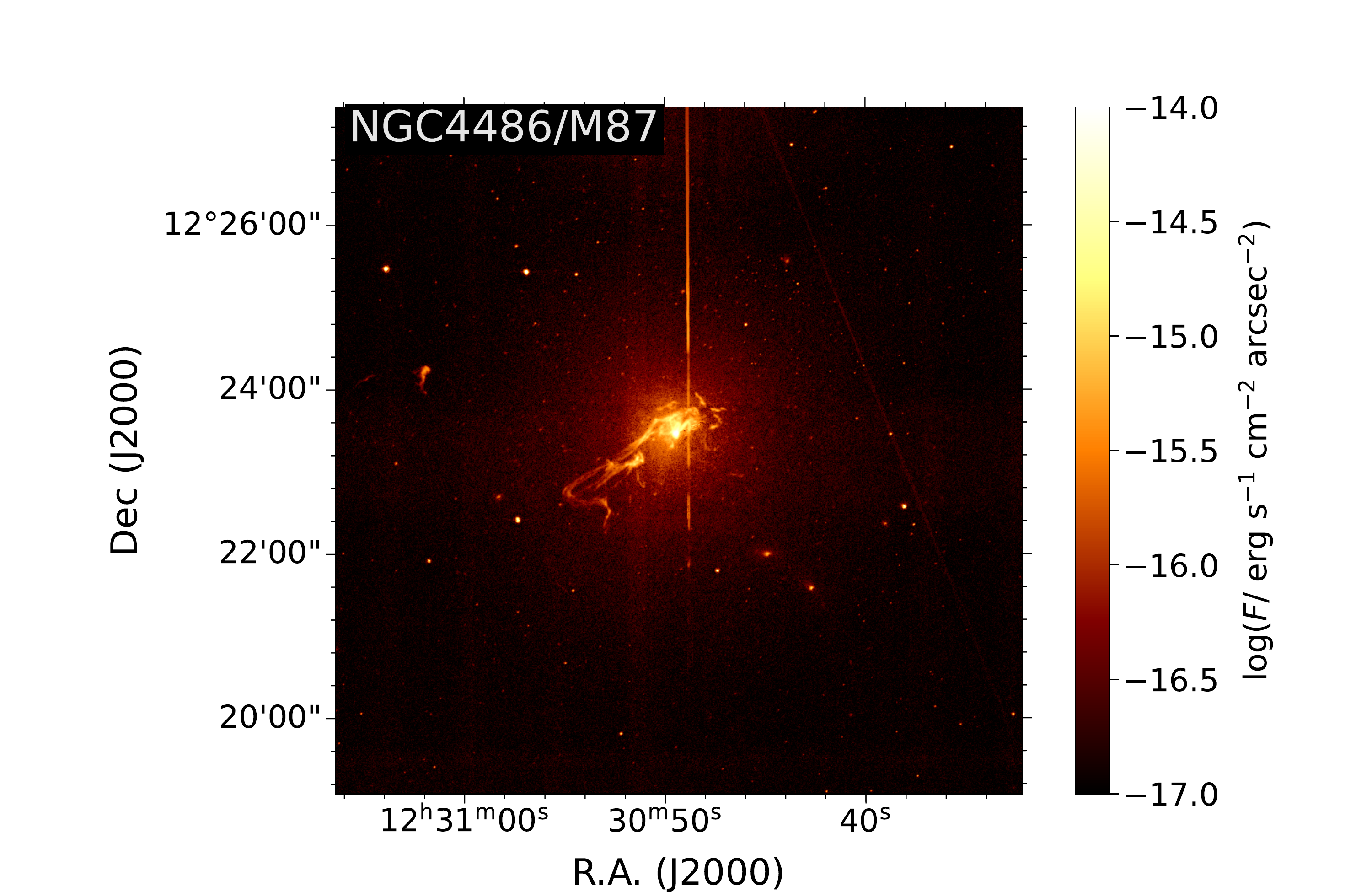}
\caption{Continuum$-$subtracted H$\alpha$ images of the target galaxies from the VESTIGE survey \citep{boselli18}.} 
\label{fig:ha}
\end{figure*} 

\subsection{GALEX imaging}
\label{subsec:galex}
No high-resolution UVIT images were available in the NUV channel for the target galaxies, therefore we used GALEX images in the NUV ($\lambda_c= 2297$\AA) from the GALEX Ultraviolet Virgo cluster survey (GUViCS, \citealt{boselli11, voyer14}). GUViCS is a blind complete survey of the Virgo cluster, covering $\sim$120 deg$^2$ in the NUV and reaching a sensitivity of $\sim22$ AB mag for the extended sources ($\sim$23 AB mag for point sources, \citealt{voyer14}). The already reduced GALEX images were downloaded from the HeDaM website\footnote{\url{https://hedam.lam.fr/HRS/}}. The target galaxies have exposure times ($t_{\rm exp}$) of $\sim2-3$ ks, except for M87 that has $t_{\rm exp}\sim$5 ks and NGC~4262 with $t_{\rm exp}\sim$28 ks. We refer the interested reader to \cite{cortese12} for information on the data reduction. The FWHM of the GALEX NUV images is $\sim 5.3-5.6''$ \citep{morrissey07}\footnote{\url{https://asd.gsfc.nasa.gov/archive/galex/Documents/instrument_summary.html}}. 

\begin{figure*}
\includegraphics[scale=0.55]{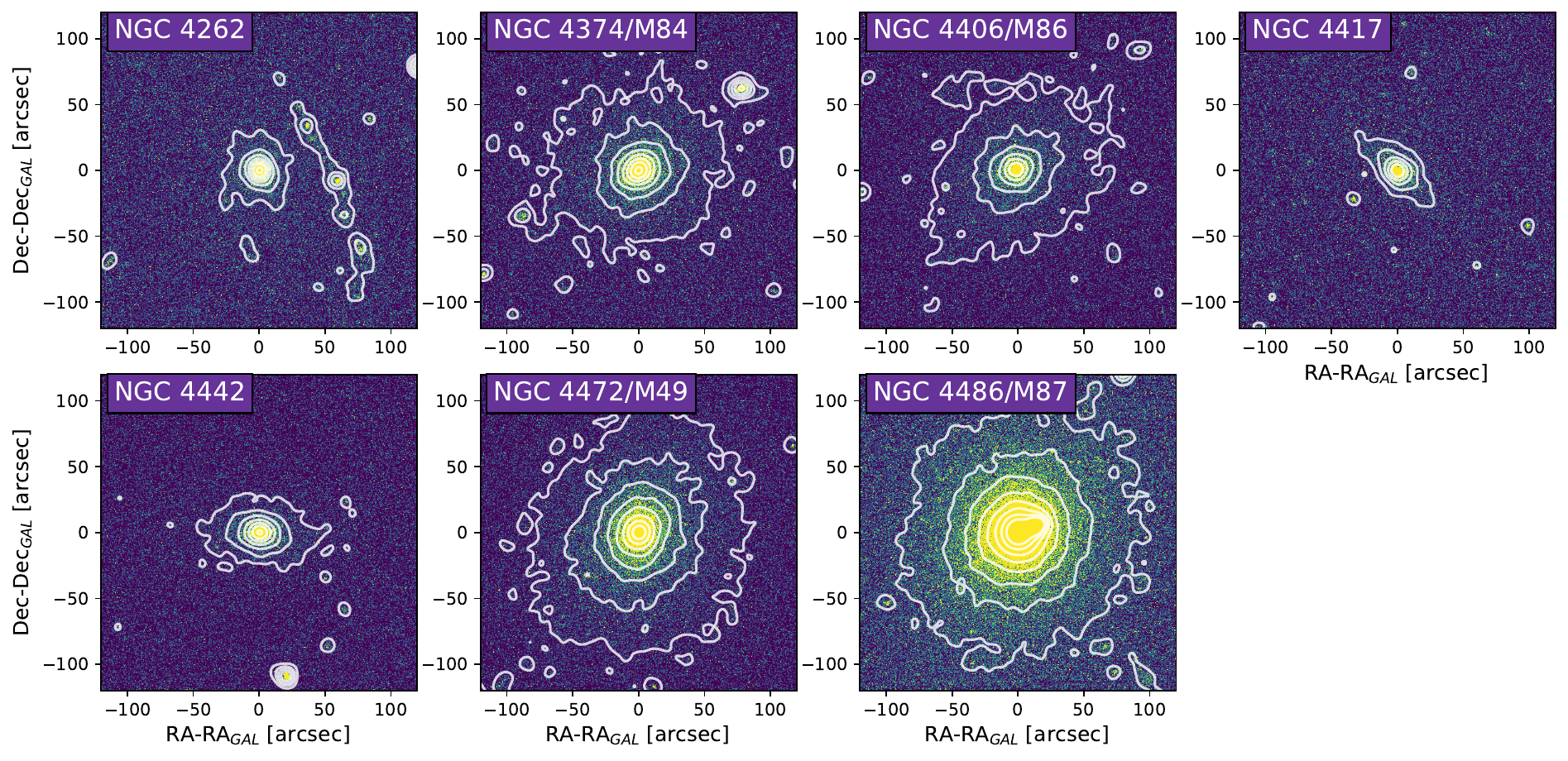}
\caption{Images of the target galaxies in the FUV BaF2 filter of Astrosat/UVIT in counts. North is up and East is to the left. The white contours are shown at levels of FUV surface brightness in the range 23.5-27 AB mag arcsec$^{-2}$ in steps of 0.5 AB mag arcsec$^{-2}$.} 
\label{fig:uvit}
\end{figure*}

\begin{table}
\centering
\caption{Main information on the archival VLT/MUSE observations for a subsample of the target galaxies.}
\begin{tabular}{cccc}
\hline
\noalign{\smallskip}
Name & Prog. ID & Exp. time & PI\\
(units) &           &  (s)    &    \\
\hline
\noalign{\smallskip}
M49 & 095.B-0295 &  6$\times$300 & C. J. Walcher \\
M84 & 0102.B-0048 & 4$\times$600 & B. Balmaverde\\
M87 & 060.A-9312 & 2$\times$1800 & Science Verification \\
\hline
\end{tabular}
\label{tab:muse}
\end{table}

\subsection{MUSE spectroscopy}
\label{subsec:muse_data}

Three galaxies from our sample have available VLT/MUSE spectroscopy in the ESO archive, namely M84, M49 and M87 \citep{sarzi18, balmaverde21}. We downloaded already reduced MUSE datacubes from the ESO Science Portal\footnote{\url{https://archive.eso.org/scienceportal/home}}, whose information are reported in Table \ref{tab:muse}. The cube of M49 was not originally corrected for the sky, thus we performed the sky subtraction by using the software \texttt{ZAP}\footnote{\url{https://zap.readthedocs.io/en/latest/\#usage}} \citep{soto16}. We report on the analysis of the MUSE data in Sect. \ref{subsec:muse_analysis}.

\subsection{Supplementary data}
\label{subsec:other_data}

Four galaxies, namely M49, M84, M86 and M87, are affected by dust, mostly in the form of filaments which are not attributable to the foreground extinction of the Milky Way. For M49, M84, M87 this is also visible from the presence of filaments in absorption in the NGVS images. In order to mask the dust emission in all the filters, we downloaded Hubble Space Telescope (HST) already reduced drizzled images from the ACSVCS (Advanced Camera for Survey Virgo Cluster Survey, \citealt{cote04}). See Sect. \ref{subsec:iso_fit} for more details on the dust masking.

M87 is well known to be a radio galaxy hosting a powerful emitting jet (e.g., \citealt{baade54, owen00}), which is clearly visible in our collected images. In Fig. \ref{fig:uvit}, the contours around the centre of M87 show the presence of the non-thermal jet.
It is then necessary to quantify and remove the contribution of the non$-$thermal emission of the jet in the UV part of the SED, as it might affect our stellar populations analysis.
In this context, we collected images of M87 to reconstruct its SED from the FUV up to the sub-mm. Details about the dataset and analysis are reported in Appendix \ref{app:m87_sed_syn}. 

All galaxies are part of the ATLAS$^{3D}$ survey \citep{cappellari11}, thus we downloaded the already calculated Voronoi binned maps  from the ATLAS$^{3D}$ website\footnote{\url{https://www-astro.physics.ox.ac.uk/atlas3d/}}, to inspect the ionised gas emission in H$\beta$ and [OIII]. 
Additionally, we downloaded and used available nuclear optical spectra (at $\lambda \sim6210-6860$\r{A}) from the red spectrograph of the 5m Hale telescope \citep{ho95}, to check for the presence of H$\alpha$ emission lines in the nuclei of our target galaxies, where the VESTIGE images are contaminated by artefacts. 
We refer to Section \ref{subsec:ha_analy} for more details.


\section{Data Analysis}
\label{sec:analysis}

\subsection{Ionised gas emission}
\label{subsec:ha_analy}
To ensure there is no contamination from young stars, and thus ongoing SF activity in the target galaxies, we made use of VESTIGE H$\alpha$ data. Thanks to their excellent image quality ($\sim0.7''$), we are able to detect the emission of individual HII regions down to $L(H\alpha) \simeq 10^{36}$ erg/s (Boselli et al. 2024, submitted).
Fig. \ref{fig:ha} shows the continuum$-$subtracted H$\alpha$ images of the target galaxies. NGC~4262, M84, M86 and M87 have an H$\alpha$ emission in the form of filaments, while the emission of the remaining galaxies is instead smooth or negligible. The lack of structured regions such as clumps or knots in the H$\alpha$ emission grants the absence of star forming complexes in the selected targets, with the exception of NGC~4262. Indeed, NGC~4262 shows the presence of an H$\alpha$ ring, extending south-west and north of the galaxy (e.g., \citealt{bettoni10, boselli22}), with visible HII regions. These will be masked in the following analysis.
The lack of SF activity in the target galaxies is further supported by the smoothness of the FUV emission (Fig. \ref{fig:uvit}), comparable to the optical one. We refer to Sect. \ref{subsec:iso_fit} for more discussion about this. 

Additionally, we estimated the [OIII] and H$\beta$ emission of each target galaxies from the ATLAS$^{3D}$ SAURON IFS data \citep{cappellari11}, within the inner $15''\times15''$. Following \cite{sarzi06}, we defined a detection threshold for the emission lines from the amplitude-to-noise ratio parameter ($A/N$), which is available in the ATLAS$^{3D}$ data. This parameter indicates how much the emission line sticks out above the noise in the stellar spectrum. Emission lines in [OIII] are detected if $A/N\geq4$, while for H$\beta$, $A/N\geq3$, even if this does not imply a 3-4$\sigma$ detection. We refer the interested reader to \cite{sarzi06} for more details about the detection threshold choices.
We also checked the nuclear spectra from \cite{ho95} to identify the presence of  H$\alpha$ emission lines within the galaxies' nuclei. In summary, we did not find a significant emission in [OIII], H$\alpha$ and H$\beta$ in M86, M49, NGC~4417 and NGC~4442. We found no significant H$\alpha$ and H$\beta$ emission in NGC~4262, however some [OIII] emission is present, although small ($f_{\rm [OIII]}\simeq 7 \times 10^{-15}$ erg s$^{-1}$ cm$^{-2}$). Emission lines are instead well detected in the central regions of M84 and M87, as these galaxies host active galactic nuclei (AGN). In the SED fitting analysis, see Sect. \ref{sec:sed}, we will mask the regions of these galaxies affected by AGN activity thanks to the MUSE data. 
We refer to Appendix \ref{app:notes_gals} for a more detailed description on the ionised gas emission of each individual galaxy within the sample. 

\begin{figure*}
\includegraphics[scale=0.55]{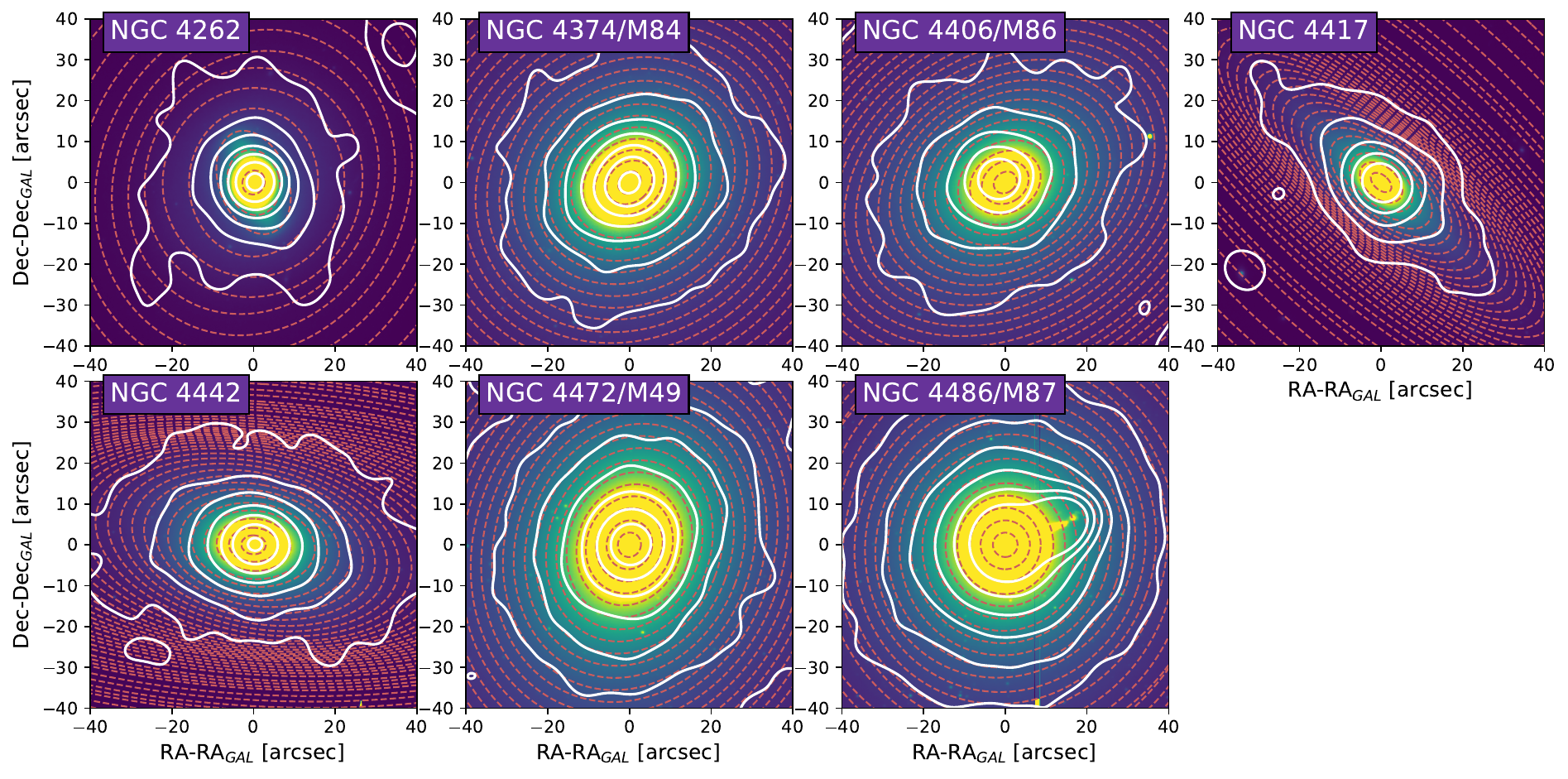}
\caption{Images of the target galaxies in the NGVS $i$ band with isophotal fits (at UVIT resolution) superimposed as red dashed ellipses. North is up and East is to the left. Contours from the FUV emission are superimposed in white, with surface brightness ranging from 23.5-27 mag arcsec$^{-2}$ in steps of 0.5 mag arcsec$^{-2}$.} 
\label{fig:ell_uvit}
\end{figure*}

\subsection{Synchrotron emission}
\label{subsec:synchr}

The spectral energy distributions of M84 and M87 are affected by synchrotron emission. 
In the following analysis, we masked the visible radio jet in M87 by defining jet regions from the NGVS $i$ image and applying the masks in each filter image. For the inner 20$''$, we adopted a different strategy, namely we corrected the UV flux for the contribution of the non-thermal emission of the jet, see Appendix \ref{app:m87_sed_syn}. 
We show that the synchrotron emission in M87 is significant in the inner 10$''$, contributing to $\sim$25\% of the flux in the FUV and $\sim$16\% of the flux in the NUV. \cite{ohl98} already reported that the non$-$thermal activity of the jet contributes to $\sim$17\% of the flux in the FUV in the inner 20$''$ of M87. Thus, our analysis is consistent with their considerations. In the following colour analysis, we then removed this contribution from the FUV and NUV fluxes.  

NGC~4374/M84 is also a radio galaxy \citep{laing_bridle87, meyer18}, however its jet is significantly less powerful than the one of M87 (see e.g. \citealt{boselli10b}), given that the AGN of M84 is $\sim$10 times less luminous than the AGN of M87 \citep{bambic23}. With this consideration in mind, we approximate the contribution from the synchrotron radiation in M84 to be $\sim$10 times less powerful than that of M87. This corresponds to a contribution of $\sim$10\% in the FUV flux of M84 and less than 1\% in the NUV flux in the inner 5$''$, where the non$-$thermal emission is expected to be the strongest. Hence, we decided not to correct for this effect in the following analysis of M84. 
Regarding the SED fitting analysis presented in Sect. \ref{sec:sed}, we decided not to fit the central parts (SMA/$R_{\rm eff}\lesssim0.1$) of M87 and M84, as they are significantly affected by their AGN, see Sect. \ref{subsubsec:dustsed}.

\subsection{Isophotal fitting}
\label{subsec:iso_fit}
We masked all the unwanted sources in each available image. To mask the foreground stars, we used the Gaia DR3 catalogue \citep{gaiadr3}, while we used the GLADE+ galaxy catalogue from \cite{dalya22} to mask the background galaxies. We used circle masks with a radius of 2.5$''$. Additional sources that were not in such catalogues as well as spikes from foreground stars were masked by hand by defining regions in \texttt{ds9}. The masking was performed starting from the $i$ image and then the same masks were applied to the other NGVS $ugz$ and VESTIGE $r-$bands. An independent, eye-based masking was applied for the FUV and NUV bands, as not all the optical sources have a counterpart in the UV and because of the different resolution in these bands.

Next, we used the HST images in the F475W filter (see Sect. \ref{subsec:other_data}) to visually check for dust filaments. We defined regions in \texttt{ds9} covering the dust filaments in absorption in M84, M87 and M49, and we masked them in each band\footnote{Dust is also present in M86 \citep{cortese12,ciesla12}, however it is not in absorption in the optical images. We refer to Sect. \ref{subsec:fitting} for the modelling of the dust absorption/emission in this galaxy.}.

We performed an isophotal fit for each target galaxy on the NGVS $i$ images, which are taken as a reference. The isophotes obtained from the fit are then used on each band to estimate the fluxes in radial bins. This is done by using isophotes in wcs coordinates and not in pixel, to take into account the different resolutions of the UV-band images.
We used the algorithm \texttt{ellipse} from \cite{ellipse} via the \texttt{Astropy} package dedicated to photometry of astronomical sources, \texttt{photutils} \citep{larry_bradley_2023_7946442}. Starting from the galaxy centre from the NGVS \citep{ferrarese12}, the algorithm \texttt{find\_center} estimates the centre for each galaxy. The centre of the ellipses is then kept fixed for the entire isophote fit. The position angle (PA) of the consecutive ellipses is left free to vary in the fit. As a test, we tried to keep the PA fixed, however this does not represent well galaxies such as NGC~4262 or NGC~4442, that show rotating isophotes. 
The starting guesses for the isophote fit parameters (PA and ellipticity) is taken by superimposing an ellipse region in \texttt{ds9} with semimajor axis (SMA) of 100 pixels on the NGVS $i$ images, for each galaxy. 
The isophote fit is then performed on consecutive ellipses of pre-defined SMAs. We carried out two different analyses based on the two different spatial resolutions that we have available in each UV band. As mentioned in Sect. \ref{sec:data}, UVIT/FUV images have a FWHM $\sim1.5''$, while GALEX/NUV images have a FWHM $\sim5.3''$. For the first type of analysis, we start fitting the ellipses from a SMA of $6''$ and define consecutive isophotes with a step of SMA$=8''$. Throughout the paper, we refer to this as the \textit{GALEX resolution} analysis, which will then form a dataset of broadband radial fluxes in the FUV, NUV and optical. For the second analysis, we start the ellipses with a SMA$=3''$ and define consecutive ellipses with a step of SMA$=3''$. This is defined as the \textit{UVIT resolution} analysis, forming a dataset which includes only the FUV and the optical broadband fluxes.

Figure \ref{fig:ell_uvit} shows a zoom-in of NGVS $i$ images centred on the target galaxies with the ellipses from the isophotal fit superimposed as red dashed curves, at UVIT resolution. Contours from the FUV emission (Fig. \ref{fig:uvit}) are superimposed in white in each panel. Overall, we observe that the FUV emission is quite smooth and consistent with the optical isophotes in PA and shape in each galaxy. Hence, no clumps or knots indicating ongoing massive SF are identified, consistently with what we found with VESTIGE data (Sect. \ref{subsec:ha_analy}). 
NGC~4262 shows clumps in emission in the FUV, see Fig. \ref{fig:uvit}, that are located west of the galaxy, spanning at least $\sim$150 arcsec on the sky - equivalent to $\sim$12 kpc at the distance of Virgo. In Appendix \ref{app:iso_galex}, Figure \ref{fig:ell_galex} shows the same NGVS $i$ images of the target galaxies at GALEX resolution. The observed clumps of NGC~4262 are not accompanied by optical emission. This was already reported in \cite{bettoni10}, as well as in \cite{boselli22}, see Sect. \ref{subsec:ha_analy} and Appendix \ref{app:notes_gals}.
Finally, we note that M86 has a FUV emission in the inner 20$''$ that is skewed towards North-East with respect to the optical isophotes. This difference in the isophotes FUV-optical was not spotted before, to the best of our knowledge. It is then possible that the FUV emission might be skewed because of the gravitational interaction with NGC~4438 (see Appendix \ref{app:notes_gals}, \citealt{kenney08, gomez10} and references therein). Alternatively, it might be due to the UV emission of the close dwarf galaxy VCC~882, located north-east of M86 within its optical halo \citep{elmegreen00}.

\subsection{Colour gradients}
\label{subsec:col_grad}

Fluxes and AB magnitudes were calculated within each concentric ellipse, for each band and each galaxy. They were corrected for the Galactic extinction using the extinction law from \cite{cardelli89} and the colour excess  $E(B-V)$ from \cite{schlafly11}\footnote{See \url{https://irsa.ipac.caltech.edu/applications/DUST/}}.
The background noise for each ellipse region is estimated by averaging the flux of 1000 regions (with the same shape of each original region), each one generated with a random position in the field of view of each image. Hereafter, we consider regions until the signal-to-noise ratio (SNR) is greater or equal than 3. 
\begin{figure}
\centering
\includegraphics[scale=0.49]{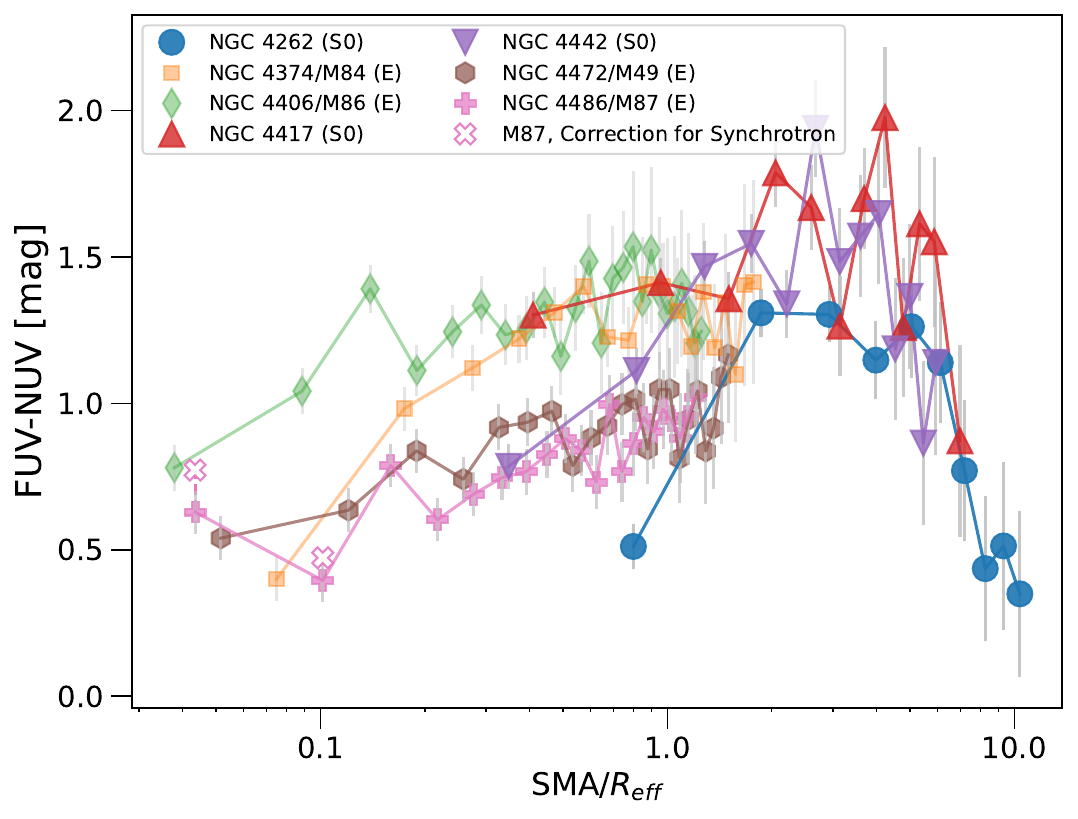}
\caption{$FUV-NUV$ colours at GALEX resolution as a function of the semimajor axis (SMA) divided by $R_{\rm eff}$. For M87, we corrected the fluxes in the inner 0.1 SMA/$R_{\rm eff}$ for synchrotron emission, see text for more details.} 
\label{fig:FUV_NUV}
\end{figure}
We adopted an error on the zeropoint of 0.053 mag for the FUV band (priv. comm. with UVIT technicians), and errors of 0.05, 0.07, 0.035 mag for the NUV \citep{morrissey07}, $u$ \citep{ferrarese12} and $griz$ bands (\citealt{ferrarese12} for NGVS, \citealt{boselli18} for VESTIGE), respectively.

\begin{figure*}
\includegraphics[scale=0.45]{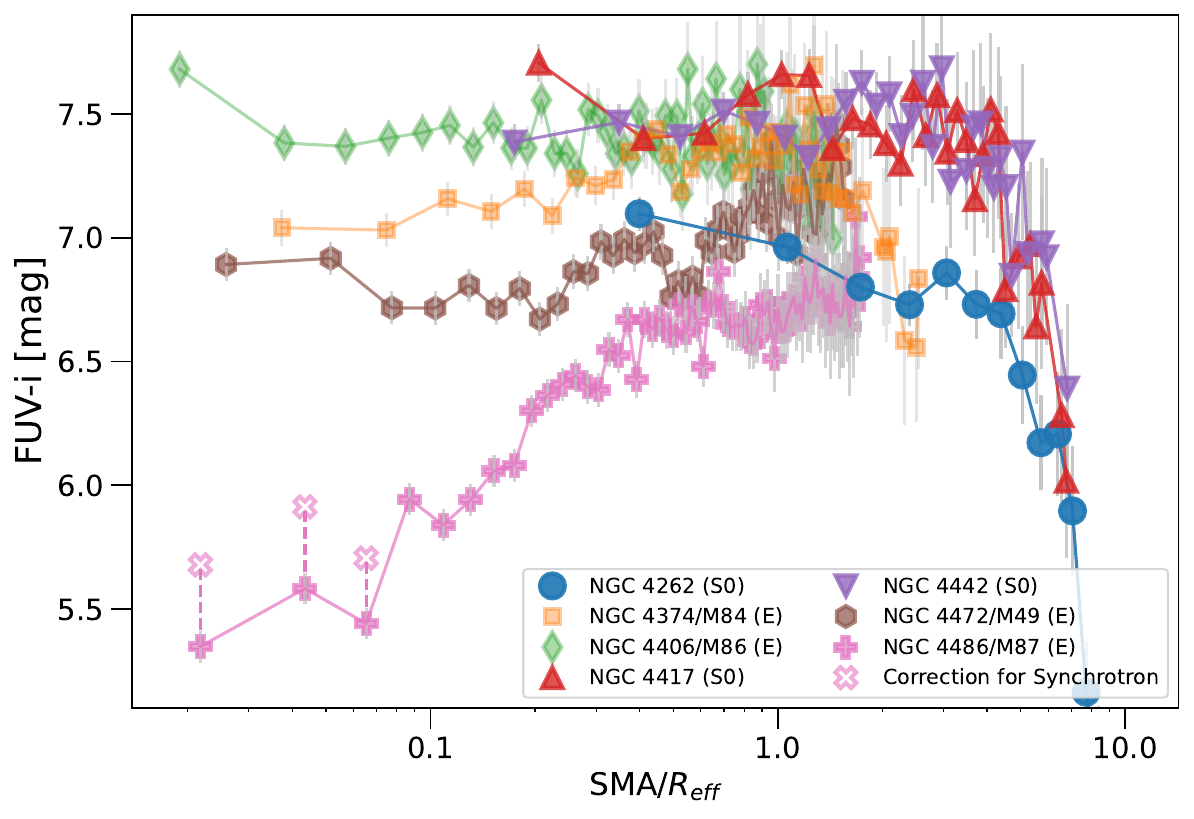}
\includegraphics[scale=0.45]{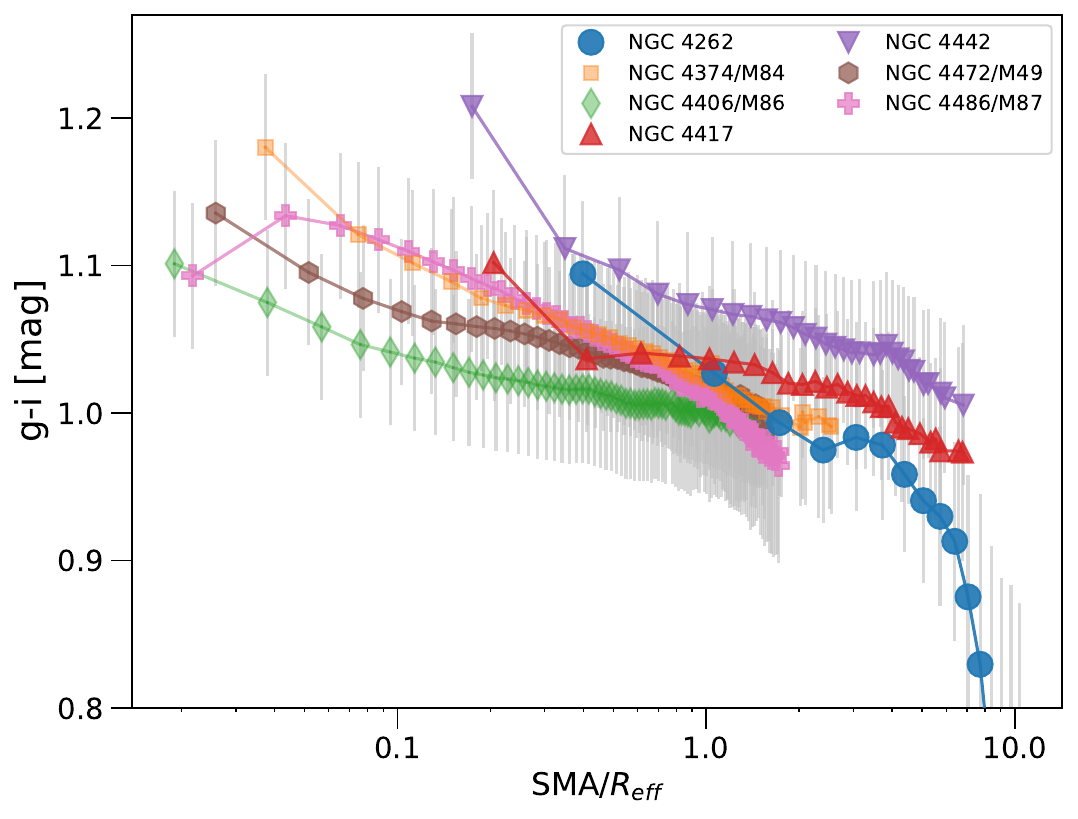}
\caption{$FUV-i$ (\textit{left panel})  and $g-i$ (\textit{right panel}) colours at UVIT resolution as a function of SMA/$R_{\rm eff}$.} 
\label{fig:FUV_i}
\end{figure*}
\begin{figure*}
\centering
\includegraphics[scale=0.57]{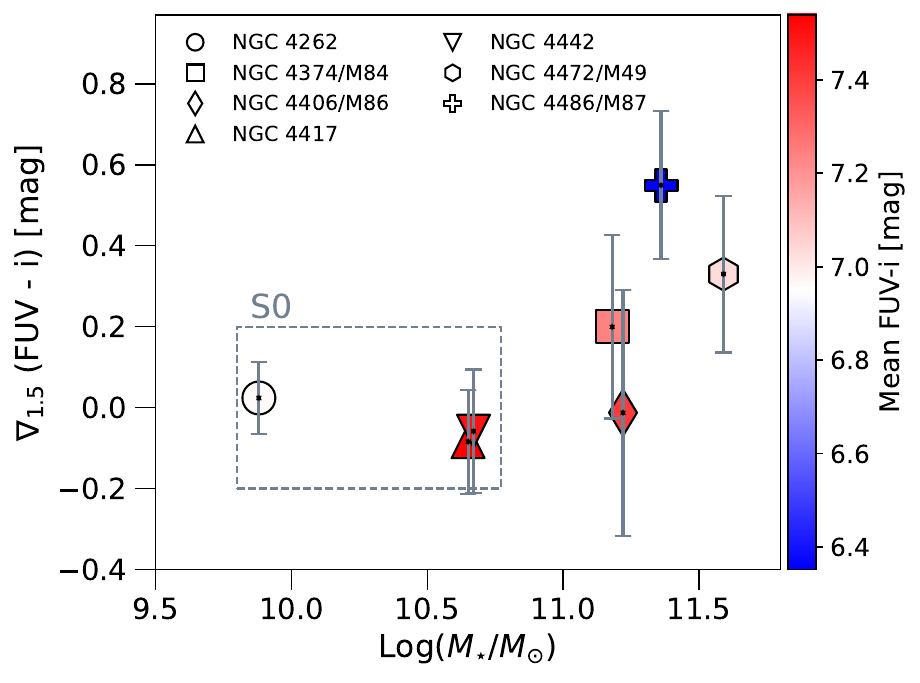}
\includegraphics[scale=0.57]{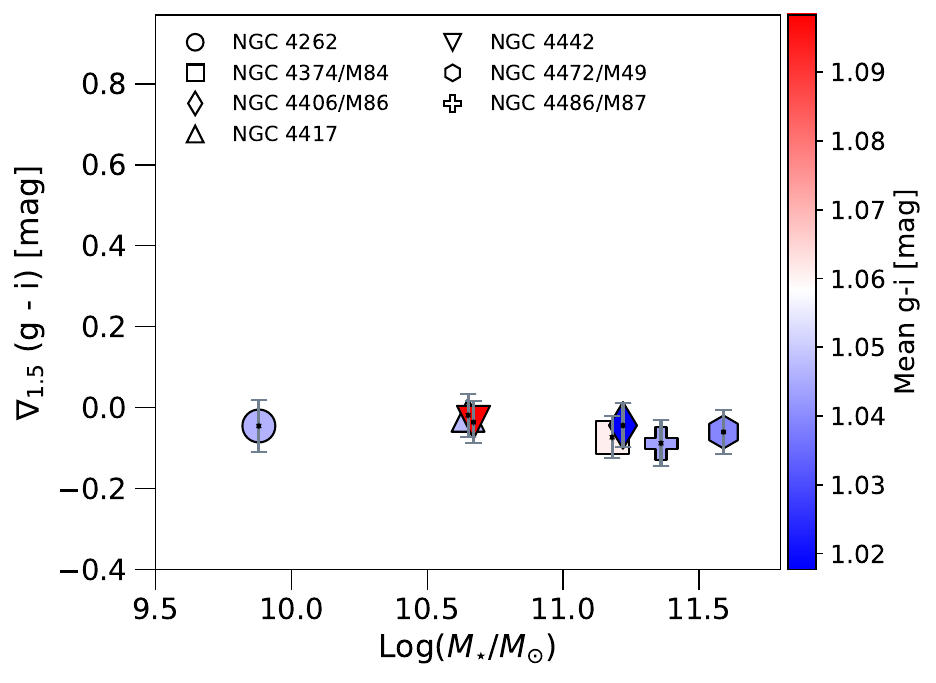}
\caption{Gradients in $FUV-i$ (\textit{left}) and $g-i$ (\textit{right}) colours as a function of the stellar mass at UVIT resolution, calculated until 1.5 SMA/$R_{\rm eff}$. Galaxies are colour-coded by their average colours at these SMAs. The grey dashed rectangle indicates the location of the S0 (lenticular) galaxies. Panels share the same y-axis. See text for more details. }
\label{fig:gradients_comparison}
\end{figure*}

Figure \ref{fig:FUV_NUV} shows the $FUV-NUV$ colours at GALEX resolution, as a function of the SMA divided by the effective radius $R_{\rm eff}$ (see Table \ref{tab:data}). This colour has been largely considered in the literature as an empirical indicator of the UV excess, because, when $FUV-NUV<1$ mag, the UV slope in the SED increases for decreasing wavelengths (e.g., \citealt{yi11}). However, this threshold depends on the assumed models and SFH, and it can occur that the UV emission from old stellar populations is also present when $FUV-NUV>1$ mag. Here we make general considerations on the colour trends, and we refer to later Sections (e.g., \ref{sec:sed}) for the identification of the UV upturn in these galaxies in comparison with the models presented in this work. 
All the galaxies in the sample are bluer in the centre with respect to their outskirts. NGC~4262 becomes very blue at SMA/$R_{\rm eff}>5$. This is due to the presence of the above mentioned HII regions in its outskirts (see Figs. \ref{fig:uvit} and \ref{fig:ell_galex}). These clumps were masked in the current analysis, however likely some residual emission in the FUV still remains.
M87, M49 and M84 are the bluest galaxies in the centre, at SMA/$R_{\rm eff}<0.2$, while the other elliptical M86 is overall redder\footnote{We note that the very blue colours in the centre of M84 are likely affected by its AGN activity, see Sect. \ref{subsubsec:dustsed}.}.
We note here that for the lenticulars (NGC~4262, 4417 and 4442), we are able to sample their outskirts down to SMA/$R_{\rm eff}\sim6-10$. For the ellipticals (M49, M84, M86 and M87) we sample their central regions, up to SMA/$R_{\rm eff}\sim0.8-1.2$ (GALEX) and SMA/$R_{\rm eff}\sim1.5-2.5$ (UVIT). This is due to the reached SNR in the FUV and especially in the NUV band. 

As reported in \cite{smithrussell12}, the $FUV-NUV$ colour is not the most suitable to interpret the origin of the UV upturn, as it is sensitive to different effects, among which main sequence turn-off (MSTO) stars, the strength of the upturn populations and the spectral slope of the upturn. This is mostly due to the NUV regime, which has some contribution due to a strong metal line blanketing from MSTO stars (e.g., \citealt{donas07}). 
On the contrary, the $FUV-i$ colour represents a pure comparison among hot old stars (strong emitters in the FUV) and old cool stars, emitters in the optical red bands. 
Figure \ref{fig:FUV_i} shows the $FUV-i$ colours as a function of the SMA at UVIT resolution (left panel). We note that M84, M87 and M49 (Es) are bluer in the centre than in the outskirts (SMA/$R_{\rm eff}\lesssim1$). This behaviour was already reported in \cite{ohl98} for M87 and M49 in the $FUV-B$ colours. We also note that on average the bluest elliptical galaxy is M87, followed by M49 and then M84. NGC~4262 is again blue in the centre at SMA/$R_{\rm eff}\sim0.8$ as M49, becoming bluer also in its outskirts. NGC~4417, NGC~4442 and M86 are the reddest galaxies in the sample.

The right panel of Figure \ref{fig:FUV_i} shows the $g-i$ colours as a function of SMA/$R_{\rm eff}$ at UVIT resolution. Very differently from the FUV colours, the optical gradients of the target galaxies are all very similar and span a small range in colours (except for NGC~4262). This is consistent with 
what was reported in the literature (e.g. \citealt{peletier90, ohl98}) and shows that the UV colours are crucial to disentangle star formation histories in different galaxies.

Figure \ref{fig:gradients_comparison} shows a comparison between the $FUV-i$ (left panel) and $g-i$ (right panel) colour gradients as a function of the galaxy stellar mass, at UVIT resolution. The colour gradient $\nabla$ (col) was defined as:
\begin{equation}
\nabla {\rm (col)} = \frac{{\rm col}_2 - {\rm col}_1}{\mid R_2-R_1 \mid},  
\end{equation}
where col$_1$ and col$_2$ represent the considered colours at $R_1$ and $R_2$, respectively. We used $R_{2}={\rm SMA}_2/R_{\rm eff}=1.5$ for all galaxies, while we used $R_{1}= {\rm SMA}_1/R_{\rm eff}=0.1$ for the ellipticals, $R_1=0.2$ for NGC~4417 and NGC~4442, $R_1=0.4$ for NGC~4262 based on their colour profiles at UVIT resolution (see Fig. \ref{fig:FUV_i}).
The very central regions were excluded to avoid contamination from the AGN in M87 and M84. 
Both panels are colour-coded by the average colour of each galaxy at these radii. 
\begin{figure}
\centering
\includegraphics[scale=0.59]{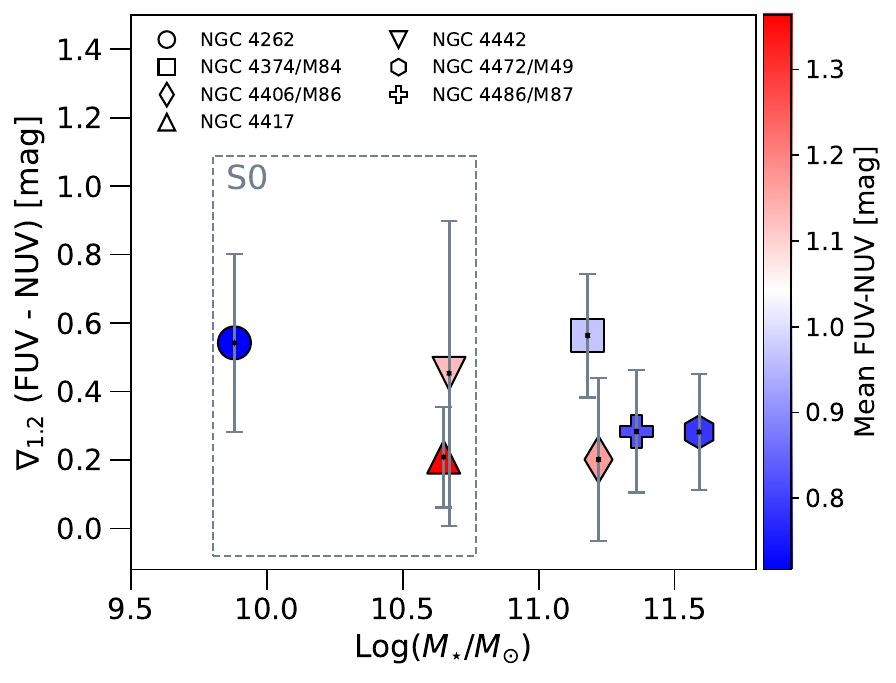}
\caption{Gradients in $FUV-NUV$ at GALEX resolution as a function of the stellar mass, calculated until 1.2 SMA/$R_{\rm eff}$. Galaxies are colour-coded by their average colours at these SMAs. The grey dashed rectangle indicates the location of the S0 (lenticular) galaxies. See text for more details. }
\label{fig:gradients_UV}
\end{figure}
The colours as a function of projected radii (see Figs. \ref{fig:FUV_NUV}, \ref{fig:FUV_i}) were interpolated with a third degree polynomial spline in order to calculate the colour value precisely at $R_{1}$ and $R_{2}$.
\begin{figure*}
\centering
\includegraphics[scale=0.49]{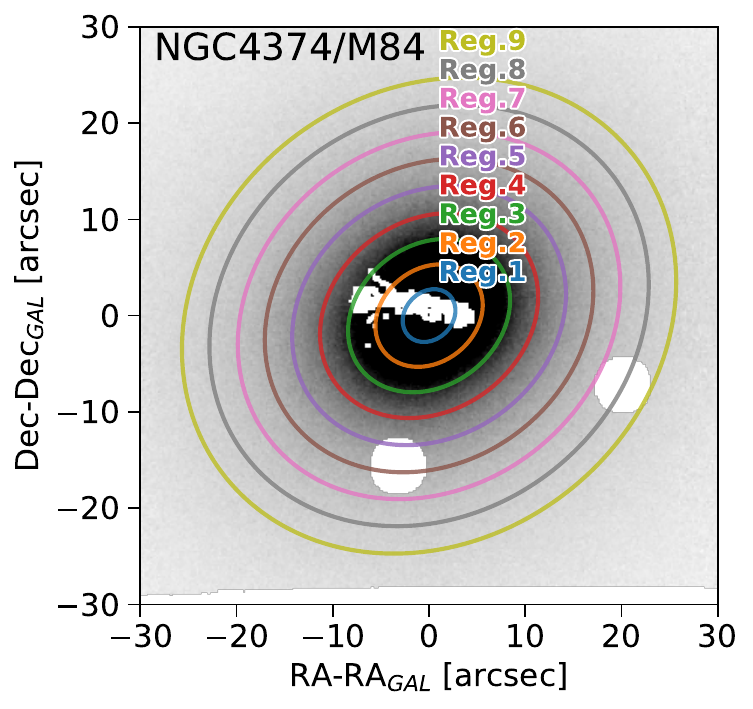}
\includegraphics[scale=0.49]{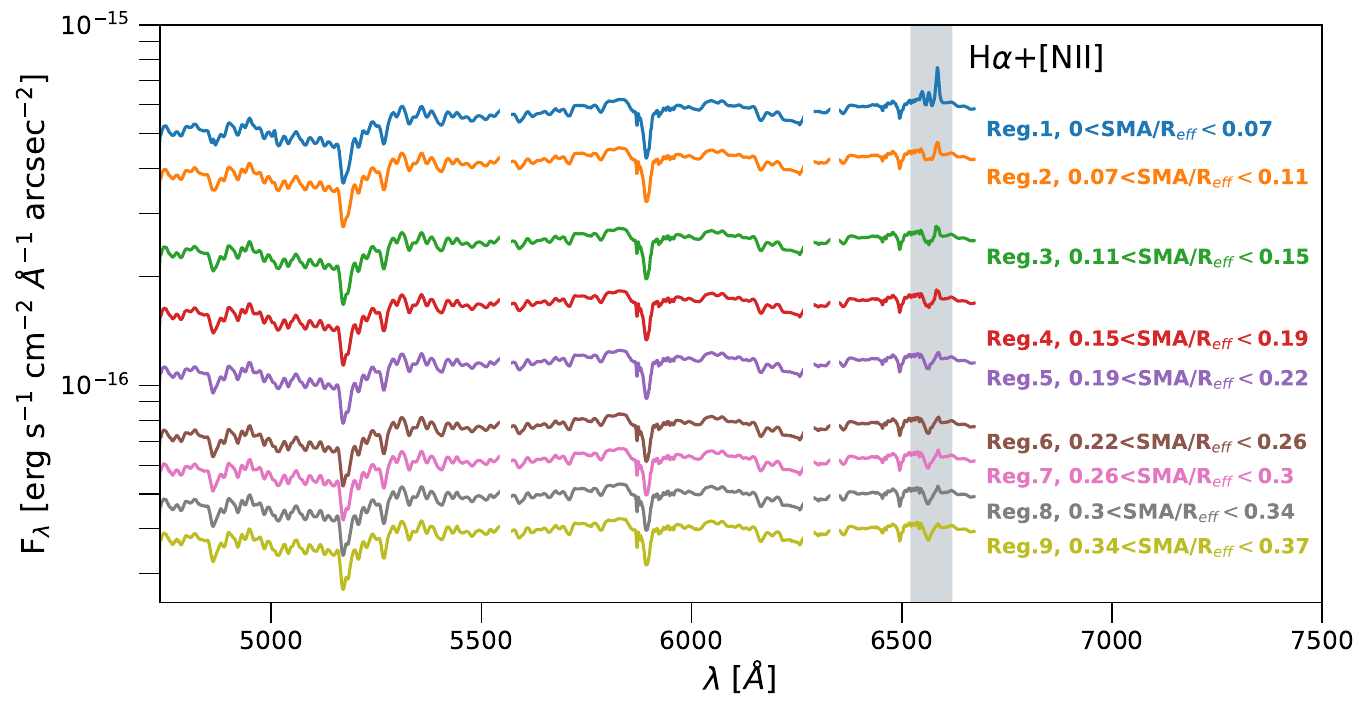}
\caption{\textit{Left Panel:} Slice of MUSE cube at $\lambda\sim5400$\AA\, for M84. The white spaxels indicate masked regions either representing dust or background/foreground sources. The superimposed coloured ellipses are the isophotes defined from the $i$ images. The number of the regions is indicated. \textit{Right Panel:} MUSE spectra representing the sum of the flux for each elliptical annulus from the left panel. The grey shaded areas indicate the H$\alpha$ + [NII] lines, while regions due to sky
and telluric residuals are omitted. See text for more details.} 
\label{fig:muse}
\end{figure*}
Note that the two panels of Fig. \ref{fig:gradients_comparison} are shown with the same limits in the y-axis. As mentioned before, gradients in the $FUV-i$ colours span a much wider range with respect to the gradients in the optical $g-i$ colours. Additionally, the mean of FUV colours spans a wider range with respect to the mean optical colours, see the colour bars. 

The left panel of Fig. \ref{fig:gradients_comparison} shows that, overall, the more massive galaxies have stronger $FUV-i$ gradients. We performed a Spearman rank's test and this gives a rank coefficient $\rho_s = 0.68$ and a probability of deviation from a random distribution $d_s \sim 9\%$. This indicates a positive correlation between $\nabla$($FUV-i$) and Log($M_{\star}/M{_\odot}$), however it is not significant. 
Indeed, we note that we are presenting results for only seven galaxies and thus it is necessary to expand the sample to robustly confirm the existence of such a correlation. 
The dashed grey rectangular shape in the left panel of Fig. \ref{fig:gradients_comparison} indicates the position of the lenticulars/S0 sample. We found that the ellipticals in our sample show overall larger gradients in $FUV-i$ than the lenticulars, and are on average bluer, as shown by the average $FUV-i$ colours on the colour bar. This is true except for NGC~4262 which is instead quite blue, with a mean $FUV-i$$\sim$6.9 mag.

Figure \ref{fig:gradients_UV} shows $\nabla$($FUV-NUV$) for SMA/$R_{\rm eff}$ up to 1.2 at GALEX resolution. The gradient was calculated for $R_1=0.1$ for all galaxies, except $R_1=0.4$ for NGC~4417 and NGC~4442 and $R_1=0.8$ for NGC~4262 (see Fig. \ref{fig:FUV_NUV}). We note that overall the gradients in $FUV-NUV$ are positive and steep, with M84 having the steeper gradient among the E galaxies and NGC~4262 among the S0 galaxies. We found no correlation among the $\nabla$ in $FUV-NUV$ colours at these radii and stellar masses.

\subsection{MUSE analysis}
\label{subsec:muse_analysis}

We used the same regions as in Sect. \ref{subsec:iso_fit} to mask the unwanted foreground and background sources in the MUSE IFU images for M49, M84, M87. We also masked the dust and the M87 jet as in Sect. \ref{subsec:synchr} and \ref{subsec:iso_fit}.
Next, we superimposed the same elliptical isophotes that we used for fitting the NGVS $i$ images (Figs. \ref{fig:ell_uvit} and \ref{fig:ell_galex}) on the MUSE cubes. The left panel of Fig. \ref{fig:muse} shows the MUSE cube of M84 at $\lambda\sim5400$\AA\,, where the coloured ellipses indicate the isopohotes from the fit, at UVIT resolution. They are labeled as numerated regions. We summed the flux in each elliptical annulus as a function of wavelength to obtain radial spectra. The right panel of Fig. \ref{fig:muse} shows the obtained radial MUSE spectra for M84 normalised by the area of the ellipse, at UVIT resolution. We performed this process for each galaxy and for the GALEX resolution as well. The same Figures as Fig. \ref{fig:muse} but for M49 and M87 are reported in the Appendix \ref{app:muse_images}.

The wavelength region of the H$\alpha$ + [NII] doublet emission lines is indicated as a grey shaded area. We used the MUSE data to select the regions that are contaminated by the AGN activity. We removed these regions in the SED fitting analysis, see Section \ref{sec:sed}.

We also note that the MUSE spectra allows for the calculation of line index strengths of several absorption lines that are sensitive to galaxy properties such as stellar ages and metallicities (e.g., \citealt{worthey94, thomas03, thomas11, worthey14}). However, the stellar models computed in this pilot work (see Sect. \ref{subsec:mod}) do not contain specific abundance ratios for these galaxies, i.e. they are solar-scaled. Hence, a comparison between model and observed line indices would not be meaningful. In a future work, we plan on updating the models by including typical abundance ratios and enhanced $\alpha$ elements in order to fit the photometry together with absorption lines.
We then postpone the analysis of the EWs to a forthcoming paper.

\begin{figure*}
\centering
\includegraphics[scale=0.54]{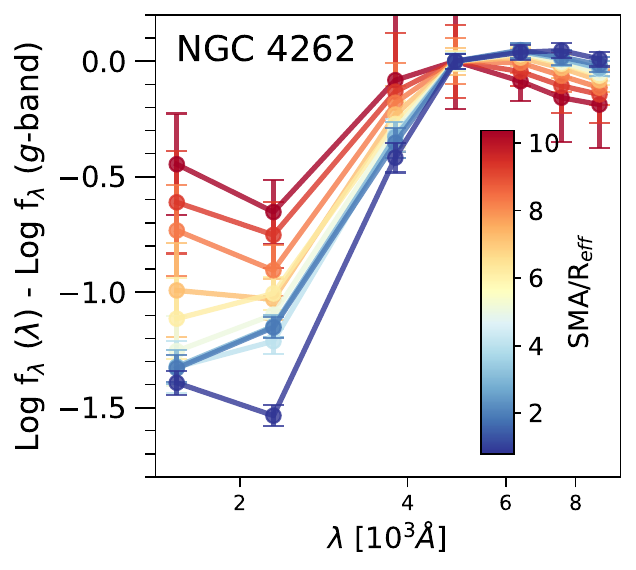}
\includegraphics[scale=0.54]{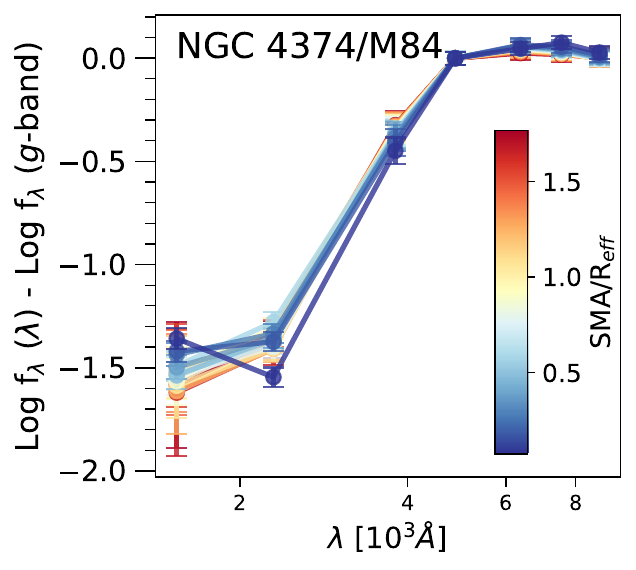}
\includegraphics[scale=0.54]{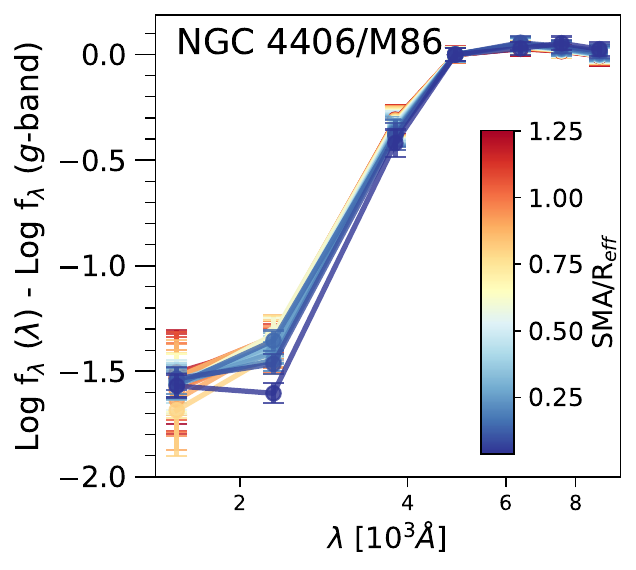}
\includegraphics[scale=0.54]{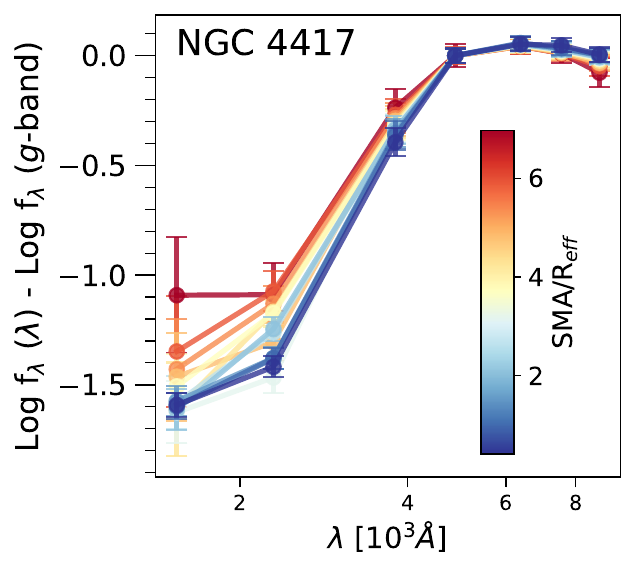}
\includegraphics[scale=0.54]{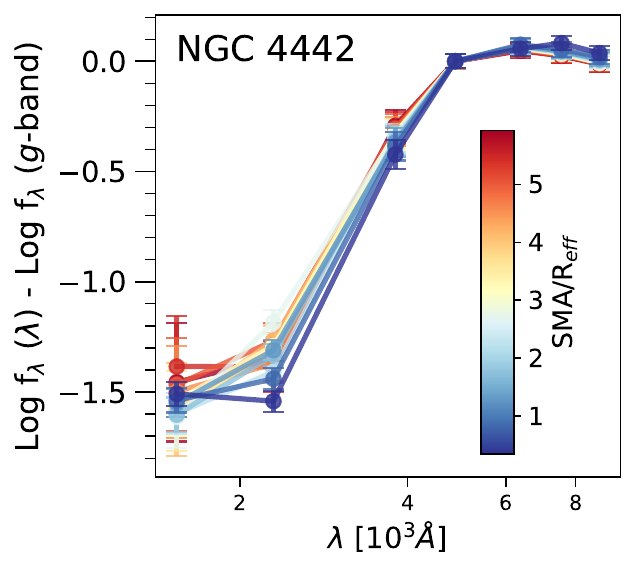}
\includegraphics[scale=0.54]{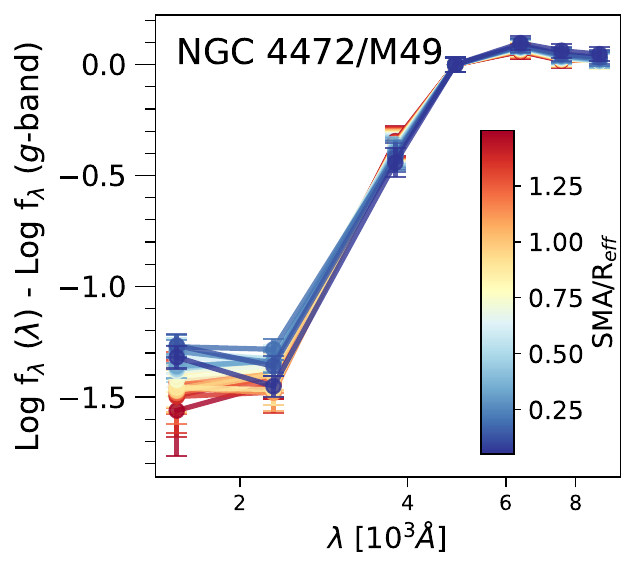}
\includegraphics[scale=0.54]{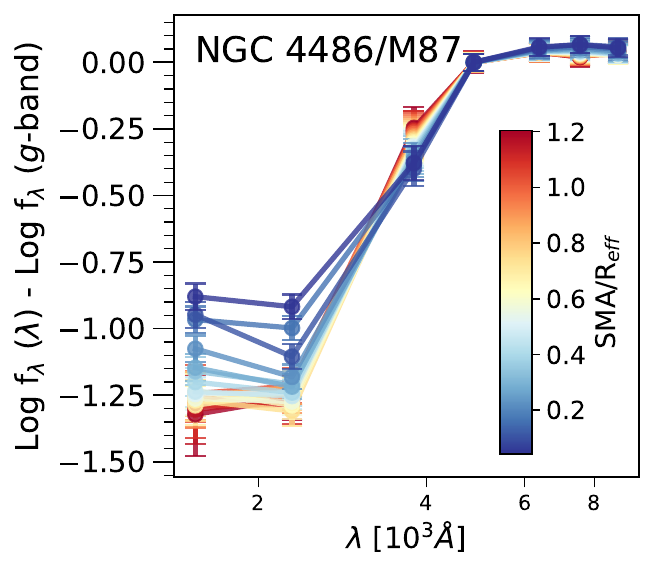}
\caption{SEDs normalised to the flux in the $g-$band, at GALEX resolution, colour-coded to the ratio SMA/$R_{\rm eff}$.} 
\label{fig:seds}
\end{figure*}

\begin{figure}
\centering
\includegraphics[scale=0.52]{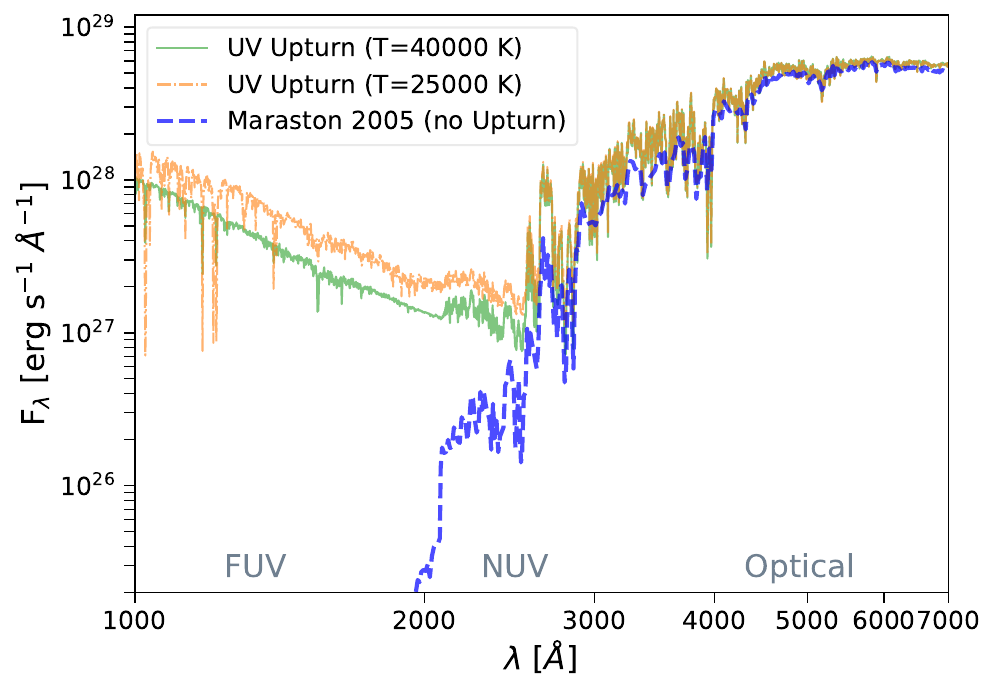}
\caption{Comparison between SSP models with an age of 13~Gyr, solar metallicity ($Z=0.02$) and Salpeter IMF. The green solid (orange dash-dotted) line indicates a model with the presence of UV upturn component at temperature T$=40000 (25000)$~K and low fuel (this work and \citealt{lecras16}), while the blue dashed line represents the SSP model without an upturn \citep{maraston05}. See text for more details.} 
\label{fig:models_ex}
\end{figure}

\section{Spectral Energy Distribution fitting}
\label{sec:sed}

We constructed the SEDs as a function of the SMAs for each galaxy and for both resolutions (GALEX and UVIT).
Each panel in Figure \ref{fig:seds} reports the radial SEDs for a target galaxy, at GALEX resolution. The SEDs are shown normalised to the $g-$band and are colour-coded by their projected distance (SMA$/R_{\rm eff}$). The normalisation is chosen to better visualise the UV upturn, as done in \cite{burstein88}. 
The SEDs show that all the galaxies in the sample have an increasing slope with decreasing wavelength in the 1500-2500\r{A} region at least in their central region, represented by the bluest lines. The only exception is NGC~4417. 
As soon as we move further out from the centre, the SEDs in the UV region flatten out for M84, M86 and M49. The two lenticulars NGC~4262 and NGC~4417 become bluer in the outskirts as previously shown.

\subsection{Stellar Population Models}
\label{subsec:mod}

To estimate the SFH and radial properties of our galaxies, we computed new single stellar population (SSP) model spectra, which include the contribution of the UV emission from evolved stars. 
The models are computed with the evolutionary population synthesis (EPS) code from \cite{maraston98, maraston05}. The code takes into account theoretical recipes, such as the fuel consumption theorem from \cite{renzini86}, to calculate the post-main sequence (PMS) phases of the stellar populations 
(see \citealt{maraston05}). The models specifically presented in this work were either published already in \cite{lecras16} or newly computed.
They include a UV upturn in the form of an old, hot stellar component at certain defined temperatures, as in \cite{mt00}, which simulates the evolution of PMS stars. This is not modeled using stellar tracks, thus the UV emission might come from hot horizontal branch stars as well as from post asymptotic giant branch (AGB), AGB-manqu\'e, and other PMS evolutionary phases.
The models span an age range of 1 Myr$-13.7$ Gyr, with the upturn starting at an age of 3 Gyr. We computed the models for 3 metallicities, at $Z=0.01$ (half-solar), 0.02 (solar), 0.04 (twice solar). Furthermore, models for the upturn are usually computed as a function of two parameters: the temperature of the old UV components and the fuel consumption at these temperatures. The fuel represents the amount of hydrogen and/or helium burned during the PMS phases. We refer the interested reader to \cite{maraston05} for more details. 
We computed models with two values of fuel, as in \cite{lecras16}. A high fuel value, $f_{\rm H}= 6.5 \times 10^{-2}$ \msun , and a low fuel $f_{\rm L}= 6.5 \times 10^{-3}$ \msun. The SED shape of our galaxies is compatible to a low fuel value, consistently with previous results in local galaxies (e.g., \citealt{mt00}). 
Each model is calculated with a temperature of 25000, 35000 and 40000~K. For $Z=0.01$ and 0.02, a temperature of 30000~K is also available.
We implemented these models in the public available SED fitting code \texttt{CIGALE} \citep{boquien19} under the new SSP module named \texttt{uvupturn}. In the same module, a model without upturn (defined as $T=0~$K) is also available for comparison. See Sect. \ref{subsec:cigale} for more details on the implementation. 
 
Figure \ref{fig:models_ex} shows the difference between a standard \cite{maraston05} SSP model without an UV upturn component (indicated with a blue dashed line) versus two UV upturn models at different temperatures. The models are shown with the same parameters, namely an age of $13$~Gyr, solar metallicity ($Z=0.02$) and a Salpeter IMF \citep{salpeter55}. However, it is remarkable how they differ at $\lambda<3000$\r{A}, where the upturn models show a powerful emission in the UV.

\subsection{New SSPs in CIGALE: the \texttt{uvupturn} module}
\label{subsec:cigale}

We used the SED fitting code \texttt{CIGALE} (Code Investigating GALaxy Evolution, e.g., \citealt{burgarella05, noll09, boquien19}), a versatile \texttt{python} code to fit the SED of galaxies in a modular and efficient way. Composite stellar populations are generated through the convolution of SSPs with parametrisations of the SFHs of galaxies. Nebular emission lines and dust attenuation laws, as well as several AGN emission laws are available. 
The advantage of \texttt{CIGALE} is the possibility to implement novel models in an easy and fast way thanks to its modular architecture. Specifically, in this work we implemented novel stellar population models from C. Maraston that include the contribution of the UV upturn (see Sect. \ref{subsec:mod}). We named the new SSP module \texttt{uvupturn}. The models have an original resolution of $R=40000$ which was downgraded to a $R=1000$ to allow a better handling of the memory space. 
As the new models span a limited wavelength range from 1000$-$7000\AA, they were merged with the already available SSP models from \cite{maraston05} for wavelengths longer than 7000\AA\, (module \texttt{m2005} in \texttt{CIGALE}, see \citealt{boquien19} for more details).

\subsection{Adopted CIGALE modules and parameters}
\label{subsec:sed_modules}

We fit each radial SED for each galaxy with a SFH modeled ``a la Sandage'' \citep{gavazzi02}, according to the \texttt{sfhdelayed} module from CIGALE, i.e. SFR$(t)\propto t/\tau^2 \times \exp(-t/\tau)$, where $t$ is the SFH range in Myr and $\tau$ the e-folding timescale of SF in Myr.
Regarding the SSP models, we used the new implemented models \texttt{uvupturn}. They are calculated with a Salpeter IMF \citep{salpeter55}. 
The models have a powerful UV emission which is degenerate with the emission produced by recent episodes of SF and young stars. For this reason, we used the VESTIGE H$\alpha$ images (see Fig. \ref{fig:ha}) to measure an upper limit on the star formation rate (SFR) in each galaxy from their H$\alpha$ emission and use it as a constraint for the SED fitting. In \texttt{CIGALE} it is indeed possible to include rest-frame galaxy properties whose values will be fitted together with the broadband photometric fluxes. As discussed in Sect. \ref{subsec:ha_analy}, the analysis of the VESTIGE images reveals that the H$\alpha$ emission in our galaxies is either lacking or it is of filamentary nature, thus it is not due to recent SF. 
For each galaxy, we estimated the surface brightness limit in H$\alpha$ by taking the standard deviation of the flux of 1000 regions (with the same shape as the original region), each one generated with a random position in the field of view. This corresponds to $\Sigma_{H\alpha}\sim (2-4) \times 10^{-18}$ erg s$^{-1}$ cm$^{-2}$ arcsec$^{-2}$ at 1$\sigma$ depending on the galaxy and region. We converted the H$\alpha$ flux at 5$\sigma$ into SFRs \citep{kennicutt98} and used these values as constraints in upper limit for the SED fitting.


\subsubsection{Dust modelling and AGN contamination}
\label{subsubsec:dustsed}

As mentioned in Sect. \ref{sec:data}, M49, M84, M86 and M87 contain dust. Even though the dust seen in absorption in the images was masked, it is important to parametrise the dust within the SED fitting, as it might affect the shape of the SED.
\begin{figure*}
 \centering
    \includegraphics[scale=0.33]{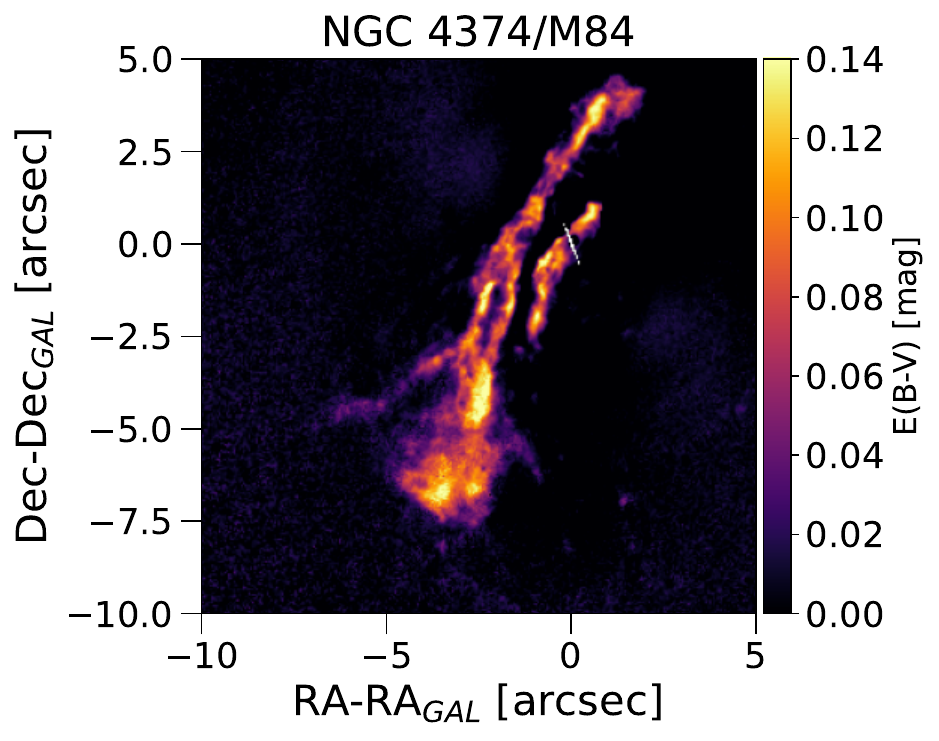}
    \includegraphics[scale=0.33]{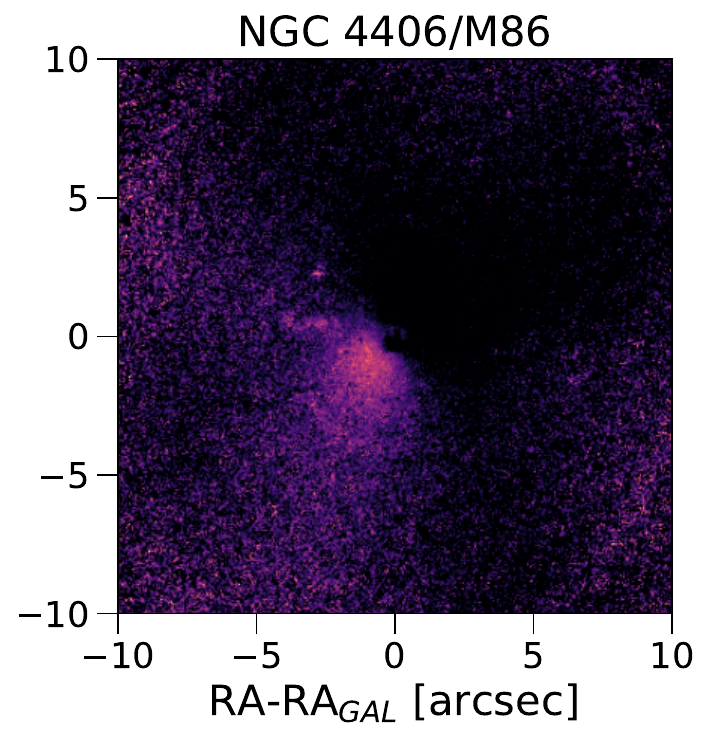}
    \includegraphics[scale=0.33]{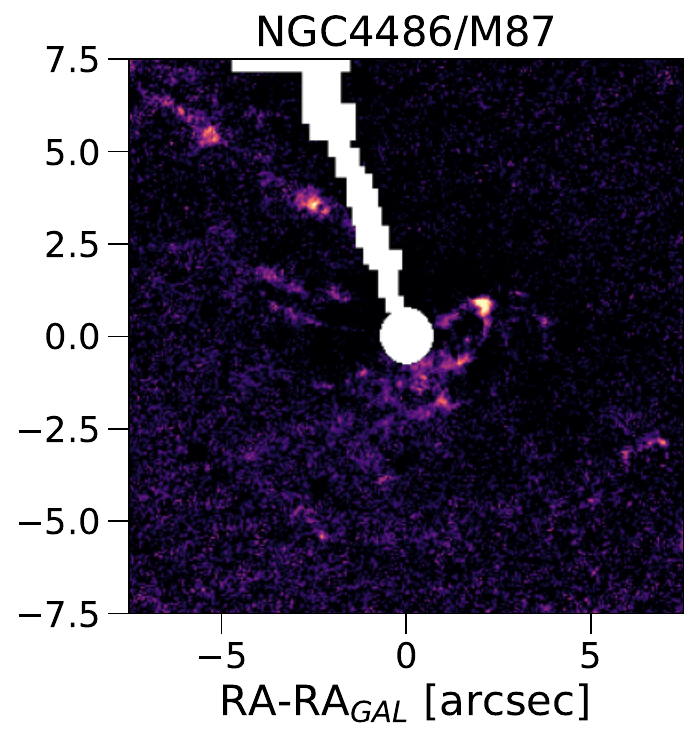}
    \includegraphics[scale=0.33]{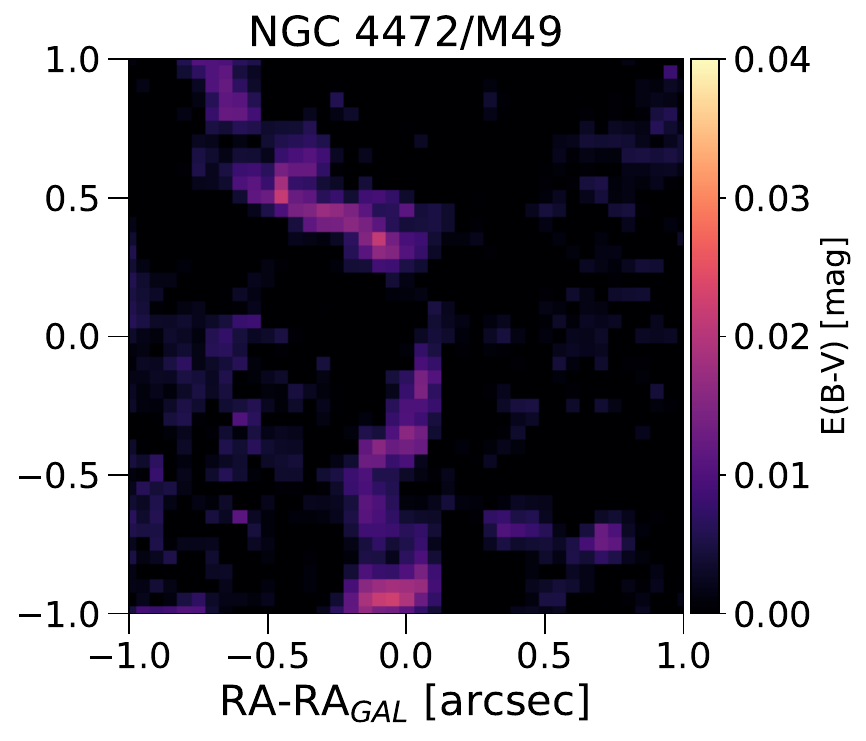}
\caption{$E(B-V)$ extinction maps for M84, M86, M87 and M49 (from left to right), constructed from the HST ACS F475W images. North is to the left and east is down. M86, M87 and M49 share the same colourbar. Note that M84 has a more extended colour bar with respect to M86, M87 and M49.}
\label{fig:dust}
\end{figure*}
\begin{table*}
\centering
\caption{CIGALE adopted parameters and properties to fit the SEDs of the target galaxies.}
\begin{tabular}{c l c l}
\hline
Parameter & Value & Units & Description \\
\hline
\multicolumn{4}{c}{SFH: \texttt{sfhdelayed} module - All galaxies} \\ 
\\
$t$ & 0$-$13500  & Myr & SFH range\\
$\tau$ & 100, 300, 500, 800, 1000, 1200 & Myr & e-folding timescale of SF\\
& 1500, 2000, 2500, 3000, 4000, 5000 & & \\
\hline
\multicolumn{4}{c}{SSPs: \texttt{uvupturn} module - All galaxies} \\
\\
$Z$ & 0.01, 0.02, 0.04 & & Metallicity\\
$T$ & 0 (no Upturn), 25000, 35000, 40000 & K & Temperature of the upturn component\\
$f$ & 0 & & Fuel parameter. 0: low fuel, 1: high fuel. \\
 & & &See Sect. \ref{subsec:mod}.\\
\hline
\multicolumn{4}{c}{Dust Attenuation: \texttt{dustatt\_modified\_starburst} module - For M84, M86, M87} \\
\\
$E(B-V)_{\rm young}$ & 0, 0.005, 0.01, 0.02, 0.05, 0.1, 0.3 & mag & Stellar colour excess of the young populations.\\
$E(B-V)_{\rm old}$ & 0, 0.0025, 0.005, 0.01, 0.025, 0.05, 0.15 & mag & Stellar colour excess of the old populations.\\
$E(B-V)$ factor &  0.5 & mag & Conversion factor between young and old.\\
uv\_bump\_amplitude &  3 &  & Amplitude of the UV bump\\
powerlaw\_slope & $-0.25$  & & Slope $\delta$ of the attenuation powerlaw\\
\hline
\multicolumn{4}{c}{Properties - All galaxies} \\
sfh.sfr & from VESTIGE images & $M_{\sun}/{\rm yr}$ & 5$\sigma$ upper limit on star formation rate.\\
\hline
\end{tabular}
\label{tab:cigale_p}
\end{table*}
To get an estimation of the stellar colour excess $E(B-V)$, we constructed a 2D model of the stellar distribution for each galaxy in the sample. This was performed on the F475W high-quality HST images, by using a similar isophote analysis as presented in Sect. \ref{subsec:iso_fit}. Next, we subtracted the stellar 2D model from the original image and we estimated the A$_{F475W}$ as in \cite{kulkarni14} and \cite{boselli22}. Finally, we transformed the A$_{F475W}$ into $E(B-V)$ by using the transformation on the dedicated HST webpages and a $R_V=3.1$. We report the obtained extinction maps for M49, M84, M86 and M87 in Figure \ref{fig:dust}. As already mentioned, NGC~4262, NGC~4417, NGC~4472 do not have any clear dust structures. In NGC~4472/M49, a very faint dust filament is visible in a 1 arcsec$^2$ region near the centre. In M86, the dust structure is diffuse and not filamentary, barely visible. It is likely that this is instead due to some residuals in the stellar continuum subtraction. In M87 and M84 the dust lanes are clearly visible, with a maximum $E(B-V)$ of $\sim$0.06 and $\sim$0.15 mag, respectively. 

The dust distribution and emission of our target galaxies have additionally been investigated in the mid- and far- infrared (MIR, FIR), within the Herschel Reference Survey (HRS) by \cite{ciesla12, smith12, cortese14}. Such authors showed that NGC~4262, NGC~4417, NGC~4442 and M49 are not detected by $Herschel$ SPIRE and PACS in the FIR ($100-500 \mu m$), therefore their dust attenuation and emission is negligible. 
This, combined with the fact that their dust absorption in the HST and NGVS images is either very faint (Fig. \ref{fig:dust} ) or absent, led us not to include dust attenuation modules in the SED fitting for these galaxies\footnote{As a test, we also performed a SED fitting on M49 and NGC~4262 including a dust attenuation module. We chose NGC~4262 as representative for the lenticular sample. When including dust, the fit finds small $E(B-V)<0.05$ mag, only in the central parts of the galaxies. While the single values of the stellar parameters in a specific region might change, the overall results of the paper stay unchanged.}. For M86, M84 and M87, we instead use the \texttt{dustatt\_modified\_starburst} module, based on the law by \cite{calzetti00}. Modelling the attenuation in galaxies is challenging, as a wide diversity in the shape of extinction laws is usually observed (e.g., \citealt{salim20}). To understand which parameters better model the dust emission in these galaxies, we performed a SED fitting with \texttt{CIGALE} on the total SED of M86, from the FUV to the FIR. We chose M86 as a test galaxy, as the SEDs of M84 and M87 in the IR is contaminated by the jet and syncrothron emission. We used the already published flux values from \cite{cortese12, ciesla12, ciesla14} in the IR, from \textit{Spitzer}/IRAC at 8$\mu m$, WISE at $12\mu m$ and $22\mu m$, \textit{Herschel}/PACS at 100 and 160$\mu m$, and \textit{Herschel}/SPIRE at 250, 350 and 500$\mu m$. For the FUV, NUV, $ugriz$ bands we used the images presented in this work and we extracted the flux as in Sect. \ref{subsec:col_grad} in a similar aperture as those reported in \cite{ciesla12, cortese12, ciesla14}. We used the \texttt{sfhdelayed} and \texttt{uvupturn} modules for the SFH and stellar populations, while we used the \texttt{dustatt\_modified\_starburst} and \texttt{themis} modules for the dust absorption and emission, respectively. Regarding the dust absorption/emission, we used the same parameter space as reported in \cite{nersesian19}. We found a very small best-fit $E(B-V)=0.0025$ mag for M86, confirming that the emission we observed in Fig. \ref{fig:dust} is likely due to some residuals in the stellar continuum subtraction. The attenuation law that best describes the total SED of M86 has a powerlaw slope $\delta=-0.25$, mildly steeper with respect to a typical MW slope (e.g., \citealt{salim18, trayford20}). The amplitude of the UV bump at 2175\AA\, that we found is instead MW-like, $\sim$3. 
\begin{figure*}
    \centering
    \includegraphics[scale=0.48]{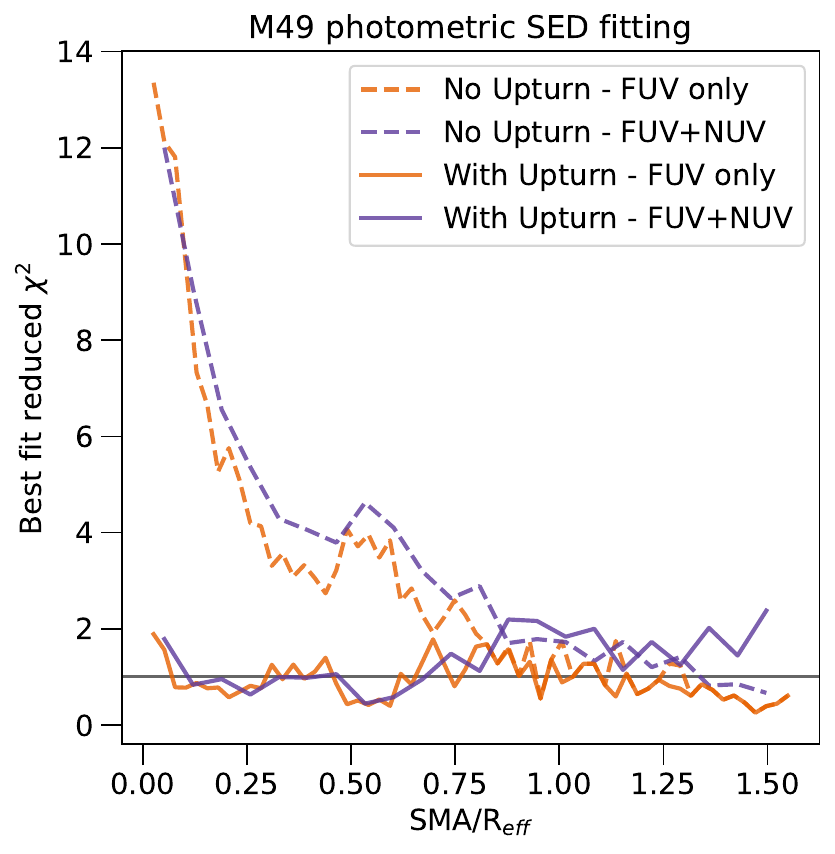}
    \includegraphics[scale=0.48]{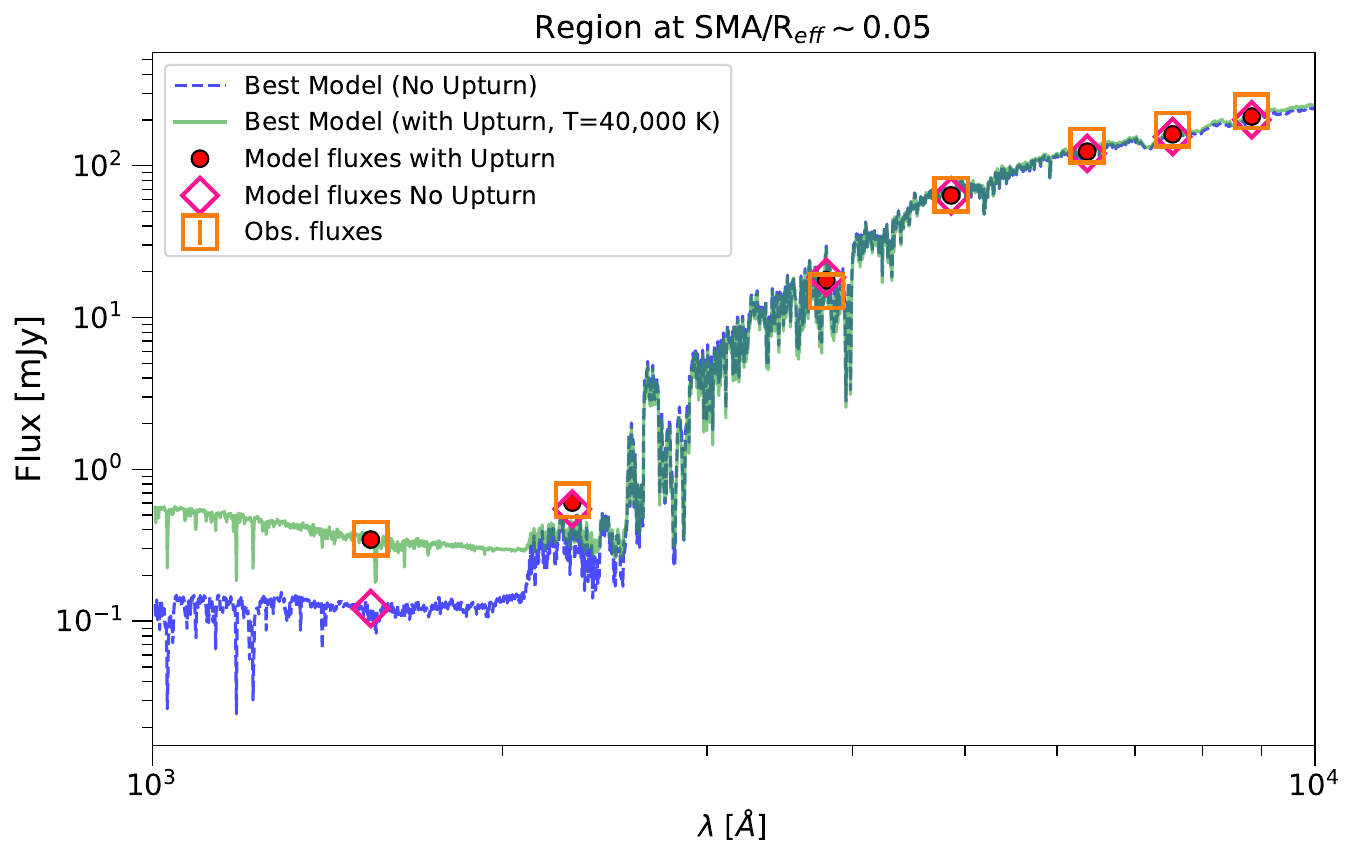}
    \caption{\textit{Left Panel}: Reduced $\chi^2$ values from \texttt{CIGALE} best fit as a function of SMA/$R_{\rm eff}$ for M49. Solid (dashed) lines indicate fits performed with (without) an UV upturn component, while orange (purple) lines represent fits performed on FUV+optical (FUV+NUV+optical) broadband data. The grey horizontal solid line marks a reduced $\chi^2=1$. \textit{Right Panel}: M49 photometric FUV+NUV+optical SED fitting. Open orange squares are the observed fluxes for the region at SMA/R$_{\rm eff}\sim 0.05$. The blue dashed line is the best model found by CIGALE when the data are fitted with no Upturn models, while the green solid line is the best fit by considering the upturn component. Red filled circles and pink open diamonds are the model fluxes for the upturn and no-upturn models, respectively. }
    \label{fig:test_chi}
\end{figure*}
To check whether our SED fitting results for M86 make sense, we compared our estimated dust mass with that estimated by \cite{diserego13}, which use a black-body fitting technique. We found $M_{\rm Dust, M86}\simeq 3.6\pm 2.0 \times 10^{6}$ \msun, while they obtained $M_{\rm Dust, M86}\simeq 2.8 \times 10^{6}$ \msun (considering the central and south east emission), consistent with our estimate.
To fit the \textit{radial} SEDs of M84, M86 and M87, we then employed such values for the bump amplitude and powerlaw slope\footnote{We nonetheless performed the radial fit also with different values for the slope, $\delta=0$ and $\delta=-0.5$ as well as with suppressed bump amplitudes ($=0, 2.21$). The final results stay unchanged.}, in the \texttt{dustatt\_modified\_starburst} module.

In the next Sections, for the radial SED fitting, we do not model the dust emission as dust re-emits at $\lambda>10,000$\AA, a wavelength range that is not covered by our radial SEDs. 

Finally, both M84 and M87 host an AGN, which also generates a radio jet. We decided to exclude from the following SED fitting analysis the regions contaminated by the AGN, as this might affect the shape of the SED in a non-trivial way. To mask the central regions, we used the MUSE data. We did not fit all the regions that show emission lines in the wavelength range $6500-6600$\AA, where the H$\alpha$+[NII] lines are present. For M87, we excluded the first 2 (6) central regions for the GALEX (UVIT) analysis; while for M84 we excluded the first 1 (2) central regions for the GALEX (UVIT) analysis. These correspond to SMA/$R_{\rm eff}\lesssim0.1$. See Figs. \ref{app:muse_images}.



All the parameters adopted in the \texttt{CIGALE} fits for the different galaxies are reported in Table \ref{tab:cigale_p}. 

\subsection{Fitting the photometric SEDs}
\label{subsec:fitting}

The left panel of Figure \ref{fig:test_chi} shows the obtained best-fit reduced $\chi^2$ values on the SEDs of M49, as a function of SMA/$R_{\rm eff}$. The solid (dashed) line represents the fit with (without) the UV upturn component, while the colour indicates the type of analysis (orange for UVIT, only FUV, and purple for GALEX, FUV+NUV).
The upturn models in both analyses significantly improve the quality of the fits in the central regions of this galaxy, lowering the values of the $\chi^2$ from $\sim12$ down to $\sim2$.
The right panel of Fig. \ref{fig:test_chi} reports the observed SED of M49 in the central region at SMA/$R_{\rm eff}\sim0.05$ and its best-fit models, when ignoring (blue dashed line) or considering (green solid line) the upturn component. We remind the reader here that a constraint on the SFR has been considered in the fit. The no-upturn model does not reproduce the FUV emission of M49, consistently with what found in the left panel, thus it is crucial to include the contribution of the UV upturn to correctly reproduce the shape of its FUV-to-NIR SED. Moreover, the no-upturn models found a best-fit mass-weighted age of Age$_{(\rm no-up)}\simeq11$ Gyr, and a SF timescale $\tau_{(\rm no-up)}=1200$ Myr. When using the upturn models, the best-fit age is Age$_{(\rm upturn)}\simeq13.3$ Gyr, and a $\tau_{(\rm upturn)}=500$ Myr. Hence, the no-upturn fit tries wrongly to interpret the FUV emission as younger stellar populations with longer SF timescales. Such fast timescales ($\tau_{(\rm upturn)}$) are remarkably close to what observed by JWST in red and quiescent galaxies at high redshift (e.g. \citealt{degraaff24, carnall24}), see Section \ref{sec:disc} for more discussion about this.

\begin{figure*}
\centering
\includegraphics[scale=0.65]{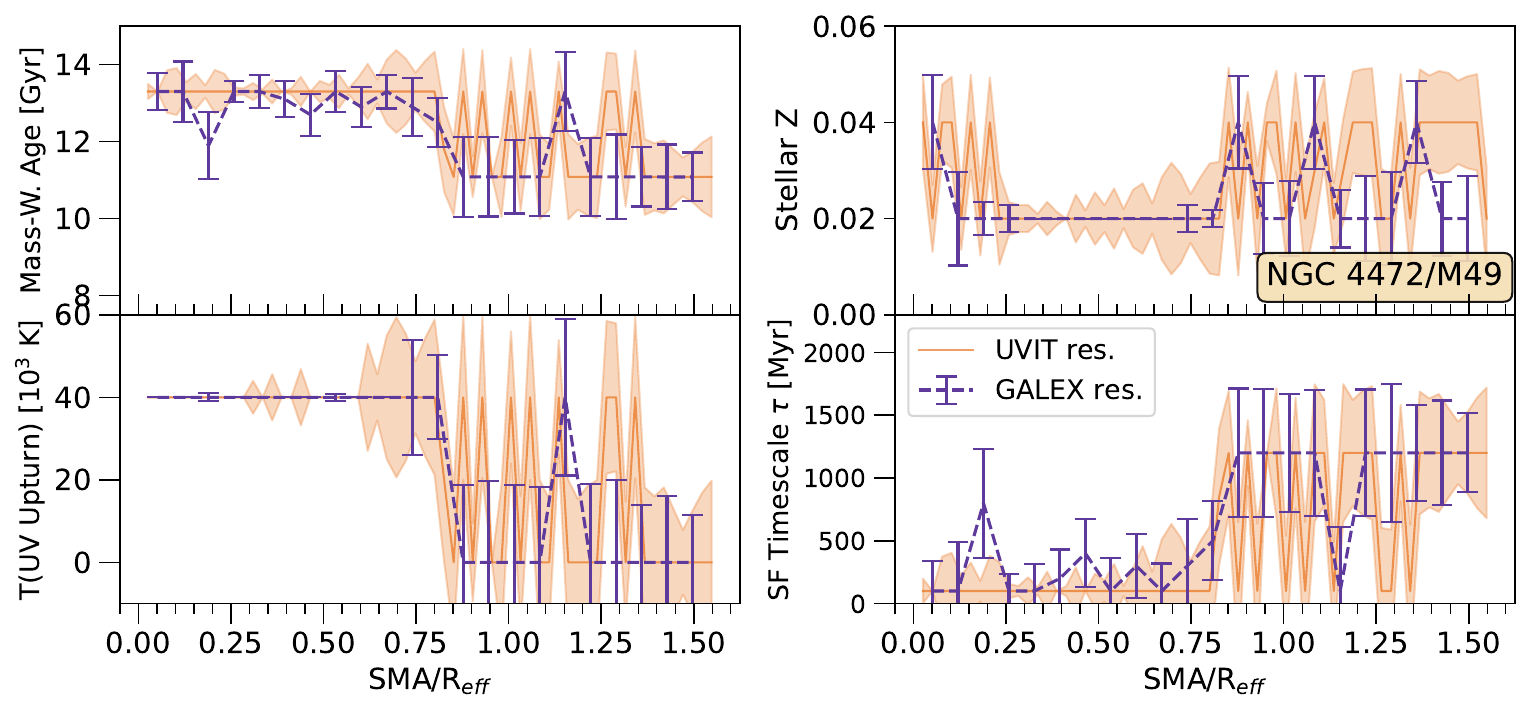}
\includegraphics[scale=0.65]{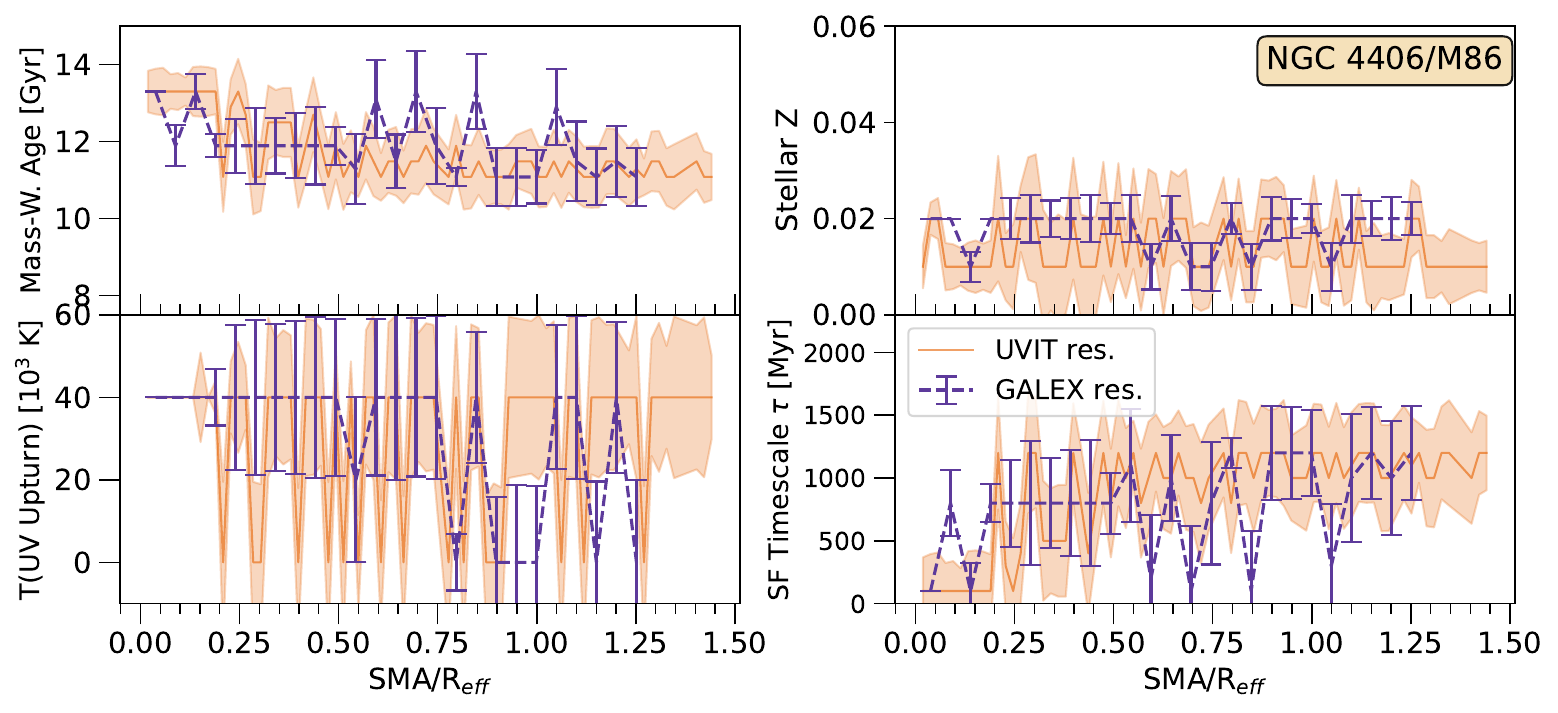}
\caption{Photometric SED fitting results for M49 (top panels) and M86 (bottom panels). For each galaxy, from the upper left to lower right: mass-weighted age, stellar metallicity, temperature of the UV upturn and SF timescale as a function of the SMA$/R_{\rm eff}$. Purple dashed lines with 1$\sigma$ errors represent the fit at GALEX resolution, while orange solid lines indicate fits at UVIT resolution, with shaded orange regions as 1$\sigma$ errors. } 
\label{fig:m49_m86_res}
\end{figure*}

We performed two kinds of SED fitting for each region in each target galaxy: one including FUV, NUV and optical data (GALEX analysis) and the other including only FUV and optical data (UVIT analysis, at higher spatial resolution). 
To obtain the best fit parameters and their errors, we performed a Monte Carlo (MC) analysis. For each region, the SED fit was repeated 100 times, where at each time the broadband fluxes were sampled from a Gaussian distribution where the mean is equal to the measured flux and the standard deviation is equal to the measured 1$\sigma$ error on the flux. The best fit parameters are the median of the 100 parameters from the MC realisations  and the $1\sigma$ error is the standard deviation of the 100 parameters. 
The final results of the SED analysis are reported in Sect. \ref{sec:res}.

\begin{figure*}
\centering
\includegraphics[scale=0.65]{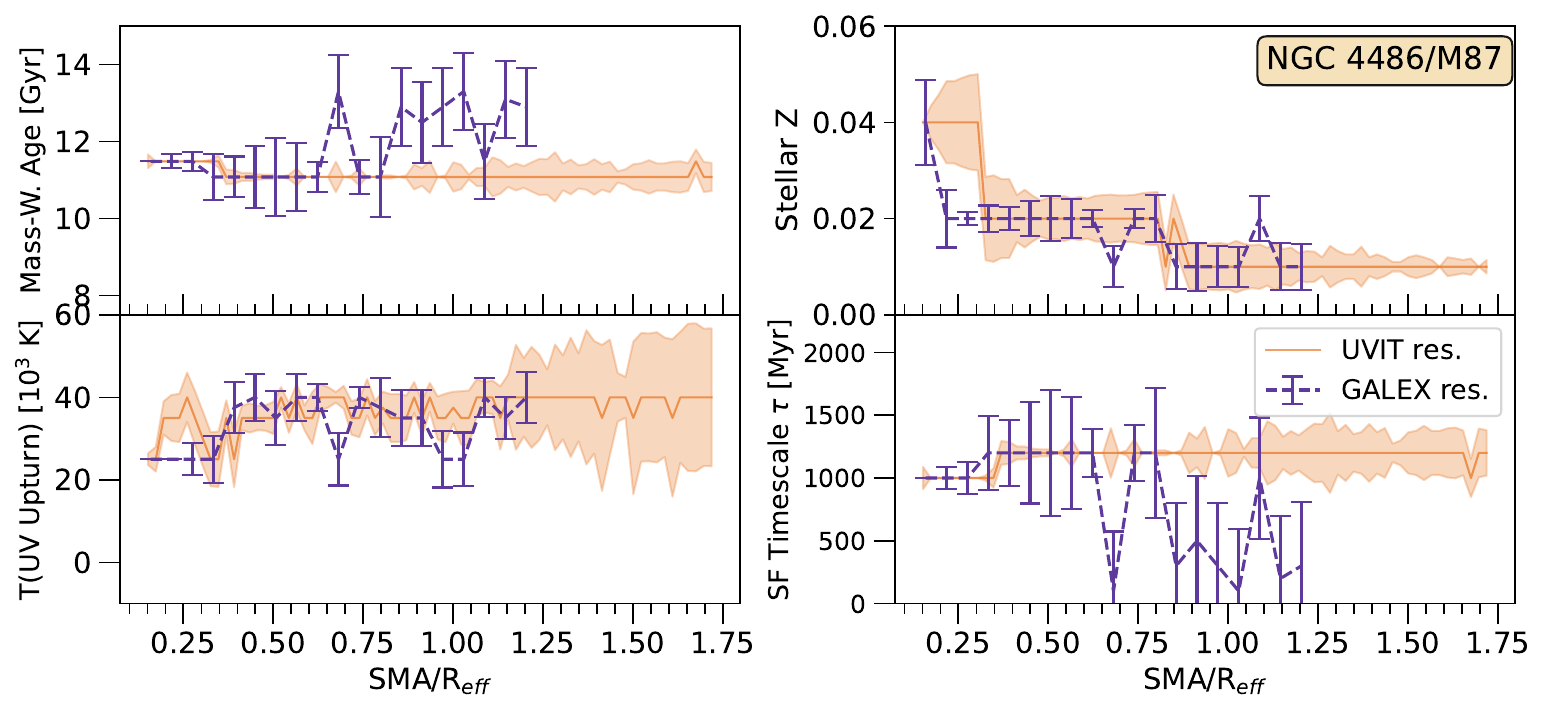}
\includegraphics[scale=0.65]{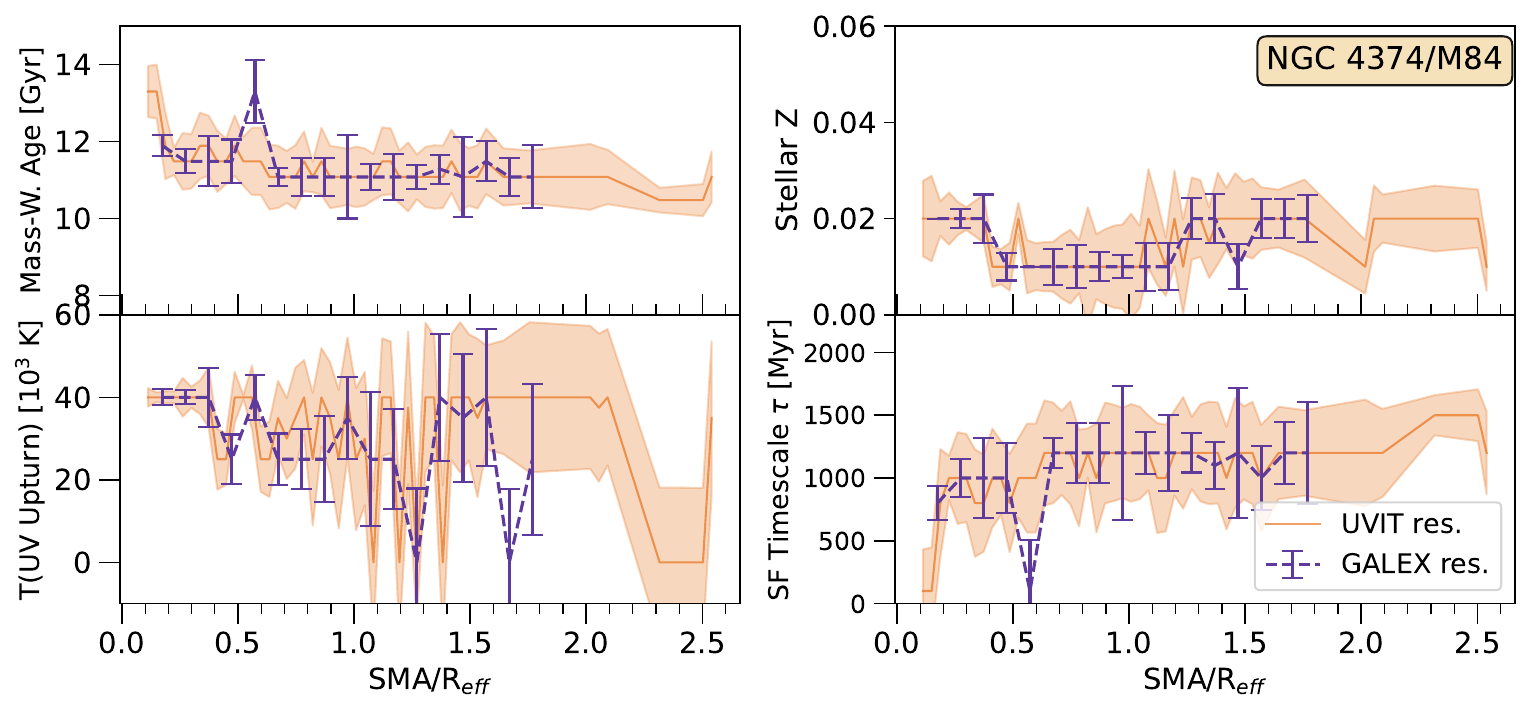}
\caption{As in Fig. \ref{fig:m49_m86_res}, but for M87 (top) and M84 (bottom).} 
\label{fig:m87_m84_res}
\end{figure*}

\subsection{Mock analysis}
\label{subsec:mock}
To assess the reliability of the physical parameters derived from the SED fitting, we generated a catalogue of simulated galaxies, which we analysed in the same fashion as the real galaxies. We used the \texttt{savefluxes} analysis mode within \texttt{CIGALE}, which integrates the galaxy models and then calculates the fluxes in the same photometric bands as the observations. The simulated galaxies span a wider range of parameters with respect to the observations in order to avoid edge effects. The mock galaxies are then generated by re-sampling the model fluxes introducing gaussian noise according to the typical observational error in each band. We then fit each SED of the mock galaxies and we obtained the estimated physical parameters, namely the mass-weighted age, the stellar metallicity Z, the temperature of the UV upturn $T_{UV}$, the e-folding timescale of star formation $\tau$ and the star formation rate SFR. We repeated the exercise by creating a new catalogue, where this time the galaxies are also extincted by dust, to represent the cases of M84, M86 and M87. In this way we also recover the colour excess of the stellar continuum $E(B-V)$.
The results of the mock analysis showed us that all parameters are very well recovered by the SED fitting analysis. This is reported in Appendix \ref{app:mock}, where Figs. \ref{fig:mock_1} and \ref{fig:mock_2} show the exact (input) parameters of the mock galaxies versus the median value of the estimated (output) best-fit parameters, when a FUV+NUV+optical fit is performed on central regions, with and without extinction, respectively. The correlation exact-estimated is very strong with a Pearson correlation coefficient $r^2\gtrsim$0.9 in all cases. Finally, we performed the same mock analysis (i) when considering a fit with only FUV and optical data, and (ii) with a different error, typical of the outer regions (at SMA/R$_{\rm eff}\sim1.5$). In both cases, the outcome stays unchanged.

\begin{figure*}
\centering
\includegraphics[scale=0.65]{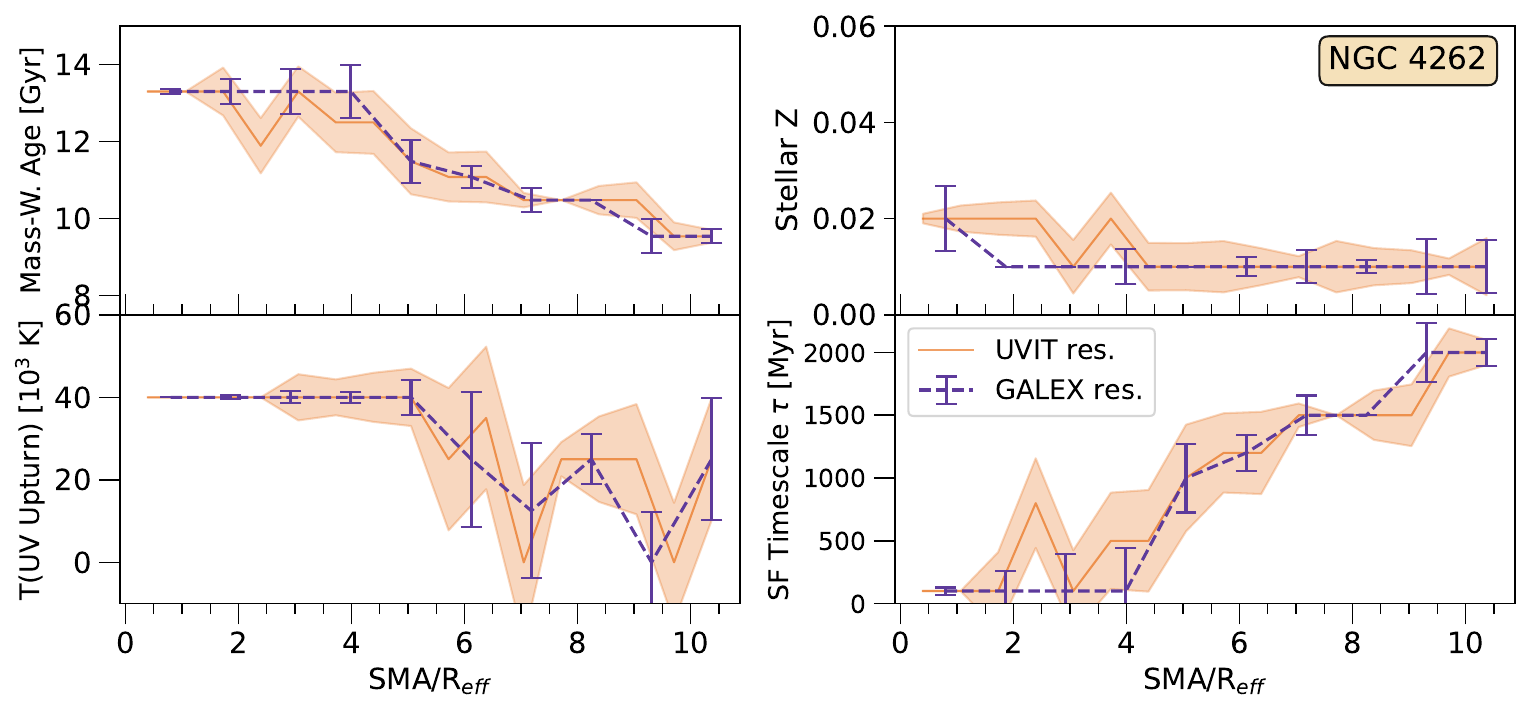}
\includegraphics[scale=0.65]{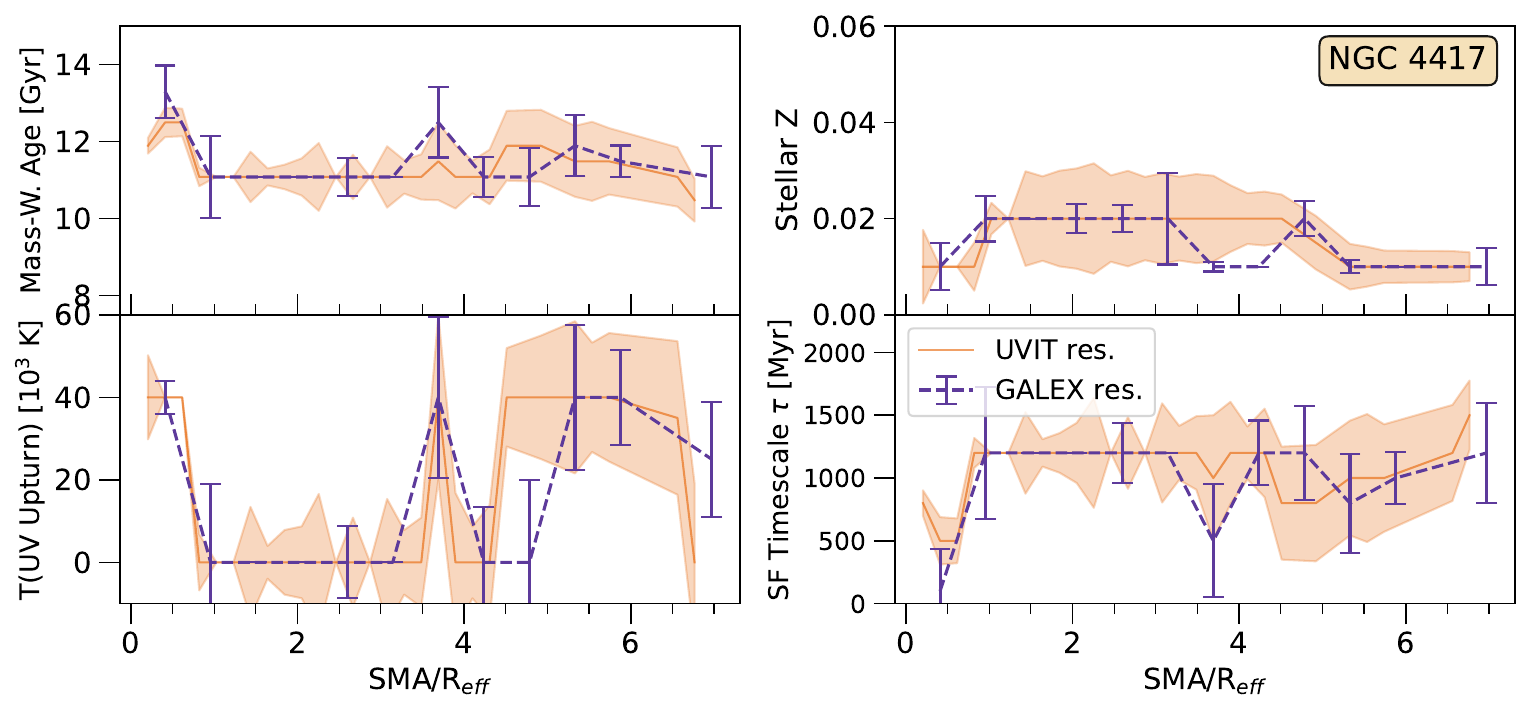}
\includegraphics[scale=0.65]{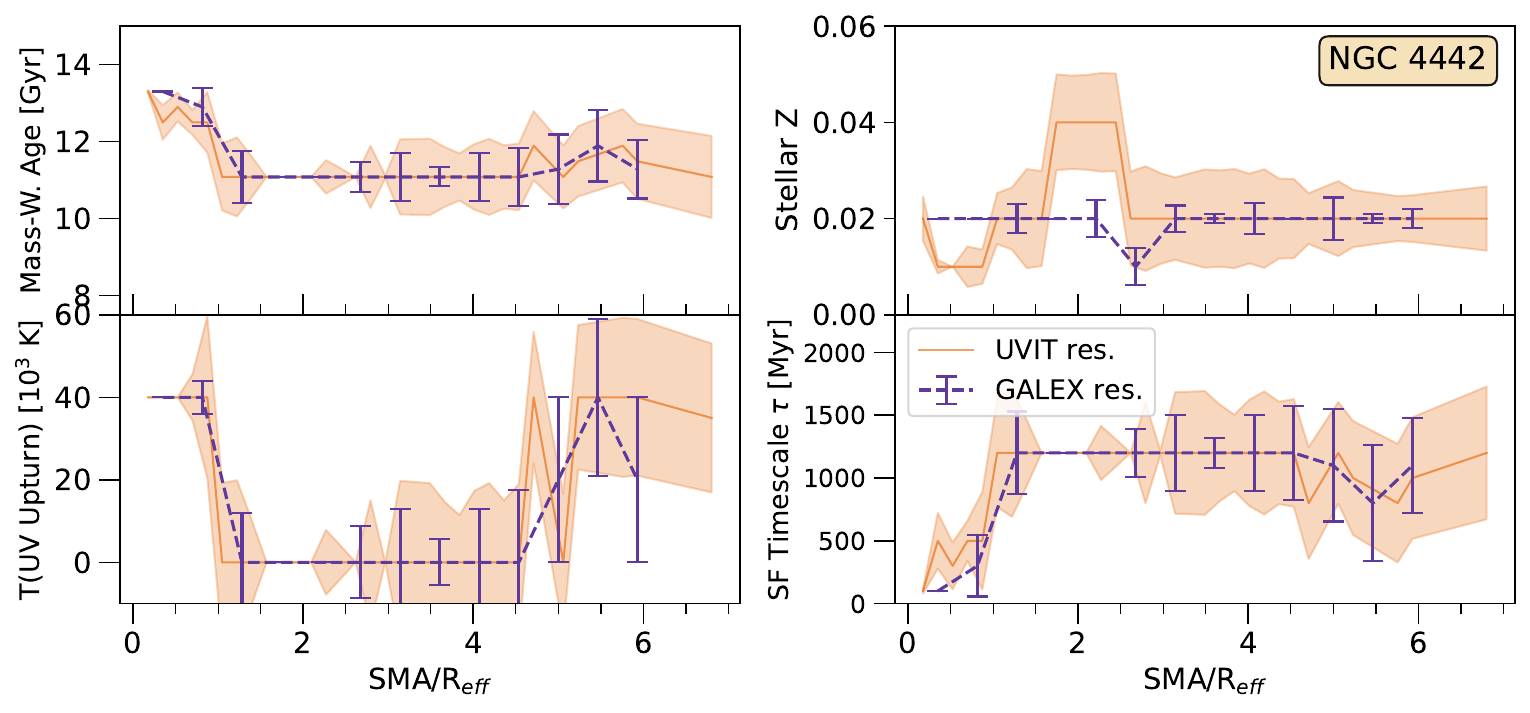}
\caption{As in Fig. \ref{fig:m49_m86_res}, but for NGC~4262 (top), NGC~4417 (central) and NGC~4442(bottom).} 
\label{fig:lent_res}
\end{figure*}

\section{Results}
\label{sec:res}

Figures \ref{fig:m49_m86_res}$-$\ref{fig:lent_res} show the results of the photometric SED fitting for the target galaxies, for both the GALEX (purple dashed lines) and UVIT (orange lines and shaded areas) resolution. From top left to bottom right each galaxy panel displays: the radial profile of the mass-weighted age in Gyr, the stellar metallicity $Z$, the temperature of the UV upturn component $T$ and the SF e-folding timescale $\tau$. The radial profile of the best-fit stellar extinction $E(B-V)$ for M84, M86 and M87 is reported in Appendix \ref{app:extinction}, Fig. \ref{fig:extinction}. Overall, there is no significant difference between the results from the two types of analysis, at 1$\sigma$. 

Our galaxies are old, with populations aged $>10$ Gyr. In the central regions, our galaxies are found to be older, reaching a mass-weighted age of 12$-$13 Gyr. M49 and M87 have super$-$solar metallicities within their inner regions, at SMA/R$_{\rm eff}<0.2$, that then decrease at solar values. The metallicity profile of the other galaxies is instead flatter. 
One of the main results of the SED fit is that the UV upturn component is necessary to describe the central UV emission of each target galaxy (down to 1$R_{\rm eff}$ except for M87 and NGC~4262\footnote{Although some contamination from the HII regions is expected in NGC~4262.}), thus meaning that the old stars responsible for this emission are more centrally concentrated. 
The preferred upturn models are those with temperatures greater than T$\sim$35000~K, except for M87 where the best fit in the central regions prefers temperatures T$\sim$25000~K. 
The recovered SF timescale for all galaxies is relatively fast, of the order of 500$-$1500 Myr, somewhat consistently with what was found in the literature (e.g., \citealt{thomas05}, see Sect. \ref{subsec:jwst_discu}). Indeed, within the central regions, the SF timescales are found to be faster with respect to the outskirts (overall at SMA/$R_{\rm eff}<0.5-1$), of the order or less than 500 Myr. 
Finally, the recovered $E(B-V)$ values from the SED fitting are qualitatively consistent with the extinction maps reported in Fig. \ref{fig:dust}. 
\begin{figure*}
    \centering
    \includegraphics[scale=0.6]{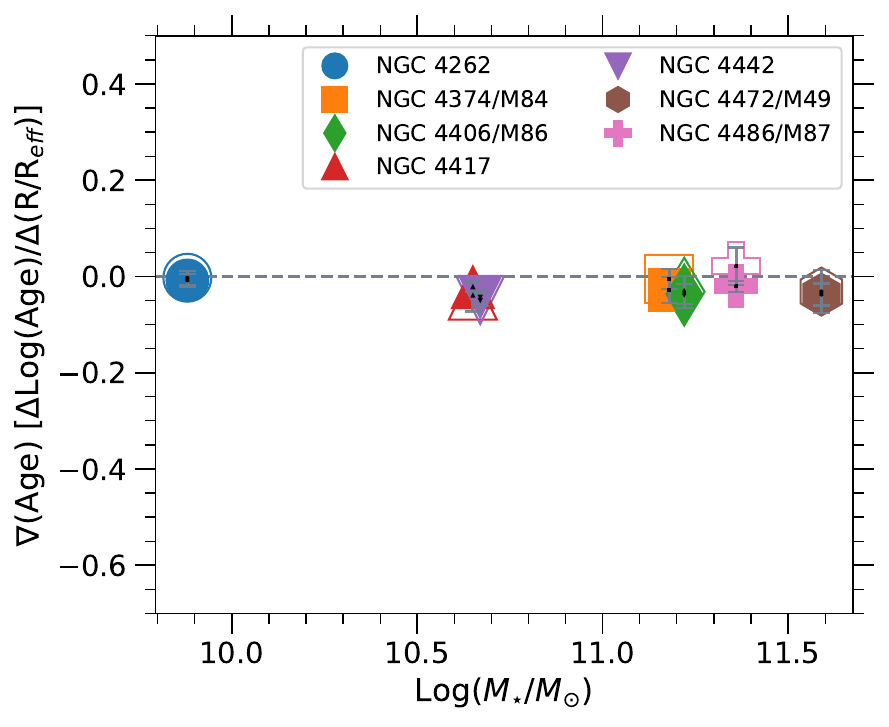}
    \includegraphics[scale=0.6]{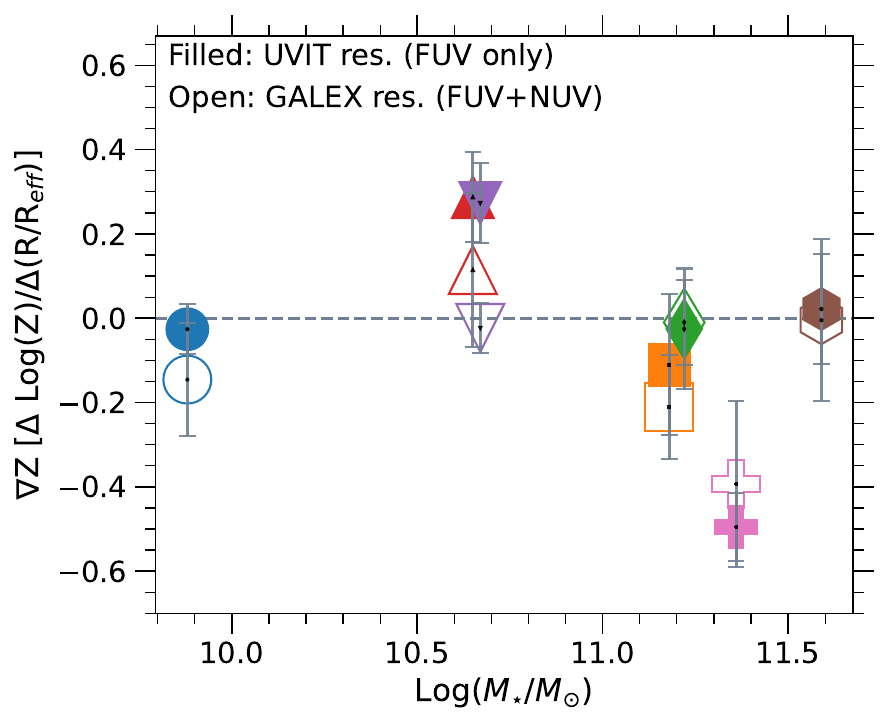}
    \caption{Mass-weighted age gradients (left) and metallicity gradients (right) as a function of galaxy stellar mass. Filled (open) symbols indicate values calculated for the UVIT (GALEX) resolution analysis. See text for more details. The horizontal grey dashed line indicates a flat gradient. Panels share the same y-axis.}
    \label{fig:age-met-grad}
\end{figure*}

Next, we calculated age and metallicity gradients, as these are very useful measurements to set constraints on the formation and evolution of ETGs (see the next Section \ref{sec:disc} for more discussion about this and references). We computed logarithmic gradients 
as $\Delta$ Log/$\Delta$(R/R$_{\rm eff}$)\footnote{We calculated the gradients between 0.1 and 1.2 (1.5) $R_{\rm eff}$ for the elliptical sample at GALEX (UVIT) resolution, between 0.4 (0.8) and 2 R$_{\rm eff}$ for NGC~4262 at UVIT (GALEX) resolution, and between 0.2 (0.4) and 2 R$_{\rm eff}$ for NGC~4417 and NGC~4442 at UVIT (GALEX) resolution.}. The gradients are reported in Figure \ref{fig:age-met-grad} as a function of galaxy stellar mass. 
We found that age gradients are flatter ($\nabla$Log(Age)$\sim -0.04 - 0.0$ dex) than metallicity gradients, on average. All the age gradients are consistent with zero, except for NGC~4442. NGC~4417, NGC~4442 and M49 have slightly negative age gradients. Within the metallicity gradients, the scatter is much higher due to larger errors ($\sim$0.16 dex). We found shallow metallicity gradients overall ($\nabla$Log($Z$)$ \gtrsim -$0.2 dex), except for M87, where $\nabla$Log($Z_{\rm M87}) \sim -$0.45 dex. Two among the most massive galaxies (i.e., M84 and M87) have steeper metallicity gradients. Nonetheless, no correlation between $Z$ gradient and stellar mass is observed, with a deviation from a random distribution $d_s\sim53\% (88\%)$ when performing the Spearman-rank test on UVIT (GALEX) resolution. 

Recently, several surveys have been carried out to test ETG formation theories (e.g., ATLAS$^{3D}$, MaNGA, CALIFA, SAMI, to cite a few), by using both imaging and spectroscopic techniques to estimate stellar population gradients that span from the very central regions to several effective radii. A very informative histogram of age and metallicity gradients from recent results is reported in Figure 22 of \cite{goddard17}. Overall, our age gradients are consistent with the literature, while our $Z$ gradients are shallower. We recall that this might be also due to the fact that we are only investigating seven (massive) galaxies while the results reported by the different surveys include more galaxies spanning several masses and environments. Nevertheless, it is possible to remark that negative metallicity gradients are observed up to 1$R_{\rm eff}$ (e.g., \citealt{mehlert03, sb07, spolaor09, kuntschner10, gonzalez-delgado15, goddard17, parikh19, zibetti20, lu23} and references therein). After 1$R_{\rm eff}$, metallicity gradients tend to flatten, which is interpreted as evidence of a dominant contribution from minor mergers (e.g., \citealt{labarbera12,oyarzun19}, see next Section \ref{sec:disc}). Our results are in agreement with e.g., \cite{gonzalez-delgado15, goddard17, zheng17}, which found shallower $Z$ gradients. 
Also, we observe that, overall, the ellipticals in our sample (M84, M86 and M87) show negative metallicity gradients (although mostly consistent with zero), while the lenticulars have either flat or positive metallicity gradients (NGC~4262, NGC~4417, NGC~4442), although the dichotomy is not striking. This might be interpreted as a sign of a different formation scenario between the two classes of galaxies, with the S0s as a transition class between spirals and ellipticals. Such a difference is also imprinted in terms of their stellar kinematics, as the lenticulars in our sample are all fast rotators (FR) while the ellipticals are slow rotators (SR, \citealt{emsellem04, emsellem11}). However,
with the current sample we cannot draw strong or significant conclusions about this proposed picture. Besides, we note that previous studies based on cosmological simulations have shown that there are no two unique formation scenarios for the origin of FRs and SRs, respectively, but this depends on several factors such as the merger mass-ratio, the timing of major mergers and gas fraction (e.g., \citealt{naab14,lagos22}).

\section{Discussion}
\label{sec:disc}

\subsection{The origin of the UV upturn}
\label{subsec:upturn}

We found a significant FUV emission within the centre of each galaxy in the sample (SMA/$R_{\rm eff}<1$) which cannot be attributed to recent SF events or young stars, given their negligible ionised gas emission. Standard stellar populations models are not able to reproduce such an UV emission, the so-called phenomenon of the UV upturn (e.g., \citealt{oconnell92}). In this work, we employed novel SSP models that take into account the UV contribution from old, hot low-mass stars in all post-MS phases, such as hot HB, post-AGB, AGB manqu\'e stars. The origin of the UV emission in ETGs is still a matter of debate. A possible explanation, given by e.g., \cite{greggio90}, is that metal-rich post-MS stars increase their opacities enough to cause enhanced mass-loss and expose the internal hot shells during their horizontal branch (HB) phase. These shells are hotter than 25000~K and might be responsible for the UV emission. Other explanations include UV emission caused by extreme HB stars that are likely He-enhanced (e.g., \citealt{brown00b, ali18, depropris22}), or formed via binary interactions (e.g., \citealt{han07, han10}). 
Here, we modeled the UV emission to be produced by a hot component of post-MS stars, 
where the evolutionary mass is small enough that mass-loss can effectively remove the stellar envelope to expose the inner hotter burning shells, resulting in a range of temperatures spanning from $T=25000$ up to 40000~K. We found that the slope of the SED of our galaxies is well described by models with the component at temperatures $T\gtrsim 25000$(M87) up to $40000$~K. This is consistent with studies that recently made a census of UV-bright stars in galactic globular clusters (GGCs) with AstroSat/UVIT (e.g. \citealt{sahu22, prabhu22}). Such authors found that GGCs host more HB stars than post-AGB stars, with the temperatures of the UV-bright stars population peaking around 25000$-$30000~K up to 100000~K. Indeed, the proximity of GGCs can be exploited to serve as template for the integrated properties of distant and unresolved extragalactic systems (e.g., \citealt{schiavon12, dalessandro12}). Nevertheless, we note that GGCs are much more metal-poor ($Z\lesssim0.003$) with respect to the galaxies studied in this work, thus the origin of the UV brightness in GC stars might be different from the upturn observed in massive ETGs (e.g., \citealt{mt00}). 
Additionally, by using deep HST/STIS FUV images of the dwarf spheroidal galaxy M32, \cite{brown08} found that the core of this galaxy is mostly devoid of UV-bright post-AGB stars, while it hosts hot HB stars capable to reproduce the UV emission. However, we note that M32 is a dwarf galaxy, while here we are investigating massive early-type galaxies (some of them among the most massive ellipticals), hence the sources for the UV upturn might also be different. 
In order to understand the origin of the UV upturn, it would be extremely helpful to resolve UV-bright stars in massive ETGs and identify the evolutionary phase that majorly contribute to their UV emission.

\subsection{Stellar population gradients and their implication on galaxy formation and evolution}
\label{subsec:grad_discu}

Gradients and radial profiles have been proven to be excellent tools in setting valuable constraints on ETGs formation and evolution scenarios. Recent cosmological simulations predict that massive ETGs form in a two-phase scenario (e.g. \citealt{oser10}). In the first phase, stars form in-situ very rapidly and through the infall of cold gas, at $z\gtrsim2$. The result is a compact galaxy with a steep metallicity gradient ($\nabla$(Fe/H) $< -0.35$ dex, e.g., \citealt{larson74, carlberg84, thomas99, kawata03, kobayashi04, pipino10}) and a positive age gradient. This is due to the fact that younger stars continue forming within the centre, as the  potential well of the galaxy retains the stellar winds. However, the presence of an AGN or a morphological quenching mechanism \citep{martig09} might instead halt star formation in the centre and flatten the inner radial profiles (e.g., \citealt{zibetti20}). In a second phase, the ETG grows in mass through dissipationless mergers, which can either steepen or flatten the pre-existing gradients. 
Stellar major mergers tend to flatten all gradients as metal-rich stars from the center of the galaxy are reshuffled in the outskirts ($\nabla$(Fe/H)$\sim -0.1$ dex, e.g., \citealt{ogando05, dimatteo09, kobayashi04}). Minor mergers tend instead to steepen metallicity gradients as the accreted satellites at larger radii are more metal-poor than the central core (e.g., \citealt{hirschmann15}).

The age gradients we found are all consistent with zero, except for NGC~4442 (lenticular), whose value is slightly negative ($\nabla$Log(Age)$\sim -0.04$ dex), similar to what is more common in late-type galaxies (e.g., \citealt{goddard17}). We note that we calculate gradients up to 2$R_{\rm eff}$, while we have access to larger effective radii for the lenticulars. We observe that the stellar populations in NGC~4262 are significantly younger within $\sim4-6R_{\rm eff}$, as expected from its ring of HII regions. 
In addition, we found shallow metallicity gradients ($\nabla$Log($Z$)$\gtrsim-0.2$ dex), except for M87 where $\nabla$Log($Z_{\rm M87}$)$=-0.44\pm0.21$ dex. Hence, in an attempt to match our results with the models and simulations we discussed earlier, we support a scenario of ETG formation where major mergers might have played an important role in the formation of these galaxies. We recall that this is constrained within the central regions of the ellipticals (up to $1.5-2R_{\rm eff}$) and further away for the lenticulars (up to $6-10R_{\rm eff}$).

We checked for other signatures of major mergers within these galaxies. 
M86 shows indeed long trails of stripped gas, which have been interpreted as evidence of an interaction with the nearby spiral galaxy NGC~4438 (e.g. \citealt{kenney08, gomez10, ehlert13}). M84 is located only 60~kpc in projected distance from M86, however, it does not appear that it is undergoing a significant interaction with M86, as shown from the lack of diffuse gas bridging between the two galaxies \citep{ehlert13}. Similarly, no tidal debris are observed in the vicinity of M84 \citep{janowiecki10}. This might be in accordance with our results, possibly indicating a lack of accretion from smaller satellite galaxies and pointing to major merger events to explain the shallow metallicity gradient. On the other side, it could be due to a recent passage through the cluster core. 
\cite{janowiecki10} also reports that M49 is surrounded by shells and tidal structures, representative of intense satellite accretion. This is corroborated by the evidence that M49 is in the process of tidally stripping the nearby dwarf galaxy VCC~1249 \citep{arrigoni12,junais22}. It might be possible that our study, covering only gradients up to $R_{\rm eff}\sim1.5$ for this particular galaxy does not represent the extensive outer halo of the very massive M49 (possibly the most massive of the Virgo cluster). 
M87, the extensively studied central galaxy of Virgo, is expected to undergo accretion of infalling smaller galaxies, which are being tidally stripped. This is corroborated by the presence of tidal streams and tails in the north-west outer halo of M87 \citep{janowiecki10}, as well as its globular clusters (GC) and planetary nebulae kinematics and spatial distributions (e.g., \citealt{larsen01, romanowsky12, longobardi15, ferrarese16, longobardi18}).
This is consistent with the negative $Z$ gradient and flat age gradient we found in this work. 

Signs of galaxy-galaxy interactions are also observed in NGC~4262, as there is evidence of filamentary infalling gas into its nucleus (see \citealt{boselli22} and references therein). As visible from its FUV emission (Fig. \ref{fig:uvit}), NGC~4262 has a polar ring structure characterised by some HII regions which has also been detected in HI (e.g., \citealt{oosterloo10}). This ring has been interpreted by \cite{bettoni10} as a signature of a merging event. Additionally, this galaxy shows a bar (see Fig. \ref{fig:ell_uvit}) which is likely to indicate gravitational perturbations, consistent with its almost flat age and $Z$ gradients reported in this work.

Furthermore, one might look at globular cluster (GC) colour distributions. Previous studies reported that our target galaxies all have bimodal (red and blue) GC populations (e.g., \citealt{peng06}), with the red population generally more centrally concentrated than the blue one (in M87, M49, M84 and M86, e.g.  \citealt{longobardi18, cote98, cote03, taylor21, lambert20}). Such a bimodality is thought to be the imprint of the two-phase scenario of ETGs. The red GCs are more metal-rich and more centrally concentrated: they represent the in-situ star formation episode(s); blue/metal-poor GCs, which extend further in the outskirts of galaxies, are thought to be accreted from mergers with smaller satellites (e.g., \citealt{cote98,tonini13}). 
Connecting this picture from the results presented here, we might speculate that the central regions of the ellipticals (up to $1.5-2R_{\rm eff}$) was dominated by major merger accretion (although see next Section, \ref{subsec:jwst_discu} for more details about this), while minor mergers might effectively dominate at larger $R_{\rm eff}$. This picture is also supported by studies of colour gradients of GC systems in massive ETGs (e.g., \citealt{liu11}).
As for the lenticulars (NGC~4417, 4442), for which we observe flat age and metallicity gradients up 6$R_{\rm eff}$ (see Fig. \ref{fig:lent_res}), it would be interesting to investigate the spatial distribution and kinematics of their GC population.

\subsection{Star formation histories and comparison with results at high-z from JWST}
\label{subsec:jwst_discu}

In addition, our results imply that these galaxies must have assembled their masses relatively fast, especially in their inner regions. The derived SF timescales $\tau$ are overall smaller than 1200~Myr down to 100~Myr, except for the outskirts of NGC~4262, where $\tau$ reaches 2000~Myr. This is consistent with other studies of present-day and intermediate redshift massive ETGs in dense environments (e.g. \citealt{bendo96, thomas05, thomas10}), as well as with the evidence that massive ETGs are enhanced in $\alpha$ elements (e.g., \citealt{worthey92, trager00a, gavazzi02, thomas05, spolaor10, parikh19}). 
Recently, results from the JWST revealed extremely rapid SF timescales for red quiescent galaxies at $z\sim5$ (e.g., \citealt{carnall23, degraaff24, carnall24}), of the order of $\sim$200~Myr. We find similar results within the inner regions of our target galaxies.
Hence, we estimated the fraction of stellar mass that each galaxy in our sample grows at this redshift ($z\sim5$) when the Universe is $\sim$1.2 Gyr old, based on the radial SFHs we obtained. 
Our galaxies at $z\sim5$ had already formed from $\sim$42\% (M84) up to $\sim$92\% (NGC~4262) of their mass in stars, according to the derived SFHs. If we apply these fractions to the masses reported in Table \ref{tab:data}, we obtain that our galaxies had stellar masses ranging from $M_{\star}\simeq7\times 10^{9}$\msun (NGC~4262) up to $M_{\star}\simeq2.9\times 10^{11}$\msun \, (M49) at $z\sim5$.
\cite{degraaff24} found a stellar mass $M_{\star}\simeq1\times 10^{11}$\msun \, for RUBIES-EGS-QG1 at $z\sim$4.9, while \cite{carnall23} find $M_{\star}\simeq4\times 10^{10}$\msun \, for GS-9209 at z$\sim$4.7\footnote{Similar results are found for other three quiescent galaxies at $z\sim$4.6 by \cite{carnall24}.}, consistent with what we found for the Virgo massive ETGs. 

Therefore, our results on low redshift galaxies are in support of observations at high redshift within a consistent picture where massive quiescent galaxies formed and quenched with extreme starbursts and rapid mass assembly. This is difficult to reconcile with predictions from hydrodynamical cosmological simulations and theoretical models of galaxy formation (e.g., \citealt{lovell23, kimmig23, delucia24}) where these types of galaxies are envisioned to be rare (comoving number densities of $1-10\times 10^{-8}$~Mpc$^{-3}$ in very massive haloes at $z=4-5$). Similarly, the inferred fast mass growth within the inner regions of ETGs is difficult to reconcile with the two-phase scenario. According to the simulations by \cite{oser10}, the mass growth happens in a slower fashion, with 80\% of the mass for high-mass galaxies being assembled at $z<4.5$ from ex-situ accreted stars.
Our derived stellar gradients (Sect. \ref{subsec:grad_discu}) support the occurrence of major mergers within these systems, and they could in principle provide a fast stellar mass accretion. However, according to the hierarchical formation of galaxies, major mergers should be rare at $z\gtrsim5$. The recent JWST studies (e.g., \citealt{degraaff24, carnall24}) invoked instead for novel prescriptions on star formation recipes, i.e. the conversion of baryons into stars $\epsilon$ should be $>0.2$ up to extreme values $\epsilon=1$, much greater than the expected value at the peak of the stellar$-$halo mass relation ($\epsilon=0.2$, e.g., \citealt{wechsler18}).

\section{Summary and conclusions}
\label{sec:concl}

We analysed the radial stellar properties of seven massive ($M_{\star}>10^{9.8}$\msun) Virgo ETGs by combining a high angular resolution, high sensitive multiwavelength dataset (FUV$-$to$-$NIR) with state-of-the-art stellar population models. The dataset consists of four elliptical and three lenticular galaxies, which are among the most massive (and most studied) galaxies in the Virgo cluster. M87 and M49 are the central of the Virgo subcluster A and B, respectively, as well as the most massive galaxies in the sample, together with M84 ($M_{\star}>10^{11}$\msun). 
The studied radial profiles extend down to $\sim1.5-2.5R_{\rm eff}$ for the ellipticals, and down to $\sim6-10 R_{\rm eff}$ for the lenticulars.
Thanks to the high quality VESTIGE narrow-band H$\alpha$ images, we selected galaxies which do not show clumpy or clustered H$\alpha$ emission, hence with no ongoing SF. We analysed radial fluxes thanks to AstroSat/UVIT, GALEX and CFHT NGVS imaging.

We found that, despite their lack of strong SF, the ETGs in our sample show flat or positive $FUV-i$ gradients, contrarily to what found in optical colours, e.g. the $g-i$ colours have flatter gradients. We also found a hint of a positive correlation between the $FUV-i$ gradient and galaxy stellar mass, although this is only based on seven galaxies, and a larger sample is needed to confirm or refute such results.  Overall, the lenticular galaxies in our sample tend to have flat $FUV-i$ gradients, while the $\nabla(FUV-i)$ for the ellipticals are positive. M87, which is also the bluest galaxy on average, has the strongest gradient with $\nabla (FUV-i)\sim 0.6$ mag between $0.1<$SMA/R$_{\rm eff}<1.5$.

Next, we analysed radial SEDs with FUV, NUV and optical data, adding constraints on the SFR from the H$\alpha$ emission. We found a significant FUV emission within the centre of each galaxy  (SMA/$R_{\rm eff}<1$) which cannot be attributed to recent SF events or young stars, given their negligible ionised gas emission. We fit the radial SEDs of our galaxies with novel stellar models which include a UV upturn component in the form of post-MS stars with various temperatures and fuels. We can then explore the main stellar parameters responsible for the UV upturn stars irregardless of their evolutionary path.
We used the SED fitting code \texttt{CIGALE}, through which we also make the models publicly available.
The SEDs of our galaxies are well described by UV upturn components with temperatures $T\gtrsim 25000-40000$~K within their central regions (SMA/$R_{\rm eff}\lesssim1$). We also found that our galaxies are old (mass-weighted ages $>10$~Gyr) and the centres of the most massive M49 and M87 are supersolar, $Z=2Z_{\odot}$ at SMA/$R_{\rm eff}\lesssim0.2$. We calculated stellar gradients in mass-weighted age and metallicity and we found that age gradients are overall flat while metallicity gradients are shallow. This might indicate that major mergers have contributed to the formation and evolution of such galaxies. Additionally, we found fast SFHs (with timescales $\tau$ spanning from 100$-$1500~Myr) for our galaxies, which assembled between $\sim40-90$\% of their stellar mass at $z\sim5$. 
Such results imply that the massive quiescent galaxies observed with JWST at high redshift are likely the progenitors of the most massive ETGs we observe in the local Universe. 

More detailed studies of stellar populations within early-type systems in the nearby Universe are crucial to set valuable constraints on how these types of galaxies form. From an observational point of view, spectroscopy in the FUV domain is needed to definitely improve estimates of stellar population properties through the characterization of the UV upturn in old and massive galaxies (e.g., \citealt{lonoce20}), spanning different redshifts and environments. Efforts in modeling typical abundance ratios in SSPs which include the UV emission of ETGs will go hand-in-hand.

\begin{acknowledgements}
We thank the anonymous referee for the careful reading of the manuscript. SM was supported by a Gliese Fellowship at the Zentrum f\"ur Astronomie, University of Heidelberg, Germany. SM also thanks the Laboratoire d'Astrophysique de Marseille, where part of this work was carried out.
J. is funded by the European Union (MSCA EDUCADO, GA 101119830 and WIDERA ExGal-Twin, GA 101158446).
We thank Eric Peng for useful discussions.
Based on observations obtained with MegaPrime/MegaCam, a joint project of CFHT and CEA/DAPNIA, at the Canada-French Hawaii Telescope (CFHT) which is operated by the National Research Council (NRC) of Canada, the Institut National des Sciences de l’Univers of the Centre National de la Recherche Scientifique (CNRS) of France and the University of Hawaii. 
We are grateful to the whole CFHT team who assisted us in the preparation and in the execution of the observations and in the calibration and data reduction: Todd Burdullis, Daniel Devost, Bill Mahoney, Nadine Manset, Andreea Petric, Simon Prunet, Kanoa Withington.
The UVIT project is a collaboration among the following institutes from India: Indian Institute of Astrophysics (IIA), Bengaluru, Inter University Centre for Astronomy and Astrophysics (IUCAA), Pune, and National Centre for Radioastrophysics
(NCRA) (TIFR), Pune, and the Canadian Space Agency (CSA).
Based on observations made with ESO Telescopes at the
La Silla Paranal Observatory under programme IDs 095.B-0295, 0102.B-0048, 060.A-9312.

\end{acknowledgements}

%
%



\bibliographystyle{aa}
\bibliography{biblio} 

\begin{thebibliography}{202}
\expandafter\ifx\csname natexlab\endcsname\relax\def\natexlab#1{#1}\fi

\bibitem[{{Agrawal}(2006)}]{agrawal06}
{Agrawal}, P.~C. 2006, Advances in Space Research, 38, 2989

\bibitem[{{Akhil} {et~al.}(2024){Akhil}, {Kartha}, {Kizhuprakkat}, {Ujjwal}, \& {P}}]{akhil24}
{Akhil}, K.~R., {Kartha}, S.~S., {Kizhuprakkat}, N., {Ujjwal}, K., \& {P}, N. 2024, \mnras, 534, 4063

\bibitem[{{Ali} {et~al.}(2018{\natexlab{a}}){Ali}, {Bremer}, {Phillipps}, \& {De Propris}}]{ali18b}
{Ali}, S.~S., {Bremer}, M.~N., {Phillipps}, S., \& {De Propris}, R. 2018{\natexlab{a}}, \mnras, 478, 541

\bibitem[{{Ali} {et~al.}(2018{\natexlab{b}}){Ali}, {Bremer}, {Phillipps}, \& {De Propris}}]{ali18}
{Ali}, S.~S., {Bremer}, M.~N., {Phillipps}, S., \& {De Propris}, R. 2018{\natexlab{b}}, \mnras, 476, 1010

\bibitem[{{Ali} {et~al.}(2019){Ali}, {Bremer}, {Phillipps}, \& {De Propris}}]{ali19}
{Ali}, S.~S., {Bremer}, M.~N., {Phillipps}, S., \& {De Propris}, R. 2019, \mnras, 487, 3021

\bibitem[{{Antwi-Danso} {et~al.}(2023){Antwi-Danso}, {Papovich}, {Esdaile}, {Nanayakkara}, {Glazebrook}, {Hutchison}, {Whitaker}, {Marsan}, {Diaz}, {Marchesini}, {Muzzin}, {Tran}, {Setton}, {Kaushal}, {Speagle}, \& {Cole}}]{antwi-danso23}
{Antwi-Danso}, J., {Papovich}, C., {Esdaile}, J., {et~al.} 2023, arXiv e-prints, arXiv:2307.09590

\bibitem[{{Arrigoni Battaia} {et~al.}(2012){Arrigoni Battaia}, {Gavazzi}, {Fumagalli}, {Boselli}, {Boissier}, {Cortese}, {Heinis}, {Ferrarese}, {C{\^o}t{\'e}}, {Mihos}, {Cuillandre}, {Duc}, {Durrell}, {Gwyn}, {Jord{\'a}n}, {Liu}, {Peng}, \& {Mei}}]{arrigoni12}
{Arrigoni Battaia}, F., {Gavazzi}, G., {Fumagalli}, M., {et~al.} 2012, \aap, 543, A112

\bibitem[{{Baade} \& {Minkowski}(1954)}]{baade54}
{Baade}, W. \& {Minkowski}, R. 1954, \apj, 119, 215

\bibitem[{{Baggen} {et~al.}(2023){Baggen}, {van Dokkum}, {Labb{\'e}}, {Brammer}, {Miller}, {Bezanson}, {Leja}, {Wang}, {Whitaker}, {Suess}, \& {Nelson}}]{baggen23}
{Baggen}, J. F.~W., {van Dokkum}, P., {Labb{\'e}}, I., {et~al.} 2023, \apjl, 955, L12

\bibitem[{{Balmaverde} {et~al.}(2021){Balmaverde}, {Capetti}, {Marconi}, {Venturi}, {Chiaberge}, {Baldi}, {Baum}, {Gilli}, {Grandi}, {Meyer}, {Miley}, {O'Dea}, {Sparks}, {Torresi}, \& {Tremblay}}]{balmaverde21}
{Balmaverde}, B., {Capetti}, A., {Marconi}, A., {et~al.} 2021, \aap, 645, A12

\bibitem[{{Bambic} {et~al.}(2023){Bambic}, {Russell}, {Reynolds}, {Fabian}, {McNamara}, \& {Nulsen}}]{bambic23}
{Bambic}, C.~J., {Russell}, H.~R., {Reynolds}, C.~S., {et~al.} 2023, \mnras, 522, 4374

\bibitem[{{Bastian} {et~al.}(2010){Bastian}, {Covey}, \& {Meyer}}]{bastian10}
{Bastian}, N., {Covey}, K.~R., \& {Meyer}, M.~R. 2010, \araa, 48, 339

\bibitem[{{Bender} {et~al.}(1996){Bender}, {Ziegler}, \& {Bruzual}}]{bendo96}
{Bender}, R., {Ziegler}, B., \& {Bruzual}, G. 1996, \apjl, 463, L51

\bibitem[{{Bettoni} {et~al.}(2010){Bettoni}, {Buson}, \& {Galletta}}]{bettoni10}
{Bettoni}, D., {Buson}, L.~M., \& {Galletta}, G. 2010, \aap, 519, A72

\bibitem[{{Binggeli} {et~al.}(1985){Binggeli}, {Sandage}, \& {Tammann}}]{binggeli85}
{Binggeli}, B., {Sandage}, A., \& {Tammann}, G.~A. 1985, \aj, 90, 1681

\bibitem[{{Boissier} {et~al.}(2018){Boissier}, {Cucciati}, {Boselli}, {Mei}, \& {Ferrarese}}]{boissier18}
{Boissier}, S., {Cucciati}, O., {Boselli}, A., {Mei}, S., \& {Ferrarese}, L. 2018, \aap, 611, A42

\bibitem[{{Boquien} {et~al.}(2019){Boquien}, {Burgarella}, {Roehlly}, {Buat}, {Ciesla}, {Corre}, {Inoue}, \& {Salas}}]{boquien19}
{Boquien}, M., {Burgarella}, D., {Roehlly}, Y., {et~al.} 2019, \aap, 622, A103

\bibitem[{{Boselli} {et~al.}(2009){Boselli}, {Boissier}, {Cortese}, {Buat}, {Hughes}, \& {Gavazzi}}]{boselli09}
{Boselli}, A., {Boissier}, S., {Cortese}, L., {et~al.} 2009, \apj, 706, 1527

\bibitem[{{Boselli} {et~al.}(2011){Boselli}, {Boissier}, {Heinis}, {Cortese}, {Ilbert}, {Hughes}, {Cucciati}, {Davies}, {Ferrarese}, {Giovanelli}, {Haynes}, {Baes}, {Balkowski}, {Brosch}, {Chapman}, {Charmandaris}, {Clemens}, {Dariush}, {De Looze}, {di Serego Alighieri}, {Duc}, {Durrell}, {Emsellem}, {Erben}, {Fritz}, {Garcia-Appadoo}, {Gavazzi}, {Grossi}, {Jord{\'a}n}, {Hess}, {Huertas-Company}, {Hunt}, {Kent}, {Lambas}, {Lan{\c{c}}on}, {MacArthur}, {Madden}, {Magrini}, {Mei}, {Momjian}, {Olowin}, {Papastergis}, {Smith}, {Solanes}, {Spector}, {Spekkens}, {Taylor}, {Valotto}, {van Driel}, {Verstappen}, {Vlahakis}, {Vollmer}, \& {Xilouris}}]{boselli11}
{Boselli}, A., {Boissier}, S., {Heinis}, S., {et~al.} 2011, \aap, 528, A107

\bibitem[{{Boselli} {et~al.}(2010{\natexlab{a}}){Boselli}, {Ciesla}, {Buat}, {Cortese}, {Auld}, {Baes}, {Bendo}, {Bianchi}, {Bock}, {Bomans}, {Bradford}, {Castro-Rodriguez}, {Chanial}, {Charlot}, {Clemens}, {Clements}, {Corbelli}, {Cooray}, {Cormier}, {Dariush}, {Davies}, {de Looze}, {di Serego Alighieri}, {Dwek}, {Eales}, {Elbaz}, {Fadda}, {Fritz}, {Galametz}, {Galliano}, {Garcia-Appadoo}, {Gavazzi}, {Gear}, {Giovanardi}, {Glenn}, {Gomez}, {Griffin}, {Grossi}, {Hony}, {Hughes}, {Hunt}, {Isaak}, {Jones}, {Levenson}, {Lu}, {Madden}, {O'Halloran}, {Okumura}, {Oliver}, {Page}, {Panuzzo}, {Papageorgiou}, {Parkin}, {Perez-Fournon}, {Pierini}, {Pohlen}, {Rangwala}, {Rigby}, {Roussel}, {Rykala}, {Sabatini}, {Sacchi}, {Sauvage}, {Schulz}, {Schirm}, {Smith}, {Spinoglio}, {Stevens}, {Sundar}, {Symeonidis}, {Trichas}, {Vaccari}, {Verstappen}, {Vigroux}, {Vlahakis}, {Wilson}, {Wozniak}, {Wright}, {Xilouris}, {Zeilinger}, \& {Zibetti}}]{boselli10b}
{Boselli}, A., {Ciesla}, L., {Buat}, V., {et~al.} 2010{\natexlab{a}}, \aap, 518, L61

\bibitem[{{Boselli} {et~al.}(2005){Boselli}, {Cortese}, {Deharveng}, {Gavazzi}, {Yi}, {Gil de Paz}, {Seibert}, {Boissier}, {Donas}, {Lee}, {Madore}, {Martin}, {Rich}, \& {Sohn}}]{boselli05}
{Boselli}, A., {Cortese}, L., {Deharveng}, J.~M., {et~al.} 2005, \apjl, 629, L29

\bibitem[{{Boselli} {et~al.}(2010{\natexlab{b}}){Boselli}, {Eales}, {Cortese}, {Bendo}, {Chanial}, {Buat}, {Davies}, {Auld}, {Rigby}, {Baes}, {Barlow}, {Bock}, {Bradford}, {Castro-Rodriguez}, {Charlot}, {Clements}, {Cormier}, {Dwek}, {Elbaz}, {Galametz}, {Galliano}, {Gear}, {Glenn}, {Gomez}, {Griffin}, {Hony}, {Isaak}, {Levenson}, {Lu}, {Madden}, {O'Halloran}, {Okamura}, {Oliver}, {Page}, {Panuzzo}, {Papageorgiou}, {Parkin}, {Perez-Fournon}, {Pohlen}, {Rangwala}, {Roussel}, {Rykala}, {Sacchi}, {Sauvage}, {Schulz}, {Schirm}, {Smith}, {Spinoglio}, {Stevens}, {Symeonidis}, {Vaccari}, {Vigroux}, {Wilson}, {Wozniak}, {Wright}, \& {Zeilinger}}]{boselli10}
{Boselli}, A., {Eales}, S., {Cortese}, L., {et~al.} 2010{\natexlab{b}}, \pasp, 122, 261

\bibitem[{{Boselli} {et~al.}(2018){Boselli}, {Fossati}, {Ferrarese}, {Boissier}, {Consolandi}, {Longobardi}, {Amram}, {Balogh}, {Barmby}, {Boquien}, {Boulanger}, {Braine}, {Buat}, {Burgarella}, {Combes}, {Contini}, {Cortese}, {C{\^o}t{\'e}}, {C{\^o}t{\'e}}, {Cuillandre}, {Drissen}, {Epinat}, {Fumagalli}, {Gallagher}, {Gavazzi}, {Gomez-Lopez}, {Gwyn}, {Harris}, {Hensler}, {Koribalski}, {Marcelin}, {McConnachie}, {Miville-Deschenes}, {Navarro}, {Patton}, {Peng}, {Plana}, {Prantzos}, {Robert}, {Roediger}, {Roehlly}, {Russeil}, {Salome}, {Sanchez-Janssen}, {Serra}, {Spekkens}, {Sun}, {Taylor}, {Tonnesen}, {Vollmer}, {Willis}, {Wozniak}, {Burdullis}, {Devost}, {Mahoney}, {Manset}, {Petric}, {Prunet}, \& {Withington}}]{boselli18}
{Boselli}, A., {Fossati}, M., {Ferrarese}, L., {et~al.} 2018, \aap, 614, A56

\bibitem[{{Boselli} {et~al.}(2019){Boselli}, {Fossati}, {Longobardi}, {Consolandi}, {Amram}, {Sun}, {Andreani}, {Boquien}, {Braine}, {Combes}, {C{\^o}t{\'e}}, {Cuillandre}, {Duc}, {Emsellem}, {Ferrarese}, {Gavazzi}, {Gwyn}, {Hensler}, {Peng}, {Plana}, {Roediger}, {Sanchez-Janssen}, {Sarzi}, {Serra}, \& {Trinchieri}}]{boselli19}
{Boselli}, A., {Fossati}, M., {Longobardi}, A., {et~al.} 2019, \aap, 623, A52

\bibitem[{{Boselli} {et~al.}(2022){Boselli}, {Fossati}, {Longobardi}, {Kianfar}, {Dametto}, {Amram}, {Anderson}, {Andreani}, {Boissier}, {Boquien}, {Buat}, {Consolandi}, {Cortese}, {C{\^o}t{\'e}}, {Cuillandre}, {Ferrarese}, {Galbany}, {Gavazzi}, {Gwyn}, {Hensler}, {Hutchings}, {Peng}, {Postma}, {Roediger}, {Roehlly}, {Serra}, \& {Trinchieri}}]{boselli22}
{Boselli}, A., {Fossati}, M., {Longobardi}, A., {et~al.} 2022, \aap, 659, A46

\bibitem[{{Boselli} {et~al.}(2014){Boselli}, {Voyer}, {Boissier}, {Cucciati}, {Consolandi}, {Cortese}, {Fumagalli}, {Gavazzi}, {Heinis}, {Roehlly}, \& {Toloba}}]{boselli14}
{Boselli}, A., {Voyer}, E., {Boissier}, S., {et~al.} 2014, \aap, 570, A69

\bibitem[{{Boyett} {et~al.}(2024){Boyett}, {Trenti}, {Leethochawalit}, {Calabr{\'o}}, {Metha}, {Roberts-Borsani}, {Dalmasso}, {Yang}, {Santini}, {Treu}, {Jones}, {Henry}, {Mason}, {Morishita}, {Nanayakkara}, {Roy}, {Wang}, {Fontana}, {Merlin}, {Castellano}, {Paris}, {Brada{\v{c}}}, {Malkan}, {Marchesini}, {Mascia}, {Glazebrook}, {Pentericci}, {Vanzella}, \& {Vulcani}}]{boyett24}
{Boyett}, K., {Trenti}, M., {Leethochawalit}, N., {et~al.} 2024, Nature Astronomy, 8, 657

\bibitem[{{Boylan-Kolchin}(2023)}]{boylan23}
{Boylan-Kolchin}, M. 2023, Nature Astronomy, 7, 731

\bibitem[{Bradley {et~al.}(2023)Bradley, Sip{\H o}cz, Robitaille, Tollerud, Vin{\'{\i}}cius, Deil, Barbary, Wilson, Busko, Donath, G{\"u}nther, Cara, Lim, Me{\ss}linger, Conseil, Bostroem, Droettboom, Bray, Bratholm, Barentsen, Craig, Rathi, Pascual, Perren, Georgiev, de~Val-Borro, Kerzendorf, Bach, Quint, \& Souchereau}]{larry_bradley_2023_7946442}
Bradley, L., Sip{\H o}cz, B., Robitaille, T., {et~al.} 2023, astropy/photutils: 1.8.0

\bibitem[{{Bressan} {et~al.}(1994){Bressan}, {Chiosi}, \& {Fagotto}}]{bressan94}
{Bressan}, A., {Chiosi}, C., \& {Fagotto}, F. 1994, \apjs, 94, 63

\bibitem[{{Brown}(2004)}]{brown04}
{Brown}, T.~M. 2004, \apss, 291, 215

\bibitem[{{Brown} {et~al.}(2000{\natexlab{a}}){Brown}, {Bowers}, {Kimble}, \& {Ferguson}}]{brown00}
{Brown}, T.~M., {Bowers}, C.~W., {Kimble}, R.~A., \& {Ferguson}, H.~C. 2000{\natexlab{a}}, \apjl, 529, L89

\bibitem[{{Brown} {et~al.}(2000{\natexlab{b}}){Brown}, {Bowers}, {Kimble}, {Sweigart}, \& {Ferguson}}]{brown00b}
{Brown}, T.~M., {Bowers}, C.~W., {Kimble}, R.~A., {Sweigart}, A.~V., \& {Ferguson}, H.~C. 2000{\natexlab{b}}, \apj, 532, 308

\bibitem[{{Brown} {et~al.}(1995){Brown}, {Ferguson}, \& {Davidsen}}]{brown95}
{Brown}, T.~M., {Ferguson}, H.~C., \& {Davidsen}, A.~F. 1995, \apjl, 454, L15

\bibitem[{{Brown} {et~al.}(2008){Brown}, {Smith}, {Ferguson}, {Sweigart}, {Kimble}, \& {Bowers}}]{brown08}
{Brown}, T.~M., {Smith}, E., {Ferguson}, H.~C., {et~al.} 2008, \apj, 682, 319

\bibitem[{{Burgarella} {et~al.}(2005){Burgarella}, {Buat}, \& {Iglesias-P{\'a}ramo}}]{burgarella05}
{Burgarella}, D., {Buat}, V., \& {Iglesias-P{\'a}ramo}, J. 2005, \mnras, 360, 1413

\bibitem[{{Burstein} {et~al.}(1988){Burstein}, {Bertola}, {Buson}, {Faber}, \& {Lauer}}]{burstein88}
{Burstein}, D., {Bertola}, F., {Buson}, L.~M., {Faber}, S.~M., \& {Lauer}, T.~R. 1988, \apj, 328, 440

\bibitem[{{Calzetti} {et~al.}(2000){Calzetti}, {Armus}, {Bohlin}, {Kinney}, {Koornneef}, \& {Storchi-Bergmann}}]{calzetti00}
{Calzetti}, D., {Armus}, L., {Bohlin}, R.~C., {et~al.} 2000, \apj, 533, 682

\bibitem[{{Cantiello} {et~al.}(2024){Cantiello}, {Blakeslee}, {Ferrarese}, {C{\^o}t{\'e}}, {Raimondo}, {Cuillandre}, {Durrell}, {Gwyn}, {Hazra}, {Peng}, {Roediger}, {S{\'a}nchez-Janssen}, \& {Kurzner}}]{cantiello24}
{Cantiello}, M., {Blakeslee}, J.~P., {Ferrarese}, L., {et~al.} 2024, \apj, 966, 145

\bibitem[{{Cantiello} {et~al.}(2018){Cantiello}, {Blakeslee}, {Ferrarese}, {C{\^o}t{\'e}}, {Roediger}, {Raimondo}, {Peng}, {Gwyn}, {Durrell}, \& {Cuillandre}}]{cantiello18}
{Cantiello}, M., {Blakeslee}, J.~P., {Ferrarese}, L., {et~al.} 2018, \apj, 856, 126

\bibitem[{{Cappellari} {et~al.}(2011){Cappellari}, {Emsellem}, {Krajnovi{\'c}}, {McDermid}, {Scott}, {Verdoes Kleijn}, {Young}, {Alatalo}, {Bacon}, {Blitz}, {Bois}, {Bournaud}, {Bureau}, {Davies}, {Davis}, {de Zeeuw}, {Duc}, {Khochfar}, {Kuntschner}, {Lablanche}, {Morganti}, {Naab}, {Oosterloo}, {Sarzi}, {Serra}, \& {Weijmans}}]{cappellari11}
{Cappellari}, M., {Emsellem}, E., {Krajnovi{\'c}}, D., {et~al.} 2011, \mnras, 413, 813

\bibitem[{{Cappellari} {et~al.}(2013){Cappellari}, {McDermid}, {Alatalo}, {Blitz}, {Bois}, {Bournaud}, {Bureau}, {Crocker}, {Davies}, {Davis}, {de Zeeuw}, {Duc}, {Emsellem}, {Khochfar}, {Krajnovi{\'c}}, {Kuntschner}, {Morganti}, {Naab}, {Oosterloo}, {Sarzi}, {Scott}, {Serra}, {Weijmans}, \& {Young}}]{cappellari13}
{Cappellari}, M., {McDermid}, R.~M., {Alatalo}, K., {et~al.} 2013, \mnras, 432, 1862

\bibitem[{{Cardelli} {et~al.}(1989){Cardelli}, {Clayton}, \& {Mathis}}]{cardelli89}
{Cardelli}, J.~A., {Clayton}, G.~C., \& {Mathis}, J.~S. 1989, \apj, 345, 245

\bibitem[{{Carlberg}(1984)}]{carlberg84}
{Carlberg}, R.~G. 1984, \apj, 286, 403

\bibitem[{{Carnall} {et~al.}(2024){Carnall}, {Cullen}, {McLure}, {McLeod}, {Begley}, {Donnan}, {Dunlop}, {Shapley}, {Rowlands}, {Almaini}, {Arellano-C{\'o}rdova}, {Barrufet}, {Cimatti}, {Ellis}, {Grogin}, {Hamadouche}, {Illingworth}, {Koekemoer}, {Leung}, {Lovell}, {P{\'e}rez-Gonz{\'a}lez}, {Santini}, {Stanton}, \& {Wild}}]{carnall24}
{Carnall}, A.~C., {Cullen}, F., {McLure}, R.~J., {et~al.} 2024, \mnras, 534, 325

\bibitem[{{Carnall} {et~al.}(2023){Carnall}, {McLure}, {Dunlop}, {McLeod}, {Wild}, {Cullen}, {Magee}, {Begley}, {Cimatti}, {Donnan}, {Hamadouche}, {Jewell}, \& {Walker}}]{carnall23}
{Carnall}, A.~C., {McLure}, R.~J., {Dunlop}, J.~S., {et~al.} 2023, \nat, 619, 716

\bibitem[{{Ciesla} {et~al.}(2014){Ciesla}, {Boquien}, {Boselli}, {Buat}, {Cortese}, {Bendo}, {Heinis}, {Galametz}, {Eales}, {Smith}, {Baes}, {Bianchi}, {De Looze}, {di Serego Alighieri}, {Galliano}, {Hughes}, {Madden}, {Pierini}, {R{\'e}my-Ruyer}, {Spinoglio}, {Vaccari}, {Viaene}, \& {Vlahakis}}]{ciesla14}
{Ciesla}, L., {Boquien}, M., {Boselli}, A., {et~al.} 2014, \aap, 565, A128

\bibitem[{{Ciesla} {et~al.}(2012){Ciesla}, {Boselli}, {Smith}, {Bendo}, {Cortese}, {Eales}, {Bianchi}, {Boquien}, {Buat}, {Davies}, {Pohlen}, {Zibetti}, {Baes}, {Cooray}, {De Looze}, {di Serego Alighieri}, {Galametz}, {Gomez}, {Lebouteiller}, {Madden}, {Pappalardo}, {Remy}, {Spinoglio}, {Vaccari}, {Auld}, \& {Clements}}]{ciesla12}
{Ciesla}, L., {Boselli}, A., {Smith}, M.~W.~L., {et~al.} 2012, \aap, 543, A161

\bibitem[{{Clemens} {et~al.}(2006){Clemens}, {Bressan}, {Nikolic}, {Alexander}, {Annibali}, \& {Rampazzo}}]{clemens06}
{Clemens}, M.~S., {Bressan}, A., {Nikolic}, B., {et~al.} 2006, \mnras, 370, 702

\bibitem[{{Code} \& {Welch}(1979)}]{code79}
{Code}, A.~D. \& {Welch}, G.~A. 1979, \apj, 228, 95

\bibitem[{{Conroy} \& {van Dokkum}(2012)}]{conroy_vandokkum}
{Conroy}, C. \& {van Dokkum}, P.~G. 2012, \apj, 760, 71

\bibitem[{{Cortese} {et~al.}(2012){Cortese}, {Boissier}, {Boselli}, {Bendo}, {Buat}, {Davies}, {Eales}, {Heinis}, {Isaak}, \& {Madden}}]{cortese12}
{Cortese}, L., {Boissier}, S., {Boselli}, A., {et~al.} 2012, \aap, 544, A101

\bibitem[{{Cortese} {et~al.}(2014){Cortese}, {Fritz}, {Bianchi}, {Boselli}, {Ciesla}, {Bendo}, {Boquien}, {Roussel}, {Baes}, {Buat}, {Clemens}, {Cooray}, {Cormier}, {Davies}, {De Looze}, {Eales}, {Fuller}, {Hunt}, {Madden}, {Munoz-Mateos}, {Pappalardo}, {Pierini}, {R{\'e}my-Ruyer}, {Sauvage}, {di Serego Alighieri}, {Smith}, {Spinoglio}, {Vaccari}, \& {Vlahakis}}]{cortese14}
{Cortese}, L., {Fritz}, J., {Bianchi}, S., {et~al.} 2014, \mnras, 440, 942

\bibitem[{{C{\^o}t{\'e}} {et~al.}(2004){C{\^o}t{\'e}}, {Blakeslee}, {Ferrarese}, {Jord{\'a}n}, {Mei}, {Merritt}, {Milosavljevi{\'c}}, {Peng}, {Tonry}, \& {West}}]{cote04}
{C{\^o}t{\'e}}, P., {Blakeslee}, J.~P., {Ferrarese}, L., {et~al.} 2004, \apjs, 153, 223

\bibitem[{{C{\^o}t{\'e}} {et~al.}(1998){C{\^o}t{\'e}}, {Marzke}, \& {West}}]{cote98}
{C{\^o}t{\'e}}, P., {Marzke}, R.~O., \& {West}, M.~J. 1998, \apj, 501, 554

\bibitem[{{C{\^o}t{\'e}} {et~al.}(2003){C{\^o}t{\'e}}, {McLaughlin}, {Cohen}, \& {Blakeslee}}]{cote03}
{C{\^o}t{\'e}}, P., {McLaughlin}, D.~E., {Cohen}, J.~G., \& {Blakeslee}, J.~P. 2003, \apj, 591, 850

\bibitem[{{Dalessandro} {et~al.}(2012){Dalessandro}, {Schiavon}, {Rood}, {Ferraro}, {Sohn}, {Lanzoni}, \& {O'Connell}}]{dalessandro12}
{Dalessandro}, E., {Schiavon}, R.~P., {Rood}, R.~T., {et~al.} 2012, \aj, 144, 126

\bibitem[{{D{\'a}lya} {et~al.}(2022){D{\'a}lya}, {D{\'\i}az}, {Bouchet}, {Frei}, {Jasche}, {Lavaux}, {Macas}, {Mukherjee}, {P{\'a}lfi}, {de Souza}, {Wandelt}, {Bilicki}, \& {Raffai}}]{dalya22}
{D{\'a}lya}, G., {D{\'\i}az}, R., {Bouchet}, F.~R., {et~al.} 2022, \mnras, 514, 1403

\bibitem[{{Dantas} {et~al.}(2020){Dantas}, {Coelho}, {de Souza}, \& {Gon{\c{c}}alves}}]{dantas20}
{Dantas}, M.~L.~L., {Coelho}, P.~R.~T., {de Souza}, R.~S., \& {Gon{\c{c}}alves}, T.~S. 2020, \mnras, 492, 2996

\bibitem[{{de Graaff} {et~al.}(2024){de Graaff}, {Setton}, {Brammer}, {Cutler}, {Suess}, {Labbe}, {Leja}, {Weibel}, {Maseda}, {Whitaker}, {Bezanson}, {Boogaard}, {Cleri}, {De Lucia}, {Franx}, {Greene}, {Hirschmann}, {Matthee}, {McConachie}, {Naidu}, {Oesch}, {Price}, {Rix}, {Valentino}, {Wang}, \& {Williams}}]{degraaff24}
{de Graaff}, A., {Setton}, D.~J., {Brammer}, G., {et~al.} 2024, arXiv e-prints, arXiv:2404.05683

\bibitem[{{De Lucia} {et~al.}(2024){De Lucia}, {Fontanot}, {Xie}, \& {Hirschmann}}]{delucia24}
{De Lucia}, G., {Fontanot}, F., {Xie}, L., \& {Hirschmann}, M. 2024, \aap, 687, A68

\bibitem[{{De Lucia} {et~al.}(2004){De Lucia}, {Kauffmann}, \& {White}}]{delucia04}
{De Lucia}, G., {Kauffmann}, G., \& {White}, S. D.~M. 2004, \mnras, 349, 1101

\bibitem[{{De Lucia} {et~al.}(2006){De Lucia}, {Springel}, {White}, {Croton}, \& {Kauffmann}}]{delucia06}
{De Lucia}, G., {Springel}, V., {White}, S. D.~M., {Croton}, D., \& {Kauffmann}, G. 2006, \mnras, 366, 499

\bibitem[{{De Propris} {et~al.}(2022){De Propris}, {Ali}, {Chung}, {Bremer}, \& {Phillipps}}]{depropris22}
{De Propris}, R., {Ali}, S.~S., {Chung}, C., {Bremer}, M.~N., \& {Phillipps}, S. 2022, \mnras, 512, 1400

\bibitem[{{Di Matteo} {et~al.}(2009){Di Matteo}, {Pipino}, {Lehnert}, {Combes}, \& {Semelin}}]{dimatteo09}
{Di Matteo}, P., {Pipino}, A., {Lehnert}, M.~D., {Combes}, F., \& {Semelin}, B. 2009, \aap, 499, 427

\bibitem[{{di Serego Alighieri} {et~al.}(2013){di Serego Alighieri}, {Bianchi}, {Pappalardo}, {Zibetti}, {Auld}, {Baes}, {Bendo}, {Corbelli}, {Davies}, {Davis}, {De Looze}, {Fritz}, {Gavazzi}, {Giovanardi}, {Grossi}, {Hunt}, {Magrini}, {Pierini}, \& {Xilouris}}]{diserego13}
{di Serego Alighieri}, S., {Bianchi}, S., {Pappalardo}, C., {et~al.} 2013, \aap, 552, A8

\bibitem[{{Donas} {et~al.}(2007){Donas}, {Deharveng}, {Rich}, {Yi}, {Lee}, {Boselli}, {Gil de Paz}, {Boissier}, {Charlot}, {Salim}, {Bianchi}, {Barlow}, {Forster}, {Friedman}, {Heckman}, {Madore}, {Martin}, {Milliard}, {Morrissey}, {Neff}, {Schiminovich}, {Seibert}, {Small}, {Szalay}, {Welsh}, \& {Wyder}}]{donas07}
{Donas}, J., {Deharveng}, J.-M., {Rich}, R.~M., {et~al.} 2007, \apjs, 173, 597

\bibitem[{{Dressler}(1980)}]{dressler80}
{Dressler}, A. 1980, \apj, 236, 351

\bibitem[{{Ehlert} {et~al.}(2013){Ehlert}, {Werner}, {Simionescu}, {Allen}, {Kenney}, {Million}, \& {Finoguenov}}]{ehlert13}
{Ehlert}, S., {Werner}, N., {Simionescu}, A., {et~al.} 2013, \mnras, 430, 2401

\bibitem[{{Elmegreen} {et~al.}(2000){Elmegreen}, {Elmegreen}, {Chromey}, \& {Fine}}]{elmegreen00}
{Elmegreen}, D.~M., {Elmegreen}, B.~G., {Chromey}, F.~R., \& {Fine}, M.~S. 2000, \aj, 120, 733

\bibitem[{{Emsellem} {et~al.}(2011){Emsellem}, {Cappellari}, {Krajnovi{\'c}}, {Alatalo}, {Blitz}, {Bois}, {Bournaud}, {Bureau}, {Davies}, {Davis}, {de Zeeuw}, {Khochfar}, {Kuntschner}, {Lablanche}, {McDermid}, {Morganti}, {Naab}, {Oosterloo}, {Sarzi}, {Scott}, {Serra}, {van de Ven}, {Weijmans}, \& {Young}}]{emsellem11}
{Emsellem}, E., {Cappellari}, M., {Krajnovi{\'c}}, D., {et~al.} 2011, \mnras, 414, 888

\bibitem[{{Emsellem} {et~al.}(2004){Emsellem}, {Cappellari}, {Peletier}, {McDermid}, {Bacon}, {Bureau}, {Copin}, {Davies}, {Krajnovi{\'c}}, {Kuntschner}, {Miller}, \& {de Zeeuw}}]{emsellem04}
{Emsellem}, E., {Cappellari}, M., {Peletier}, R.~F., {et~al.} 2004, \mnras, 352, 721

\bibitem[{{Eskenasy} {et~al.}(2024){Eskenasy}, {Olivares}, {Su}, \& {Li}}]{eskenasy24}
{Eskenasy}, R., {Olivares}, V., {Su}, Y., \& {Li}, Y. 2024, \mnras, 527, 1317

\bibitem[{{Ferrarese} {et~al.}(2012){Ferrarese}, {C{\^o}t{\'e}}, {Cuillandre}, {Gwyn}, {Peng}, {MacArthur}, {Duc}, {Boselli}, {Mei}, {Erben}, {McConnachie}, {Durrell}, {Mihos}, {Jord{\'a}n}, {Lan{\c{c}}on}, {Puzia}, {Emsellem}, {Balogh}, {Blakeslee}, {van Waerbeke}, {Gavazzi}, {Vollmer}, {Kavelaars}, {Woods}, {Ball}, {Boissier}, {Courteau}, {Ferriere}, {Gavazzi}, {Hildebrandt}, {Hudelot}, {Huertas-Company}, {Liu}, {McLaughlin}, {Mellier}, {Milkeraitis}, {Schade}, {Balkowski}, {Bournaud}, {Carlberg}, {Chapman}, {Hoekstra}, {Peng}, {Sawicki}, {Simard}, {Taylor}, {Tully}, {van Driel}, {Wilson}, {Burdullis}, {Mahoney}, \& {Manset}}]{ferrarese12}
{Ferrarese}, L., {C{\^o}t{\'e}}, P., {Cuillandre}, J.-C., {et~al.} 2012, \apjs, 200, 4

\bibitem[{{Ferrarese} {et~al.}(2016){Ferrarese}, {C{\^o}t{\'e}}, {S{\'a}nchez-Janssen}, {Roediger}, {McConnachie}, {Durrell}, {MacArthur}, {Blakeslee}, {Duc}, {Boissier}, {Boselli}, {Courteau}, {Cuillandre}, {Emsellem}, {Gwyn}, {Guhathakurta}, {Jord{\'a}n}, {Lan{\c{c}}on}, {Liu}, {Mei}, {Mihos}, {Navarro}, {Peng}, {Puzia}, {Taylor}, {Toloba}, \& {Zhang}}]{ferrarese16}
{Ferrarese}, L., {C{\^o}t{\'e}}, P., {S{\'a}nchez-Janssen}, R., {et~al.} 2016, \apj, 824, 10

\bibitem[{{Gaia Collaboration} {et~al.}(2023){Gaia Collaboration}, {Vallenari}, {Brown}, {Prusti}, {de Bruijne}, {Arenou}, {Babusiaux}, {Biermann}, {Creevey}, {Ducourant}, {Evans}, {Eyer}, {Guerra}, {Hutton}, {Jordi}, {Klioner}, {Lammers}, {Lindegren}, {Luri}, {Mignard}, {Panem}, {Pourbaix}, {Randich}, {Sartoretti}, {Soubiran}, {Tanga}, {Walton}, {Bailer-Jones}, {Bastian}, {Drimmel}, {Jansen}, {Katz}, {Lattanzi}, {van Leeuwen}, {Bakker}, {Cacciari}, {Casta{\~n}eda}, {De Angeli}, {Fabricius}, {Fouesneau}, {Fr{\'e}mat}, {Galluccio}, {Guerrier}, {Heiter}, {Masana}, {Messineo}, {Mowlavi}, {Nicolas}, {Nienartowicz}, {Pailler}, {Panuzzo}, {Riclet}, {Roux}, {Seabroke}, {Sordo}, {Th{\'e}venin}, {Gracia-Abril}, {Portell}, {Teyssier}, {Altmann}, {Andrae}, {Audard}, {Bellas-Velidis}, {Benson}, {Berthier}, {Blomme}, {Burgess}, {Busonero}, {Busso}, {C{\'a}novas}, {Carry}, {Cellino}, {Cheek}, {Clementini}, {Damerdji}, {Davidson}, {de Teodoro}, {Nu{\~n}ez Campos}, {Delchambre}, {Dell'Oro}, {Esquej},
  {Fern{\'a}ndez-Hern{\'a}ndez}, {Fraile}, {Garabato}, {Garc{\'\i}a-Lario}, {Gosset}, {Haigron}, {Halbwachs}, {Hambly}, {Harrison}, {Hern{\'a}ndez}, {Hestroffer}, {Hodgkin}, {Holl}, {Jan{\ss}en}, {Jevardat de Fombelle}, {Jordan}, {Krone-Martins}, {Lanzafame}, {L{\"o}ffler}, {Marchal}, {Marrese}, {Moitinho}, {Muinonen}, {Osborne}, {Pancino}, {Pauwels}, {Recio-Blanco}, {Reyl{\'e}}, {Riello}, {Rimoldini}, {Roegiers}, {Rybizki}, {Sarro}, {Siopis}, {Smith}, {Sozzetti}, {Utrilla}, {van Leeuwen}, {Abbas}, {{\'A}brah{\'a}m}, {Abreu Aramburu}, {Aerts}, {Aguado}, {Ajaj}, {Aldea-Montero}, {Altavilla}, {{\'A}lvarez}, {Alves}, {Anders}, {Anderson}, {Anglada Varela}, {Antoja}, {Baines}, {Baker}, {Balaguer-N{\'u}{\~n}ez}, {Balbinot}, {Balog}, {Barache}, {Barbato}, {Barros}, {Barstow}, {Bartolom{\'e}}, {Bassilana}, {Bauchet}, {Becciani}, {Bellazzini}, {Berihuete}, {Bernet}, {Bertone}, {Bianchi}, {Binnenfeld}, {Blanco-Cuaresma}, {Blazere}, {Boch}, {Bombrun}, {Bossini}, {Bouquillon}, {Bragaglia}, {Bramante}, {Breedt},
  {Bressan}, {Brouillet}, {Brugaletta}, {Bucciarelli}, {Burlacu}, {Butkevich}, {Buzzi}, {Caffau}, {Cancelliere}, {Cantat-Gaudin}, {Carballo}, {Carlucci}, {Carnerero}, {Carrasco}, {Casamiquela}, {Castellani}, {Castro-Ginard}, {Chaoul}, {Charlot}, {Chemin}, {Chiaramida}, {Chiavassa}, {Chornay}, {Comoretto}, {Contursi}, {Cooper}, {Cornez}, {Cowell}, {Crifo}, {Cropper}, {Crosta}, {Crowley}, {Dafonte}, {Dapergolas}, {David}, {David}, {de Laverny}, {De Luise}, {De March}, {De Ridder}, {de Souza}, {de Torres}, {del Peloso}, {del Pozo}, {Delbo}, {Delgado}, {Delisle}, {Demouchy}, {Dharmawardena}, {Di Matteo}, {Diakite}, {Diener}, {Distefano}, {Dolding}, {Edvardsson}, {Enke}, {Fabre}, {Fabrizio}, {Faigler}, {Fedorets}, {Fernique}, {Fienga}, {Figueras}, {Fournier}, {Fouron}, {Fragkoudi}, {Gai}, {Garcia-Gutierrez}, {Garcia-Reinaldos}, {Garc{\'\i}a-Torres}, {Garofalo}, {Gavel}, {Gavras}, {Gerlach}, {Geyer}, {Giacobbe}, {Gilmore}, {Girona}, {Giuffrida}, {Gomel}, {Gomez}, {Gonz{\'a}lez-N{\'u}{\~n}ez},
  {Gonz{\'a}lez-Santamar{\'\i}a}, {Gonz{\'a}lez-Vidal}, {Granvik}, {Guillout}, {Guiraud}, {Guti{\'e}rrez-S{\'a}nchez}, {Guy}, {Hatzidimitriou}, {Hauser}, {Haywood}, {Helmer}, {Helmi}, {Sarmiento}, {Hidalgo}, {Hilger}, {H{\l}adczuk}, {Hobbs}, {Holland}, {Huckle}, {Jardine}, {Jasniewicz}, {Jean-Antoine Piccolo}, {Jim{\'e}nez-Arranz}, {Jorissen}, {Juaristi Campillo}, {Julbe}, {Karbevska}, {Kervella}, {Khanna}, {Kontizas}, {Kordopatis}, {Korn}, {K{\'o}sp{\'a}l}, {Kostrzewa-Rutkowska}, {Kruszy{\'n}ska}, {Kun}, {Laizeau}, {Lambert}, {Lanza}, {Lasne}, {Le Campion}, {Lebreton}, {Lebzelter}, {Leccia}, {Leclerc}, {Lecoeur-Taibi}, {Liao}, {Licata}, {Lindstr{\o}m}, {Lister}, {Livanou}, {Lobel}, {Lorca}, {Loup}, {Madrero Pardo}, {Magdaleno Romeo}, {Managau}, {Mann}, {Manteiga}, {Marchant}, {Marconi}, {Marcos}, {Marcos Santos}, {Mar{\'\i}n Pina}, {Marinoni}, {Marocco}, {Marshall}, {Martin Polo}, {Mart{\'\i}n-Fleitas}, {Marton}, {Mary}, {Masip}, {Massari}, {Mastrobuono-Battisti}, {Mazeh}, {McMillan}, {Messina}, {Michalik},
  {Millar}, {Mints}, {Molina}, {Molinaro}, {Moln{\'a}r}, {Monari}, {Mongui{\'o}}, {Montegriffo}, {Montero}, {Mor}, {Mora}, {Morbidelli}, {Morel}, {Morris}, {Muraveva}, {Murphy}, {Musella}, {Nagy}, {Noval}, {Oca{\~n}a}, {Ogden}, {Ordenovic}, {Osinde}, {Pagani}, {Pagano}, {Palaversa}, {Palicio}, {Pallas-Quintela}, {Panahi}, {Payne-Wardenaar}, {Pe{\~n}alosa Esteller}, {Penttil{\"a}}, {Pichon}, {Piersimoni}, {Pineau}, {Plachy}, {Plum}, {Poggio}, {Pr{\v{s}}a}, {Pulone}, {Racero}, {Ragaini}, {Rainer}, {Raiteri}, {Rambaux}, {Ramos}, {Ramos-Lerate}, {Re Fiorentin}, {Regibo}, {Richards}, {Rios Diaz}, {Ripepi}, {Riva}, {Rix}, {Rixon}, {Robichon}, {Robin}, {Robin}, {Roelens}, {Rogues}, {Rohrbasser}, {Romero-G{\'o}mez}, {Rowell}, {Royer}, {Ruz Mieres}, {Rybicki}, {Sadowski}, {S{\'a}ez N{\'u}{\~n}ez}, {Sagrist{\`a} Sell{\'e}s}, {Sahlmann}, {Salguero}, {Samaras}, {Sanchez Gimenez}, {Sanna}, {Santove{\~n}a}, {Sarasso}, {Schultheis}, {Sciacca}, {Segol}, {Segovia}, {S{\'e}gransan}, {Semeux}, {Shahaf}, {Siddiqui}, {Siebert},
  {Siltala}, {Silvelo}, {Slezak}, {Slezak}, {Smart}, {Snaith}, {Solano}, {Solitro}, {Souami}, {Souchay}, {Spagna}, {Spina}, {Spoto}, {Steele}, {Steidelm{\"u}ller}, {Stephenson}, {S{\"u}veges}, {Surdej}, {Szabados}, {Szegedi-Elek}, {Taris}, {Taylor}, {Teixeira}, {Tolomei}, {Tonello}, {Torra}, {Torra}, {Torralba Elipe}, {Trabucchi}, {Tsounis}, {Turon}, {Ulla}, {Unger}, {Vaillant}, {van Dillen}, {van Reeven}, {Vanel}, {Vecchiato}, {Viala}, {Vicente}, {Voutsinas}, {Weiler}, {Wevers}, {Wyrzykowski}, {Yoldas}, {Yvard}, {Zhao}, {Zorec}, {Zucker}, \& {Zwitter}}]{gaiadr3}
{Gaia Collaboration}, {Vallenari}, A., {Brown}, A.~G.~A., {et~al.} 2023, \aap, 674, A1

\bibitem[{{Gavazzi} {et~al.}(2002){Gavazzi}, {Bonfanti}, {Sanvito}, {Boselli}, \& {Scodeggio}}]{gavazzi02}
{Gavazzi}, G., {Bonfanti}, C., {Sanvito}, G., {Boselli}, A., \& {Scodeggio}, M. 2002, \apj, 576, 135

\bibitem[{{Gavazzi} {et~al.}(1999){Gavazzi}, {Boselli}, {Scodeggio}, {Pierini}, \& {Belsole}}]{gavazzi99}
{Gavazzi}, G., {Boselli}, A., {Scodeggio}, M., {Pierini}, D., \& {Belsole}, E. 1999, \mnras, 304, 595

\bibitem[{{Gavazzi} {et~al.}(2018){Gavazzi}, {Consolandi}, {Pedraglio}, {Fossati}, {Fumagalli}, \& {Boselli}}]{gavazzi18}
{Gavazzi}, G., {Consolandi}, G., {Pedraglio}, S., {et~al.} 2018, \aap, 611, A28

\bibitem[{{Gavazzi} {et~al.}(2013){Gavazzi}, {Savorgnan}, {Fossati}, {Dotti}, {Fumagalli}, {Boselli}, {Guti{\'e}rrez}, {Hern{\'a}ndez Toledo}, {Giovanelli}, \& {Haynes}}]{gavazzi13}
{Gavazzi}, G., {Savorgnan}, G., {Fossati}, M., {et~al.} 2013, \aap, 553, A90

\bibitem[{{Gil de Paz} {et~al.}(2007){Gil de Paz}, {Boissier}, {Madore}, {Seibert}, {Joe}, {Boselli}, {Wyder}, {Thilker}, {Bianchi}, {Rey}, {Rich}, {Barlow}, {Conrow}, {Forster}, {Friedman}, {Martin}, {Morrissey}, {Neff}, {Schiminovich}, {Small}, {Donas}, {Heckman}, {Lee}, {Milliard}, {Szalay}, \& {Yi}}]{gildepaz07}
{Gil de Paz}, A., {Boissier}, S., {Madore}, B.~F., {et~al.} 2007, \apjs, 173, 185

\bibitem[{{Glazebrook} {et~al.}(2024){Glazebrook}, {Nanayakkara}, {Schreiber}, {Lagos}, {Kawinwanichakij}, {Jacobs}, {Chittenden}, {Brammer}, {Kacprzak}, {Labbe}, {Marchesini}, {Marsan}, {Oesch}, {Papovich}, {Remus}, {Tran}, {Esdaile}, \& {Chandro-Gomez}}]{glazebrook24}
{Glazebrook}, K., {Nanayakkara}, T., {Schreiber}, C., {et~al.} 2024, \nat, 628, 277

\bibitem[{{Goddard} {et~al.}(2017){Goddard}, {Thomas}, {Maraston}, {Westfall}, {Etherington}, {Riffel}, {Mallmann}, {Zheng}, {Argudo-Fern{\'a}ndez}, {Lian}, {Bershady}, {Bundy}, {Drory}, {Law}, {Yan}, {Wake}, {Weijmans}, {Bizyaev}, {Brownstein}, {Lane}, {Maiolino}, {Masters}, {Merrifield}, {Nitschelm}, {Pan}, {Roman-Lopes}, {Storchi-Bergmann}, \& {Schneider}}]{goddard17}
{Goddard}, D., {Thomas}, D., {Maraston}, C., {et~al.} 2017, \mnras, 466, 4731

\bibitem[{{Gomez} {et~al.}(2010){Gomez}, {Baes}, {Cortese}, {Smith}, {Boselli}, {Ciesla}, {Bendo}, {Pohlen}, {di Serego Alighieri}, {Auld}, {Barlow}, {Bock}, {Bradford}, {Buat}, {Castro-Rodriguez}, {Chanial}, {Charlot}, {Clements}, {Cooray}, {Cormier}, {Davies}, {Dwek}, {Eales}, {Elbaz}, {Galametz}, {Galliano}, {Gear}, {Glenn}, {Griffin}, {Hony}, {Isaak}, {Levenson}, {Lu}, {Madden}, {O'Halloran}, {Okumura}, {Oliver}, {Page}, {Panuzzo}, {Papageorgiou}, {Parkin}, {Perez-Fournon}, {Rangwala}, {Rigby}, {Roussel}, {Rykala}, {Sacchi}, {Sauvage}, {Schirm}, {Schulz}, {Spinoglio}, {Srinivasan}, {Stevens}, {Symeonidis}, {Trichas}, {Vaccari}, {Vigroux}, {Wilson}, {Wozniak}, {Wright}, \& {Zeilinger}}]{gomez10}
{Gomez}, H.~L., {Baes}, M., {Cortese}, L., {et~al.} 2010, \aap, 518, L45

\bibitem[{{Gonz{\'a}lez Delgado} {et~al.}(2015){Gonz{\'a}lez Delgado}, {Garc{\'\i}a-Benito}, {P{\'e}rez}, {Cid Fernandes}, {de Amorim}, {Cortijo-Ferrero}, {Lacerda}, {L{\'o}pez Fern{\'a}ndez}, {Vale-Asari}, {S{\'a}nchez}, {Moll{\'a}}, {Ruiz-Lara}, {S{\'a}nchez-Bl{\'a}zquez}, {Walcher}, {Alves}, {Aguerri}, {Bekerait{\'e}}, {Bland-Hawthorn}, {Galbany}, {Gallazzi}, {Husemann}, {Iglesias-P{\'a}ramo}, {Kalinova}, {L{\'o}pez-S{\'a}nchez}, {Marino}, {M{\'a}rquez}, {Masegosa}, {Mast}, {M{\'e}ndez-Abreu}, {Mendoza}, {del Olmo}, {P{\'e}rez}, {Quirrenbach}, \& {Zibetti}}]{gonzalez-delgado15}
{Gonz{\'a}lez Delgado}, R.~M., {Garc{\'\i}a-Benito}, R., {P{\'e}rez}, E., {et~al.} 2015, \aap, 581, A103

\bibitem[{{Greene} {et~al.}(2013){Greene}, {Murphy}, {Graves}, {Gunn}, {Raskutti}, {Comerford}, \& {Gebhardt}}]{greene13}
{Greene}, J.~E., {Murphy}, J.~D., {Graves}, G.~J., {et~al.} 2013, \apj, 776, 64

\bibitem[{{Greggio} \& {Renzini}(1990)}]{greggio90}
{Greggio}, L. \& {Renzini}, A. 1990, \apj, 364, 35

\bibitem[{{Han} {et~al.}(2010){Han}, {Podsiadlowski}, \& {Lynas-Gray}}]{han10}
{Han}, Z., {Podsiadlowski}, P., \& {Lynas-Gray}, A. 2010, \apss, 329, 41

\bibitem[{{Han} {et~al.}(2007){Han}, {Podsiadlowski}, \& {Lynas-Gray}}]{han07}
{Han}, Z., {Podsiadlowski}, P., \& {Lynas-Gray}, A.~E. 2007, \mnras, 380, 1098

\bibitem[{{Hansen} {et~al.}(1985){Hansen}, {Norgaard-Nielsen}, \& {Jorgensen}}]{hansen85}
{Hansen}, L., {Norgaard-Nielsen}, H.~U., \& {Jorgensen}, H.~E. 1985, \aap, 149, 442

\bibitem[{{Hern{\'a}ndez-P{\'e}rez} \& {Bruzual}(2013)}]{hernandez13}
{Hern{\'a}ndez-P{\'e}rez}, F. \& {Bruzual}, G. 2013, \mnras, 431, 2612

\bibitem[{{Hirschmann} {et~al.}(2015){Hirschmann}, {Naab}, {Ostriker}, {Forbes}, {Duc}, {Dav{\'e}}, {Oser}, \& {Karabal}}]{hirschmann15}
{Hirschmann}, M., {Naab}, T., {Ostriker}, J.~P., {et~al.} 2015, \mnras, 449, 528

\bibitem[{{Ho} {et~al.}(1995){Ho}, {Filippenko}, \& {Sargent}}]{ho95}
{Ho}, L.~C., {Filippenko}, A.~V., \& {Sargent}, W.~L. 1995, \apjs, 98, 477

\bibitem[{{Hubble}(1926)}]{hubble26}
{Hubble}, E.~P. 1926, \apj, 64, 321

\bibitem[{{Janowiecki} {et~al.}(2010){Janowiecki}, {Mihos}, {Harding}, {Feldmeier}, {Rudick}, \& {Morrison}}]{janowiecki10}
{Janowiecki}, S., {Mihos}, J.~C., {Harding}, P., {et~al.} 2010, \apj, 715, 972

\bibitem[{{Jedrzejewski}(1987)}]{ellipse}
{Jedrzejewski}, R.~I. 1987, \mnras, 226, 747

\bibitem[{{Junais} {et~al.}(2022){Junais}, {Boissier}, {Boselli}, {Ferrarese}, {C{\^o}t{\'e}}, {Gwyn}, {Roediger}, {Lim}, {Peng}, {Cuillandre}, {Longobardi}, {Fossati}, {Hensler}, {Koda}, {Bautista}, {Boquien}, {Ma{\l}ek}, {Amram}, \& {Roehlly}}]{junais22}
{Junais}, {Boissier}, S., {Boselli}, A., {et~al.} 2022, \aap, 667, A76

\bibitem[{{Kaviraj} {et~al.}(2007){Kaviraj}, {Schawinski}, {Devriendt}, {Ferreras}, {Khochfar}, {Yoon}, {Yi}, {Deharveng}, {Boselli}, {Barlow}, {Conrow}, {Forster}, {Friedman}, {Martin}, {Morrissey}, {Neff}, {Schiminovich}, {Seibert}, {Small}, {Wyder}, {Bianchi}, {Donas}, {Heckman}, {Lee}, {Madore}, {Milliard}, {Rich}, \& {Szalay}}]{kaviraj07}
{Kaviraj}, S., {Schawinski}, K., {Devriendt}, J.~E.~G., {et~al.} 2007, \apjs, 173, 619

\bibitem[{{Kawata} \& {Gibson}(2003)}]{kawata03}
{Kawata}, D. \& {Gibson}, B.~K. 2003, \mnras, 346, 135

\bibitem[{{Kenney} {et~al.}(2008){Kenney}, {Tal}, {Crowl}, {Feldmeier}, \& {Jacoby}}]{kenney08}
{Kenney}, J. D.~P., {Tal}, T., {Crowl}, H.~H., {Feldmeier}, J., \& {Jacoby}, G.~H. 2008, \apjl, 687, L69

\bibitem[{{Kennicutt}(1998)}]{kennicutt98}
{Kennicutt}, Robert~C., J. 1998, \araa, 36, 189

\bibitem[{{Kimmig} {et~al.}(2023){Kimmig}, {Remus}, {Seidel}, {Valenzuela}, {Dolag}, \& {Burkert}}]{kimmig23}
{Kimmig}, L.~C., {Remus}, R.-S., {Seidel}, B., {et~al.} 2023, arXiv e-prints, arXiv:2310.16085

\bibitem[{{Kobayashi}(2004)}]{kobayashi04}
{Kobayashi}, C. 2004, \mnras, 347, 740

\bibitem[{{Kulkarni} {et~al.}(2014){Kulkarni}, {Sahu}, {Chaware}, {Chakradhari}, \& {Pandey}}]{kulkarni14}
{Kulkarni}, S., {Sahu}, D.~K., {Chaware}, L., {Chakradhari}, N.~K., \& {Pandey}, S.~K. 2014, \na, 30, 51

\bibitem[{{Kuntschner} {et~al.}(2010){Kuntschner}, {Emsellem}, {Bacon}, {Cappellari}, {Davies}, {de Zeeuw}, {Falc{\'o}n-Barroso}, {Krajnovi{\'c}}, {McDermid}, {Peletier}, {Sarzi}, {Shapiro}, {van den Bosch}, \& {van de Ven}}]{kuntschner10}
{Kuntschner}, H., {Emsellem}, E., {Bacon}, R., {et~al.} 2010, \mnras, 408, 97

\bibitem[{{La Barbera} {et~al.}(2012){La Barbera}, {Ferreras}, {de Carvalho}, {Bruzual}, {Charlot}, {Pasquali}, \& {Merlin}}]{labarbera12}
{La Barbera}, F., {Ferreras}, I., {de Carvalho}, R.~R., {et~al.} 2012, \mnras, 426, 2300

\bibitem[{{Labb{\'e}} {et~al.}(2023){Labb{\'e}}, {van Dokkum}, {Nelson}, {Bezanson}, {Suess}, {Leja}, {Brammer}, {Whitaker}, {Mathews}, {Stefanon}, \& {Wang}}]{labbe23}
{Labb{\'e}}, I., {van Dokkum}, P., {Nelson}, E., {et~al.} 2023, \nat, 616, 266

\bibitem[{{Lagos} {et~al.}(2022){Lagos}, {Emsellem}, {van de Sande}, {Harborne}, {Cortese}, {Davison}, {Foster}, \& {Wright}}]{lagos22}
{Lagos}, C. d.~P., {Emsellem}, E., {van de Sande}, J., {et~al.} 2022, \mnras, 509, 4372

\bibitem[{{Laing} \& {Bridle}(1987)}]{laing_bridle87}
{Laing}, R.~A. \& {Bridle}, A.~H. 1987, \mnras, 228, 557

\bibitem[{{Lambert} {et~al.}(2020){Lambert}, {Rhode}, \& {Vesperini}}]{lambert20}
{Lambert}, R.~A., {Rhode}, K.~L., \& {Vesperini}, E. 2020, \apj, 900, 45

\bibitem[{{Laporte} {et~al.}(2013){Laporte}, {White}, {Naab}, \& {Gao}}]{laporte13}
{Laporte}, C. F.~P., {White}, S. D.~M., {Naab}, T., \& {Gao}, L. 2013, \mnras, 435, 901

\bibitem[{{Larsen} {et~al.}(2001){Larsen}, {Brodie}, {Huchra}, {Forbes}, \& {Grillmair}}]{larsen01}
{Larsen}, S.~S., {Brodie}, J.~P., {Huchra}, J.~P., {Forbes}, D.~A., \& {Grillmair}, C.~J. 2001, \aj, 121, 2974

\bibitem[{{Larson}(1974)}]{larson74}
{Larson}, R.~B. 1974, \mnras, 166, 585

\bibitem[{{Le Cras} {et~al.}(2016){Le Cras}, {Maraston}, {Thomas}, \& {York}}]{lecras16}
{Le Cras}, C., {Maraston}, C., {Thomas}, D., \& {York}, D.~G. 2016, \mnras, 461, 766

\bibitem[{{Lequeux}(1988)}]{lequeux88}
{Lequeux}, J. 1988, in Millimetre and Submillimetre Astronomy, ed. R.~D. {Wolstencroft} \& W.~B. {Burton}, Vol. 147, 249

\bibitem[{{Liu} {et~al.}(2011){Liu}, {Peng}, {Jord{\'a}n}, {Ferrarese}, {Blakeslee}, {C{\^o}t{\'e}}, \& {Mei}}]{liu11}
{Liu}, C., {Peng}, E.~W., {Jord{\'a}n}, A., {et~al.} 2011, \apj, 728, 116

\bibitem[{{Longobardi} {et~al.}(2015){Longobardi}, {Arnaboldi}, {Gerhard}, \& {Mihos}}]{longobardi15}
{Longobardi}, A., {Arnaboldi}, M., {Gerhard}, O., \& {Mihos}, J.~C. 2015, \aap, 579, L3

\bibitem[{{Longobardi} {et~al.}(2018){Longobardi}, {Peng}, {C{\^o}t{\'e}}, {Mihos}, {Ferrarese}, {Puzia}, {Lan{\c{c}}on}, {Zhang}, {Mu{\~n}oz}, {Blakeslee}, {Guhathakurta}, {Durrell}, {S{\'a}nchez-Janssen}, {Toloba}, {Jord{\'a}n}, {Eyheramendy}, {Cuillandre}, {Gwyn}, {Boselli}, {Duc}, {Liu}, {Alamo-Mart{\'\i}nez}, {Powalka}, \& {Lim}}]{longobardi18}
{Longobardi}, A., {Peng}, E.~W., {C{\^o}t{\'e}}, P., {et~al.} 2018, \apj, 864, 36

\bibitem[{{Lonoce} {et~al.}(2020){Lonoce}, {Maraston}, {Thomas}, {Longhetti}, {Parikh}, {Guarnieri}, \& {Comparat}}]{lonoce20}
{Lonoce}, I., {Maraston}, C., {Thomas}, D., {et~al.} 2020, \mnras, 492, 326

\bibitem[{{Lovell} {et~al.}(2023){Lovell}, {Roper}, {Vijayan}, {Seeyave}, {Irodotou}, {Wilkins}, {Conselice}, {Fortuni}, {Kuusisto}, {Merlin}, {Santini}, \& {Thomas}}]{lovell23}
{Lovell}, C.~C., {Roper}, W., {Vijayan}, A.~P., {et~al.} 2023, \mnras, 525, 5520

\bibitem[{{Lu} {et~al.}(2023){Lu}, {Zhu}, {Cappellari}, {Li}, {Mao}, \& {Xu}}]{lu23}
{Lu}, S., {Zhu}, K., {Cappellari}, M., {et~al.} 2023, \mnras, 526, 1022

\bibitem[{{Man} \& {Belli}(2018)}]{man18}
{Man}, A. \& {Belli}, S. 2018, Nature Astronomy, 2, 695

\bibitem[{{Maraston}(1998)}]{maraston98}
{Maraston}, C. 1998, \mnras, 300, 872

\bibitem[{{Maraston}(2005)}]{maraston05}
{Maraston}, C. 2005, \mnras, 362, 799

\bibitem[{{Maraston} \& {Thomas}(2000)}]{mt00}
{Maraston}, C. \& {Thomas}, D. 2000, \apj, 541, 126

\bibitem[{{Martig} {et~al.}(2009){Martig}, {Bournaud}, {Teyssier}, \& {Dekel}}]{martig09}
{Martig}, M., {Bournaud}, F., {Teyssier}, R., \& {Dekel}, A. 2009, \apj, 707, 250

\bibitem[{{Mart{\'\i}n-Navarro} {et~al.}(2015){Mart{\'\i}n-Navarro}, {La Barbera}, {Vazdekis}, {Falc{\'o}n-Barroso}, \& {Ferreras}}]{martinnavarro15}
{Mart{\'\i}n-Navarro}, I., {La Barbera}, F., {Vazdekis}, A., {Falc{\'o}n-Barroso}, J., \& {Ferreras}, I. 2015, \mnras, 447, 1033

\bibitem[{{Mehlert} {et~al.}(2003){Mehlert}, {Thomas}, {Saglia}, {Bender}, \& {Wegner}}]{mehlert03}
{Mehlert}, D., {Thomas}, D., {Saglia}, R.~P., {Bender}, R., \& {Wegner}, G. 2003, \aap, 407, 423

\bibitem[{{Mei} {et~al.}(2007){Mei}, {Blakeslee}, {C{\^o}t{\'e}}, {Tonry}, {West}, {Ferrarese}, {Jord{\'a}n}, {Peng}, {Anthony}, \& {Merritt}}]{mei07}
{Mei}, S., {Blakeslee}, J.~P., {C{\^o}t{\'e}}, P., {et~al.} 2007, \apj, 655, 144

\bibitem[{{Menci} {et~al.}(2022){Menci}, {Castellano}, {Santini}, {Merlin}, {Fontana}, \& {Shankar}}]{menci22}
{Menci}, N., {Castellano}, M., {Santini}, P., {et~al.} 2022, \apjl, 938, L5

\bibitem[{{Meyer} {et~al.}(2018){Meyer}, {Petropoulou}, {Georganopoulos}, {Chiaberge}, {Breiding}, \& {Sparks}}]{meyer18}
{Meyer}, E.~T., {Petropoulou}, M., {Georganopoulos}, M., {et~al.} 2018, \apj, 860, 9

\bibitem[{{Morrissey} {et~al.}(2007){Morrissey}, {Conrow}, {Barlow}, {Small}, {Seibert}, {Wyder}, {Budav{\'a}ri}, {Arnouts}, {Friedman}, {Forster}, {Martin}, {Neff}, {Schiminovich}, {Bianchi}, {Donas}, {Heckman}, {Lee}, {Madore}, {Milliard}, {Rich}, {Szalay}, {Welsh}, \& {Yi}}]{morrissey07}
{Morrissey}, P., {Conrow}, T., {Barlow}, T.~A., {et~al.} 2007, \apjs, 173, 682

\bibitem[{{Naab} {et~al.}(2014){Naab}, {Oser}, {Emsellem}, {Cappellari}, {Krajnovi{\'c}}, {McDermid}, {Alatalo}, {Bayet}, {Blitz}, {Bois}, {Bournaud}, {Bureau}, {Crocker}, {Davies}, {Davis}, {de Zeeuw}, {Duc}, {Hirschmann}, {Johansson}, {Khochfar}, {Kuntschner}, {Morganti}, {Oosterloo}, {Sarzi}, {Scott}, {Serra}, {van de Ven}, {Weijmans}, \& {Young}}]{naab14}
{Naab}, T., {Oser}, L., {Emsellem}, E., {et~al.} 2014, \mnras, 444, 3357

\bibitem[{{Naab} \& {Ostriker}(2017)}]{naab17}
{Naab}, T. \& {Ostriker}, J.~P. 2017, \araa, 55, 59

\bibitem[{{Nanayakkara} {et~al.}(2024){Nanayakkara}, {Glazebrook}, {Jacobs}, {Kawinwanichakij}, {Schreiber}, {Brammer}, {Esdaile}, {Kacprzak}, {Labbe}, {Lagos}, {Marchesini}, {Marsan}, {Oesch}, {Papovich}, {Remus}, \& {Tran}}]{nanayakkara24}
{Nanayakkara}, T., {Glazebrook}, K., {Jacobs}, C., {et~al.} 2024, Scientific Reports, 14, 3724

\bibitem[{{Nersesian} {et~al.}(2019){Nersesian}, {Xilouris}, {Bianchi}, {Galliano}, {Jones}, {Baes}, {Casasola}, {Cassar{\`a}}, {Clark}, {Davies}, {Decleir}, {Dobbels}, {De Looze}, {De Vis}, {Fritz}, {Galametz}, {Madden}, {Mosenkov}, {Tr{\v{c}}ka}, {Verstocken}, {Viaene}, \& {Lianou}}]{nersesian19}
{Nersesian}, A., {Xilouris}, E.~M., {Bianchi}, S., {et~al.} 2019, \aap, 624, A80

\bibitem[{{Noll} {et~al.}(2009){Noll}, {Burgarella}, {Giovannoli}, {Buat}, {Marcillac}, \& {Mu{\~n}oz-Mateos}}]{noll09}
{Noll}, S., {Burgarella}, D., {Giovannoli}, E., {et~al.} 2009, \aap, 507, 1793

\bibitem[{{O'Connell}(1999)}]{oconnell99}
{O'Connell}, R.~W. 1999, \araa, 37, 603

\bibitem[{{O'Connell} {et~al.}(1992){O'Connell}, {Bohlin}, {Collins}, {Cornett}, {Hill}, {Hill}, {Landsman}, {Roberts}, {Smith}, \& {Stecher}}]{oconnell92}
{O'Connell}, R.~W., {Bohlin}, R.~C., {Collins}, N.~R., {et~al.} 1992, \apjl, 395, L45

\bibitem[{{Ogando} {et~al.}(2005){Ogando}, {Maia}, {Chiappini}, {Pellegrini}, {Schiavon}, \& {da Costa}}]{ogando05}
{Ogando}, R. L.~C., {Maia}, M. A.~G., {Chiappini}, C., {et~al.} 2005, \apjl, 632, L61

\bibitem[{{Ohl} {et~al.}(1998){Ohl}, {O'Connell}, {Bohlin}, {Collins}, {Dorman}, {Fanelli}, {Neff}, {Roberts}, {Smith}, \& {Stecher}}]{ohl98}
{Ohl}, R.~G., {O'Connell}, R.~W., {Bohlin}, R.~C., {et~al.} 1998, \apjl, 505, L11

\bibitem[{{Oosterloo} {et~al.}(2010){Oosterloo}, {Morganti}, {Crocker}, {J{\"u}tte}, {Cappellari}, {de Zeeuw}, {Krajnovi{\'c}}, {McDermid}, {Kuntschner}, {Sarzi}, \& {Weijmans}}]{oosterloo10}
{Oosterloo}, T., {Morganti}, R., {Crocker}, A., {et~al.} 2010, \mnras, 409, 500

\bibitem[{{Oser} {et~al.}(2010){Oser}, {Ostriker}, {Naab}, {Johansson}, \& {Burkert}}]{oser10}
{Oser}, L., {Ostriker}, J.~P., {Naab}, T., {Johansson}, P.~H., \& {Burkert}, A. 2010, \apj, 725, 2312

\bibitem[{{Owen} {et~al.}(2000){Owen}, {Eilek}, \& {Kassim}}]{owen00}
{Owen}, F.~N., {Eilek}, J.~A., \& {Kassim}, N.~E. 2000, \apj, 543, 611

\bibitem[{{Oyarz{\'u}n} {et~al.}(2019){Oyarz{\'u}n}, {Bundy}, {Westfall}, {Belfiore}, {Thomas}, {Maraston}, {Lian}, {Arag{\'o}n-Salamanca}, {Zheng}, {Gonzalez-Perez}, {Law}, {Drory}, \& {Andrews}}]{oyarzun19}
{Oyarz{\'u}n}, G.~A., {Bundy}, K., {Westfall}, K.~B., {et~al.} 2019, \apj, 880, 111

\bibitem[{{Parikh} {et~al.}(2019){Parikh}, {Thomas}, {Maraston}, {Westfall}, {Lian}, {Fraser-McKelvie}, {Andrews}, {Drory}, \& {Meneses-Goytia}}]{parikh19}
{Parikh}, T., {Thomas}, D., {Maraston}, C., {et~al.} 2019, \mnras, 483, 3420

\bibitem[{{Pasquali} {et~al.}(2019){Pasquali}, {Smith}, {Gallazzi}, {De Lucia}, {Zibetti}, {Hirschmann}, \& {Yi}}]{pasquali19}
{Pasquali}, A., {Smith}, R., {Gallazzi}, A., {et~al.} 2019, \mnras, 484, 1702

\bibitem[{{Peletier} {et~al.}(1990){Peletier}, {Davies}, {Illingworth}, {Davis}, \& {Cawson}}]{peletier90}
{Peletier}, R.~F., {Davies}, R.~L., {Illingworth}, G.~D., {Davis}, L.~E., \& {Cawson}, M. 1990, \aj, 100, 1091

\bibitem[{{Peng} {et~al.}(2006){Peng}, {Jord{\'a}n}, {C{\^o}t{\'e}}, {Blakeslee}, {Ferrarese}, {Mei}, {West}, {Merritt}, {Milosavljevi{\'c}}, \& {Tonry}}]{peng06}
{Peng}, E.~W., {Jord{\'a}n}, A., {C{\^o}t{\'e}}, P., {et~al.} 2006, \apj, 639, 95

\bibitem[{{Phillipps} {et~al.}(2020){Phillipps}, {Ali}, {Bremer}, {De Propris}, {Sansom}, {Cluver}, {Alpaslan}, {Brough}, {Brown}, {Davies}, {Driver}, {Grootes}, {Holwerda}, {Hopkins}, {James}, {Pimbblet}, {Robotham}, {Taylor}, \& {Wang}}]{phillips20}
{Phillipps}, S., {Ali}, S.~S., {Bremer}, M.~N., {et~al.} 2020, \mnras, 492, 2128

\bibitem[{{Pipino} {et~al.}(2010){Pipino}, {D'Ercole}, {Chiappini}, \& {Matteucci}}]{pipino10}
{Pipino}, A., {D'Ercole}, A., {Chiappini}, C., \& {Matteucci}, F. 2010, \mnras, 407, 1347

\bibitem[{{Prabhu} {et~al.}(2022){Prabhu}, {Subramaniam}, {Sahu}, {Chung}, {Leigh}, {Dalessandro}, {Chatterjee}, {Rao}, {Shara}, {C{\^o}t{\'e}}, {Choudhury}, {Pandey}, {Valcarce}, {Singh}, {Postma}, {Rani}, {Bandyopadhyay}, {Geller}, {Hutchings}, {Puzia}, {Simunovic}, {Sohn}, {Thirupathi}, \& {Yadav}}]{prabhu22}
{Prabhu}, D.~S., {Subramaniam}, A., {Sahu}, S., {et~al.} 2022, \apjl, 939, L20

\bibitem[{{Rampazzo} {et~al.}(2017){Rampazzo}, {Mazzei}, {Marino}, {Uslenghi}, {Trinchieri}, \& {Wolter}}]{rampazzo17}
{Rampazzo}, R., {Mazzei}, P., {Marino}, A., {et~al.} 2017, \aap, 602, A97

\bibitem[{{Rampazzo} {et~al.}(2013){Rampazzo}, {Panuzzo}, {Vega}, {Marino}, {Bressan}, \& {Clemens}}]{rampazzo13}
{Rampazzo}, R., {Panuzzo}, P., {Vega}, O., {et~al.} 2013, \mnras, 432, 374

\bibitem[{{Ree} {et~al.}(2007){Ree}, {Lee}, {Yi}, {Yoon}, {Rich}, {Deharveng}, {Sohn}, {Kaviraj}, {Rhee}, {Sheen}, {Schawinski}, {Rey}, {Boselli}, {Rhee}, {Donas}, {Seibert}, {Wyder}, {Barlow}, {Bianchi}, {Forster}, {Friedman}, {Heckman}, {Madore}, {Martin}, {Milliard}, {Morrissey}, {Neff}, {Schiminovich}, {Small}, {Szalay}, \& {Welsh}}]{ree07}
{Ree}, C.~H., {Lee}, Y.-W., {Yi}, S.~K., {et~al.} 2007, \apjs, 173, 607

\bibitem[{{Renzini} \& {Buzzoni}(1986)}]{renzini86}
{Renzini}, A. \& {Buzzoni}, A. 1986, in Astrophysics and Space Science Library, Vol. 122, Spectral Evolution of Galaxies, ed. C.~{Chiosi} \& A.~{Renzini}, 195--231

\bibitem[{{Romanowsky} {et~al.}(2012){Romanowsky}, {Strader}, {Brodie}, {Mihos}, {Spitler}, {Forbes}, {Foster}, \& {Arnold}}]{romanowsky12}
{Romanowsky}, A.~J., {Strader}, J., {Brodie}, J.~P., {et~al.} 2012, \apj, 748, 29

\bibitem[{{Sahu} {et~al.}(2022){Sahu}, {Subramaniam}, {Singh}, {Yadav}, {Valcarce}, {Choudhury}, {Rani}, {Prabhu}, {Chung}, {C{\^o}t{\'e}}, {Leigh}, {Geller}, {Chatterjee}, {Kameswara Rao}, {Bandyopadhyay}, {Shara}, {Dalessandro}, {Pandey}, {Postma}, {Hutchings}, {Simunovic}, {Stetson}, {Thirupathi}, {Puzia}, \& {Sohn}}]{sahu22}
{Sahu}, S., {Subramaniam}, A., {Singh}, G., {et~al.} 2022, \mnras, 514, 1122

\bibitem[{{Salim} {et~al.}(2018){Salim}, {Boquien}, \& {Lee}}]{salim18}
{Salim}, S., {Boquien}, M., \& {Lee}, J.~C. 2018, \apj, 859, 11

\bibitem[{{Salim} \& {Narayanan}(2020)}]{salim20}
{Salim}, S. \& {Narayanan}, D. 2020, \araa, 58, 529

\bibitem[{{Salpeter}(1955)}]{salpeter55}
{Salpeter}, E.~E. 1955, \apj, 121, 161

\bibitem[{{Salvador-Rusi{\~n}ol} {et~al.}(2022){Salvador-Rusi{\~n}ol}, {Ferr{\'e}-Mateu}, {Vazdekis}, \& {Beasley}}]{sr22}
{Salvador-Rusi{\~n}ol}, N., {Ferr{\'e}-Mateu}, A., {Vazdekis}, A., \& {Beasley}, M.~A. 2022, \mnras, 515, 4514

\bibitem[{{S{\'a}nchez-Bl{\'a}zquez} {et~al.}(2007){S{\'a}nchez-Bl{\'a}zquez}, {Forbes}, {Strader}, {Brodie}, \& {Proctor}}]{sb07}
{S{\'a}nchez-Bl{\'a}zquez}, P., {Forbes}, D.~A., {Strader}, J., {Brodie}, J., \& {Proctor}, R. 2007, \mnras, 377, 759

\bibitem[{{Sarzi} {et~al.}(2006){Sarzi}, {Falc{\'o}n-Barroso}, {Davies}, {Bacon}, {Bureau}, {Cappellari}, {de Zeeuw}, {Emsellem}, {Fathi}, {Krajnovi{\'c}}, {Kuntschner}, {McDermid}, \& {Peletier}}]{sarzi06}
{Sarzi}, M., {Falc{\'o}n-Barroso}, J., {Davies}, R.~L., {et~al.} 2006, \mnras, 366, 1151

\bibitem[{{Sarzi} {et~al.}(2018){Sarzi}, {Spiniello}, {La Barbera}, {Krajnovi{\'c}}, \& {van den Bosch}}]{sarzi18}
{Sarzi}, M., {Spiniello}, C., {La Barbera}, F., {Krajnovi{\'c}}, D., \& {van den Bosch}, R. 2018, \mnras, 478, 4084

\bibitem[{{Schiavon} {et~al.}(2012){Schiavon}, {Dalessandro}, {Sohn}, {Rood}, {O'Connell}, {Ferraro}, {Lanzoni}, {Beccari}, {Rey}, {Rhee}, {Rich}, {Yoon}, \& {Lee}}]{schiavon12}
{Schiavon}, R.~P., {Dalessandro}, E., {Sohn}, S.~T., {et~al.} 2012, \aj, 143, 121

\bibitem[{{Schlafly} \& {Finkbeiner}(2011)}]{schlafly11}
{Schlafly}, E.~F. \& {Finkbeiner}, D.~P. 2011, \apj, 737, 103

\bibitem[{{Serra} {et~al.}(2012){Serra}, {Oosterloo}, {Morganti}, {Alatalo}, {Blitz}, {Bois}, {Bournaud}, {Bureau}, {Cappellari}, {Crocker}, {Davies}, {Davis}, {de Zeeuw}, {Duc}, {Emsellem}, {Khochfar}, {Krajnovi{\'c}}, {Kuntschner}, {Lablanche}, {McDermid}, {Naab}, {Sarzi}, {Scott}, {Trager}, {Weijmans}, \& {Young}}]{serra12}
{Serra}, P., {Oosterloo}, T., {Morganti}, R., {et~al.} 2012, \mnras, 422, 1835

\bibitem[{{Simionescu} {et~al.}(2018){Simionescu}, {Tremblay}, {Werner}, {Canning}, {Allen}, \& {Oonk}}]{simionescu18}
{Simionescu}, A., {Tremblay}, G., {Werner}, N., {et~al.} 2018, \mnras, 475, 3004

\bibitem[{{Smith} {et~al.}(2012{\natexlab{a}}){Smith}, {Gomez}, {Eales}, {Ciesla}, {Boselli}, {Cortese}, {Bendo}, {Baes}, {Bianchi}, {Clemens}, {Clements}, {Cooray}, {Davies}, {De Looze}, {di Serego Alighieri}, {Fritz}, {Gavazzi}, {Gear}, {Madden}, {Mentuch}, {Panuzzo}, {Pohlen}, {Spinoglio}, {Verstappen}, {Vlahakis}, {Wilson}, \& {Xilouris}}]{smith12}
{Smith}, M.~W.~L., {Gomez}, H.~L., {Eales}, S.~A., {et~al.} 2012{\natexlab{a}}, \apj, 748, 123

\bibitem[{{Smith} {et~al.}(2012{\natexlab{b}}){Smith}, {Lucey}, \& {Carter}}]{smithrussell12}
{Smith}, R.~J., {Lucey}, J.~R., \& {Carter}, D. 2012{\natexlab{b}}, \mnras, 421, 2982

\bibitem[{{Soto} {et~al.}(2016){Soto}, {Lilly}, {Bacon}, {Richard}, \& {Conseil}}]{soto16}
{Soto}, K.~T., {Lilly}, S.~J., {Bacon}, R., {Richard}, J., \& {Conseil}, S. 2016, \mnras, 458, 3210

\bibitem[{{Spolaor} {et~al.}(2010){Spolaor}, {Kobayashi}, {Forbes}, {Couch}, \& {Hau}}]{spolaor10}
{Spolaor}, M., {Kobayashi}, C., {Forbes}, D.~A., {Couch}, W.~J., \& {Hau}, G. K.~T. 2010, \mnras, 408, 272

\bibitem[{{Spolaor} {et~al.}(2009){Spolaor}, {Proctor}, {Forbes}, \& {Couch}}]{spolaor09}
{Spolaor}, M., {Proctor}, R.~N., {Forbes}, D.~A., \& {Couch}, W.~J. 2009, \apjl, 691, L138

\bibitem[{{Springel} {et~al.}(2005){Springel}, {White}, {Jenkins}, {Frenk}, {Yoshida}, {Gao}, {Navarro}, {Thacker}, {Croton}, {Helly}, {Peacock}, {Cole}, {Thomas}, {Couchman}, {Evrard}, {Colberg}, \& {Pearce}}]{springel05}
{Springel}, V., {White}, S. D.~M., {Jenkins}, A., {et~al.} 2005, \nat, 435, 629

\bibitem[{{Tandon} {et~al.}(2020){Tandon}, {Postma}, {Joseph}, {Devaraj}, {Subramaniam}, {Barve}, {George}, {Ghosh}, {Girish}, {Hutchings}, {Kamath}, {Kathiravan}, {Kumar}, {Lancelot}, {Leahy}, {Mahesh}, {Mohan}, {Nagabhushana}, {Pati}, {Rao}, {Sankarasubramanian}, {Sriram}, \& {Stalin}}]{tandon20}
{Tandon}, S.~N., {Postma}, J., {Joseph}, P., {et~al.} 2020, \aj, 159, 158

\bibitem[{{Taylor} {et~al.}(2021){Taylor}, {Ko}, {C{\^o}t{\'e}}, {Ferrarese}, {Peng}, {Zabludoff}, {Roediger}, {S{\'a}nchez-Janssen}, {Hendel}, {Chilingarian}, {Liu}, {Spengler}, \& {Zhang}}]{taylor21}
{Taylor}, M.~A., {Ko}, Y., {C{\^o}t{\'e}}, P., {et~al.} 2021, \apj, 915, 83

\bibitem[{{Thomas} {et~al.}(1999){Thomas}, {Greggio}, \& {Bender}}]{thomas99}
{Thomas}, D., {Greggio}, L., \& {Bender}, R. 1999, \mnras, 302, 537

\bibitem[{{Thomas} {et~al.}(2003){Thomas}, {Maraston}, \& {Bender}}]{thomas03}
{Thomas}, D., {Maraston}, C., \& {Bender}, R. 2003, \mnras, 339, 897

\bibitem[{{Thomas} {et~al.}(2005){Thomas}, {Maraston}, {Bender}, \& {Mendes de Oliveira}}]{thomas05}
{Thomas}, D., {Maraston}, C., {Bender}, R., \& {Mendes de Oliveira}, C. 2005, \apj, 621, 673

\bibitem[{{Thomas} {et~al.}(2011){Thomas}, {Maraston}, \& {Johansson}}]{thomas11}
{Thomas}, D., {Maraston}, C., \& {Johansson}, J. 2011, \mnras, 412, 2183

\bibitem[{{Thomas} {et~al.}(2010){Thomas}, {Maraston}, {Schawinski}, {Sarzi}, \& {Silk}}]{thomas10}
{Thomas}, D., {Maraston}, C., {Schawinski}, K., {Sarzi}, M., \& {Silk}, J. 2010, \mnras, 404, 1775

\bibitem[{{Tonini}(2013)}]{tonini13}
{Tonini}, C. 2013, \apj, 762, 39

\bibitem[{{Tortora} {et~al.}(2013){Tortora}, {Romanowsky}, \& {Napolitano}}]{tortora13}
{Tortora}, C., {Romanowsky}, A.~J., \& {Napolitano}, N.~R. 2013, \apj, 765, 8

\bibitem[{{Trager} {et~al.}(2000){Trager}, {Faber}, {Worthey}, \& {Gonz{\'a}lez}}]{trager00a}
{Trager}, S.~C., {Faber}, S.~M., {Worthey}, G., \& {Gonz{\'a}lez}, J.~J. 2000, \aj, 119, 1645

\bibitem[{{Trayford} {et~al.}(2020){Trayford}, {Lagos}, {Robotham}, \& {Obreschkow}}]{trayford20}
{Trayford}, J.~W., {Lagos}, C. d.~P., {Robotham}, A. S.~G., \& {Obreschkow}, D. 2020, \mnras, 491, 3937

\bibitem[{{Valentino} {et~al.}(2023){Valentino}, {Brammer}, {Gould}, {Kokorev}, {Fujimoto}, {Jespersen}, {Vijayan}, {Weaver}, {Ito}, {Tanaka}, {Ilbert}, {Magdis}, {Whitaker}, {Faisst}, {Gallazzi}, {Gillman}, {Gim{\'e}nez-Arteaga}, {G{\'o}mez-Guijarro}, {Kubo}, {Heintz}, {Hirschmann}, {Oesch}, {Onodera}, {Rizzo}, {Lee}, {Strait}, \& {Toft}}]{valentino23}
{Valentino}, F., {Brammer}, G., {Gould}, K. M.~L., {et~al.} 2023, \apj, 947, 20

\bibitem[{{Vazdekis} {et~al.}(2016){Vazdekis}, {Koleva}, {Ricciardelli}, {R{\"o}ck}, \& {Falc{\'o}n-Barroso}}]{vazdekis16}
{Vazdekis}, A., {Koleva}, M., {Ricciardelli}, E., {R{\"o}ck}, B., \& {Falc{\'o}n-Barroso}, J. 2016, \mnras, 463, 3409

\bibitem[{{Voyer} {et~al.}(2014){Voyer}, {Boselli}, {Boissier}, {Heinis}, {Cortese}, {Ferrarese}, {Cote}, {Cuillandre}, {Gwyn}, {Peng}, {Zhang}, \& {Liu}}]{voyer14}
{Voyer}, E.~N., {Boselli}, A., {Boissier}, S., {et~al.} 2014, \aap, 569, A124

\bibitem[{{Wechsler} \& {Tinker}(2018)}]{wechsler18}
{Wechsler}, R.~H. \& {Tinker}, J.~L. 2018, \araa, 56, 435

\bibitem[{{Werle} {et~al.}(2020){Werle}, {Cid Fernandes}, {Vale Asari}, {Coelho}, {Bruzual}, {Charlot}, {de Carvalho}, {Herpich}, {Mendes de Oliveira}, {Sodr{\'e}}, {Ruschel-Dutra}, {de Amorim}, \& {Sampaio}}]{werle20}
{Werle}, A., {Cid Fernandes}, R., {Vale Asari}, N., {et~al.} 2020, \mnras, 497, 3251

\bibitem[{{Whitmore} {et~al.}(1993){Whitmore}, {Gilmore}, \& {Jones}}]{whitmore93}
{Whitmore}, B.~C., {Gilmore}, D.~M., \& {Jones}, C. 1993, \apj, 407, 489

\bibitem[{{Worthey} {et~al.}(1992){Worthey}, {Faber}, \& {Gonzalez}}]{worthey92}
{Worthey}, G., {Faber}, S.~M., \& {Gonzalez}, J.~J. 1992, \apj, 398, 69

\bibitem[{{Worthey} {et~al.}(1994){Worthey}, {Faber}, {Gonzalez}, \& {Burstein}}]{worthey94}
{Worthey}, G., {Faber}, S.~M., {Gonzalez}, J.~J., \& {Burstein}, D. 1994, \apjs, 94, 687

\bibitem[{{Worthey} {et~al.}(2014){Worthey}, {Tang}, \& {Serven}}]{worthey14}
{Worthey}, G., {Tang}, B., \& {Serven}, J. 2014, \apj, 783, 20

\bibitem[{{Yan} {et~al.}(2021){Yan}, {Je{\v{r}}{\'a}bkov{\'a}}, \& {Kroupa}}]{yan21}
{Yan}, Z., {Je{\v{r}}{\'a}bkov{\'a}}, T., \& {Kroupa}, P. 2021, \aap, 655, A19

\bibitem[{{Yi}(2008)}]{yi08}
{Yi}, S.~K. 2008, in Astronomical Society of the Pacific Conference Series, Vol. 392, Hot Subdwarf Stars and Related Objects, ed. U.~{Heber}, C.~S. {Jeffery}, \& R.~{Napiwotzki}, 3

\bibitem[{{Yi} {et~al.}(2011){Yi}, {Lee}, {Sheen}, {Jeong}, {Suh}, \& {Oh}}]{yi11}
{Yi}, S.~K., {Lee}, J., {Sheen}, Y.-K., {et~al.} 2011, \apjs, 195, 22

\bibitem[{{Young} {et~al.}(2011){Young}, {Bureau}, {Davis}, {Combes}, {McDermid}, {Alatalo}, {Blitz}, {Bois}, {Bournaud}, {Cappellari}, {Davies}, {de Zeeuw}, {Emsellem}, {Khochfar}, {Krajnovi{\'c}}, {Kuntschner}, {Lablanche}, {Morganti}, {Naab}, {Oosterloo}, {Sarzi}, {Scott}, {Serra}, \& {Weijmans}}]{young11}
{Young}, L.~M., {Bureau}, M., {Davis}, T.~A., {et~al.} 2011, \mnras, 414, 940

\bibitem[{{Young} {et~al.}(2014){Young}, {Scott}, {Serra}, {Alatalo}, {Bayet}, {Blitz}, {Bois}, {Bournaud}, {Bureau}, {Crocker}, {Cappellari}, {Davies}, {Davis}, {de Zeeuw}, {Duc}, {Emsellem}, {Khochfar}, {Krajnovi{\'c}}, {Kuntschner}, {McDermid}, {Morganti}, {Naab}, {Oosterloo}, {Sarzi}, \& {Weijmans}}]{young14}
{Young}, L.~M., {Scott}, N., {Serra}, P., {et~al.} 2014, \mnras, 444, 3408

\bibitem[{{Zheng} {et~al.}(2017){Zheng}, {Wang}, {Ge}, {Mao}, {Li}, {Li}, {Mo}, {Goddard}, {Bundy}, {Li}, {Nair}, {Lin}, {Long}, {Riffel}, {Thomas}, {Masters}, {Bizyaev}, {Brownstein}, {Zhang}, {Law}, {Drory}, {Roman Lopes}, \& {Malanushenko}}]{zheng17}
{Zheng}, Z., {Wang}, H., {Ge}, J., {et~al.} 2017, \mnras, 465, 4572

\bibitem[{{Zibetti} {et~al.}(2020){Zibetti}, {Gallazzi}, {Hirschmann}, {Consolandi}, {Falc{\'o}n-Barroso}, {van de Ven}, \& {Lyubenova}}]{zibetti20}
{Zibetti}, S., {Gallazzi}, A.~R., {Hirschmann}, M., {et~al.} 2020, \mnras, 491, 3562

\end{thebibliography}

\appendix 

\section{Notes on individual galaxies.}
\label{app:notes_gals}

In this Appendix, we complement the information on the ionised gas emission of the target galaxies from Section \ref{subsec:ha_analy}.

\textit{NGC~4262.} The H$\alpha$ emission in NGC~4262 was already investigated in \cite{boselli22}, where the authors found that the gas is shock-ionised and infalling into the nucleus of the galaxy, likely after a merger event. This is corroborated by the presence of an H$\alpha$ ring, extending from south-west and north of the galaxy, which was also observed in the UV by \cite{bettoni10} (see also Section \ref{subsec:iso_fit}). These are star forming HII regions.
Regarding the nucleus, no emission in H$\alpha$ emission is present \cite{ho95}.  We found a significant emission in [OIII], with a flux $f_{\rm [OIII]} \simeq 7\times10^{-15}$ erg s$^{-1}$ cm$^{-2}$, while gas emission in H$\beta$ is not statistically significant (we can only put an upper limit $f_{\rm H\beta} < 10^{-14}$ erg s$^{-1}$ cm$^{-2}$). 
The lack of both star forming complexes and of hydrogen emission in the disk, along with the filamentary nature of the gas suggest that in the central regions of this galaxy there is no active SF. 

\textit{NGC~4374/M84.} The H$\alpha$ emission in M84 is also of filamentary nature, extending east-west of the nucleus and with no visible clumpy structures. This is consistent with what was recently reported by \cite{eskenasy24}, who analysed the H$\alpha$ emission map of non-central ETGs with MUSE data. 
Finally, this galaxy is well known to host an active galactic nucleus (AGN), (e.g., \citealt{hansen85}), thus the central H$\alpha$ emission might be dominated by its AGN activity. We also observe the H$\alpha$ line in emission in the MUSE central spectrum (shown in Fig. \ref{fig:muse}), as well as significant H$\beta$ and [OIII] emission from the SAURON IFU data ($f_{\rm H\beta} \simeq 2\times 10^{-14}$ erg s$^{-1}$ cm$^{-2}$ and $f_{\rm [OIII]} \simeq 3\times 10^{-14}$ erg s$^{-1}$ cm$^{-2}$). 

\textit{NGC~4406/M86.} Evidence of stripped H$\alpha$ filaments was already reported in M86, due to its interaction with the spiral galaxy NGC~4438, located $\sim$23 arcmin north-east of M86 \citep{kenney08, gomez10}. The filaments in Fig. \ref{fig:ha} do not show clumps or knots, thus indicating lack of ongoing SF. Indeed, such a collision have likely heated the gas and prevented it to cool down to form stars (\citealt{kenney08} and references therein). Within the centre of M86, no emission H$\alpha$ line is present \citep{ho95}. Similarly, both [OIII] and H$\beta$ emission are not significantly detected in the SAURON data, and we were only able to put upper limits of $f_{\rm [OIII]} < 6\times 10^{-14}$ erg s$^{-1}$ cm$^{-2}$, $f_{\rm H\beta} < 3\times 10^{-14}$ erg s$^{-1}$ cm$^{-2}$.

\textit{NGC~4417 and NGC~4442.} These two lenticular galaxies show a weak H$\alpha$ emission in a disk-like structure that follow the stellar disk. This is very likely a residual artefact from the continuum subtraction (see Sect. \ref{subsec:vestige}),
indeed there are no prominent filaments nor clumps. The stellar component, fast rotator according to \cite{emsellem11}, is dominant in these galaxies. Gas emission from H$\beta$ and [OIII] from the SAURON data is not significant ($f < 5\times 10^{-14}$ erg s$^{-1}$ cm$^{-2}$ for both lines). Additionally, there is no significant detection of molecular gas in these galaxies \citep{young11}, as well as no emission line from H$\alpha$ is observed within their nuclei \citep{ho95}.

\textit{NGC~4472/M49.} No significant emission in H$\alpha$ is observed in M49 (with a surface brightness limit of $\Sigma(H\alpha) \sim 4\times 10^{-18}$ erg s$^{-1}$ cm$^{-2}$ arcsec$^{-2}$ from the VESTIGE image), except in its centre where artefacts might arise (see Sect. \ref{subsec:vestige}). For this reason, we also checked the spectrum in its nucleus from \cite{ho95}, where no H$\alpha$ in emission is observed. In addition, the emission of H$\beta$ and [OIII] is not statistically significant from the SAURON data ($f < 10^{-13}$ erg s$^{-1}$ cm$^{-2}$). This confirms the absence of gas or recent SF activity, which is also corroborated by our own MUSE analysis (see Sect. \ref{subsec:muse_analysis}).

\textit{NGC~4486/M87.} M87 is well-known to host H$\alpha$ filaments, which are also visible in the continuum-subtracted image in Fig. \ref{fig:ha}. The VESTIGE H$\alpha$ analysis of M87 was recently reported by \cite{boselli19} in detail. They concluded that the filaments are not photoionised by young stars, which is consistent with the results obtained from the molecular gas by \cite{simionescu18}. We refer the interested reader to \cite{boselli19} and references therein for more information on the nature of the ionised gas filament in M87. The stellar population analysis and interpretation for M87 is undoubtedly challenging, due to the presence of multi-phase gas, powerful AGN activity and non-thermal jet emission (see Section \ref{subsec:synchr}). However, these latest results give us confidence that the presence of young stars within M87 is likely excluded.


\section{Estimation of Synchrotron emission in M87}
\label{app:m87_sed_syn}

\begin{figure*}
\centering
\includegraphics[scale=0.58]{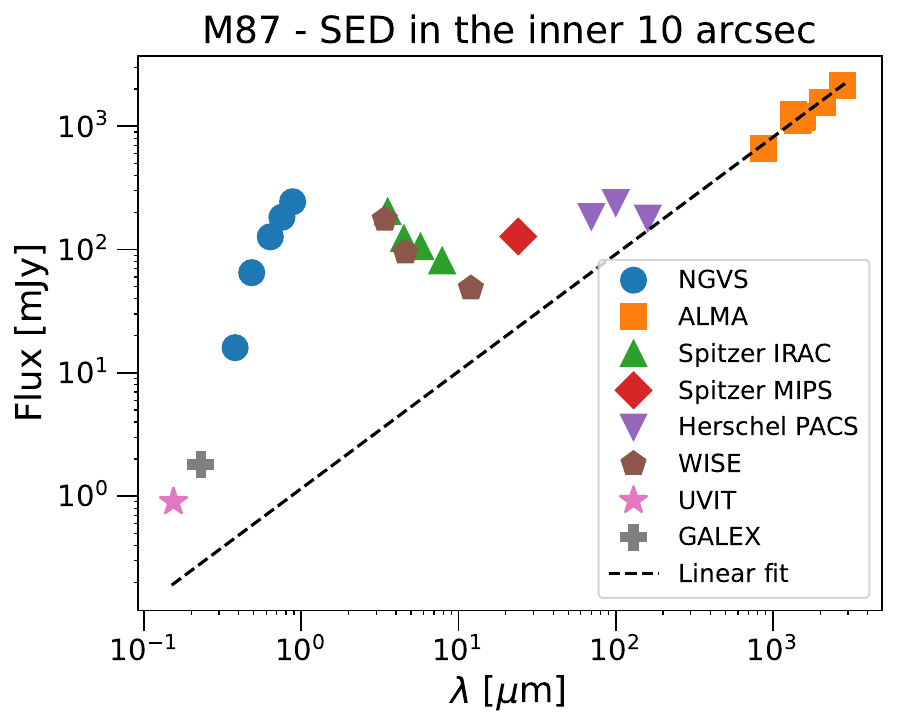}
\caption{SED in the inner 10$''$ of M87, spanning from the FUV (UVIT) up to the sub-mm (ALMA). The different data used are indicated in the legend. The black dashed line represents the linear fit to the ALMA data.} 
\label{fig:M87_sed_10}
\end{figure*}

For the FUV, NUV and optical we used the images from Sect. \ref{subsec:uvit}, \ref{subsec:galex}, \ref{subsec:ngvs}, respectively. For the IR, we used \textit{Spitzer} IRAC and MIPS, \textit{WISE} (W1, W2 and W3) and \textit{Herschel} PACS already reduced images, downloaded from the respective archives. Finally, we retrieved already processed ALMA images (from \url{https://almascience.nrao.edu/aq/}), in several bands (from band 3 to 7). We chose continuum data, with a resolution higher than 5$''$. We then defined three circular regions with a radius of 5, 10 and 20$''$ , respectively, centred on M87, and we calculated the flux in mJy in each region in each band. The fluxes and zeropoints are calculated with the prescriptions specified in the dedicated webpages. The SED for the inner 10$''$ is shown in Fig. \ref{fig:M87_sed_10}. We performed a log-log linear fit on the ALMA fluxes for the three defined radial SEDs. This is reported as a dashed black line in Fig. \ref{fig:M87_sed_10}. 
We estimated how much flux in the FUV and NUV might arise from the central synchrotron emission by extrapolating the flux at 1541\angstrom \, and 2297\angstrom \, from the linear fit. For the inner 5, 10 and 20$''$, we have that the flux contribution from the synchrotron radiation is $\sim$26\%, 22\% and 1\% for the FUV, and $\sim$16\%, 16\%, 1\% for the NUV, respectively. We then corrected the FUV and NUV magnitudes of M87 up to the inner 10$''$ by removing the contribution of the synchrotron emission.  

\section{Isophotal fit at GALEX resolution.}
\label{app:iso_galex}

\begin{figure*}
\includegraphics[scale=0.55]{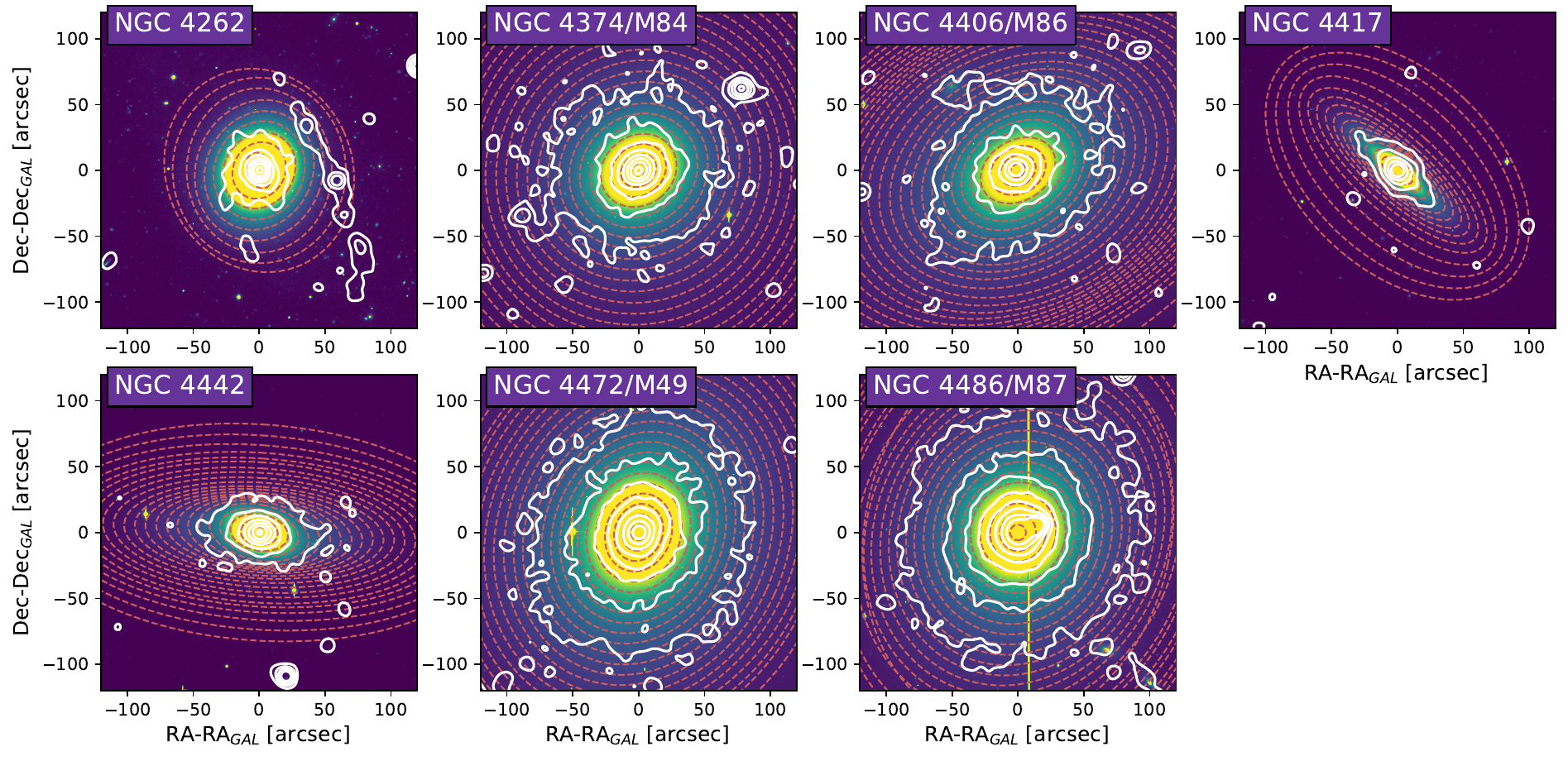}
\caption{As in Fig. \ref{fig:ell_uvit} but for the GALEX resolution analysis. See main text for more details. } 
\label{fig:ell_galex}
\end{figure*}

Figure \ref{fig:ell_galex} displays the images of the target galaxies in the NGVS $i-$band with isophotal fits at GALEX resolution superimposed as red dashed ellipses. Symbols are as in Fig. \ref{fig:ell_uvit}.

\section{VLT/MUSE radial spectra of M49 and M87}
\label{app:muse_images}

\begin{figure*}
\centering
\includegraphics[scale=0.5]{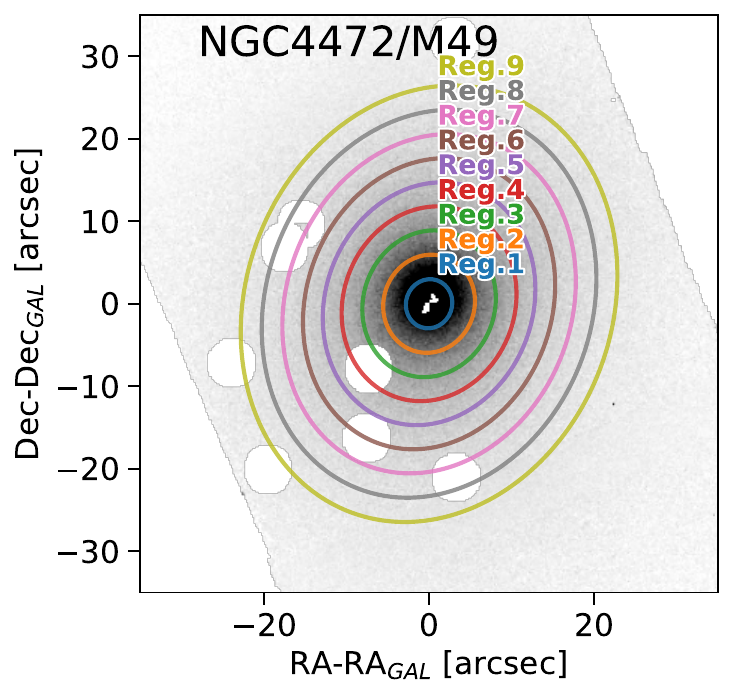}
\includegraphics[scale=0.5]{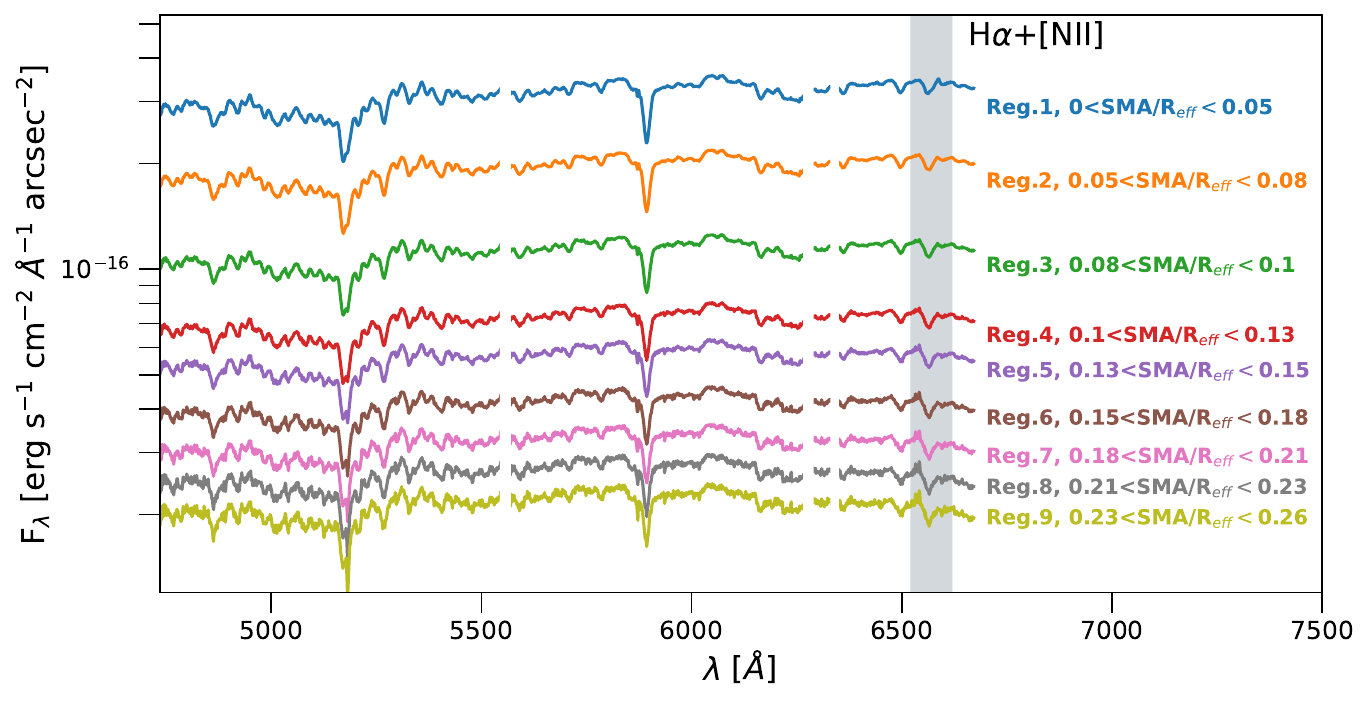}
\caption{Same as in Fig. \ref{fig:muse} but for M49.} 
\label{fig:muse_m49}
\end{figure*}

\begin{figure*}
\centering
\includegraphics[scale=0.5]{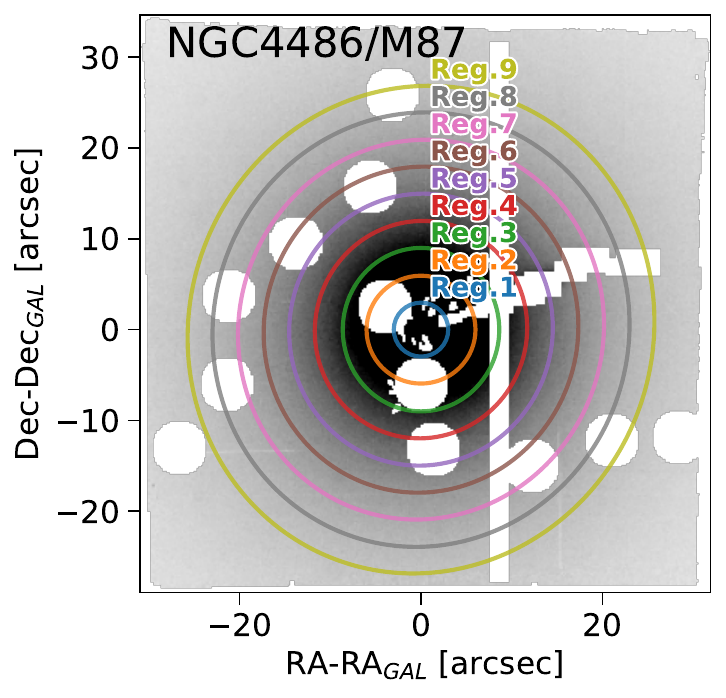}
\includegraphics[scale=0.5]{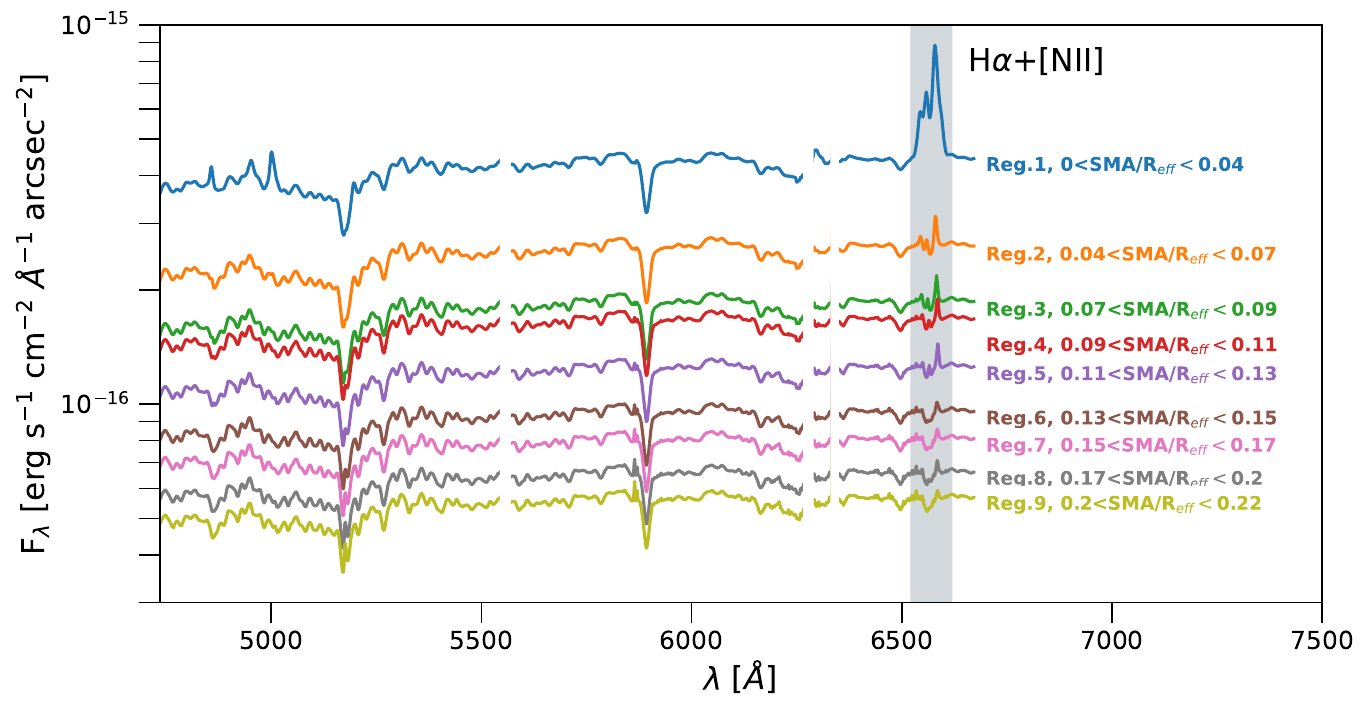}
\caption{Same as in Fig. \ref{fig:muse} but for M87.} 
\label{fig:muse_m87}
\end{figure*}

Figures \ref{fig:muse_m49} and \ref{fig:muse_m87} report the slice of the MUSE cube ($\lambda\sim5400$\AA) (left), and the radial MUSE spectra (right) for M49 and M87, respectively. Symbols as in Fig. \ref{fig:muse}.

\section{Mock analysis figures}
\label{app:mock}

\begin{figure*}
\centering
\includegraphics[scale=0.48]{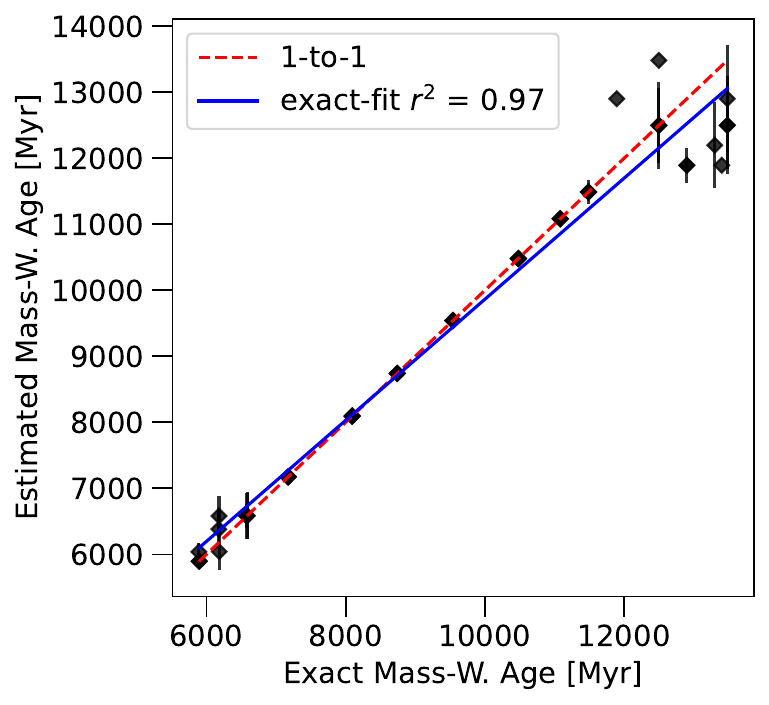}
\includegraphics[scale=0.48]{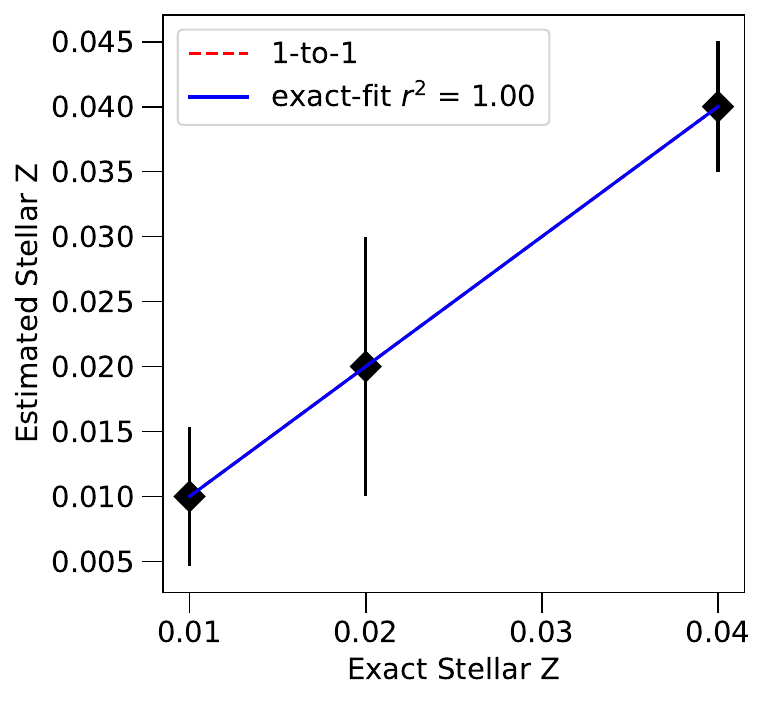}
\includegraphics[scale=0.48]{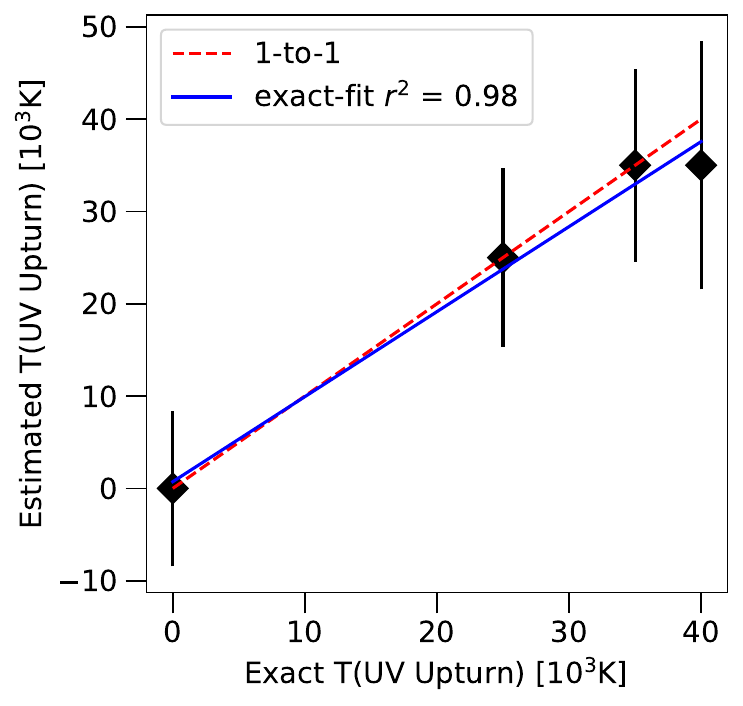}
\includegraphics[scale=0.48]{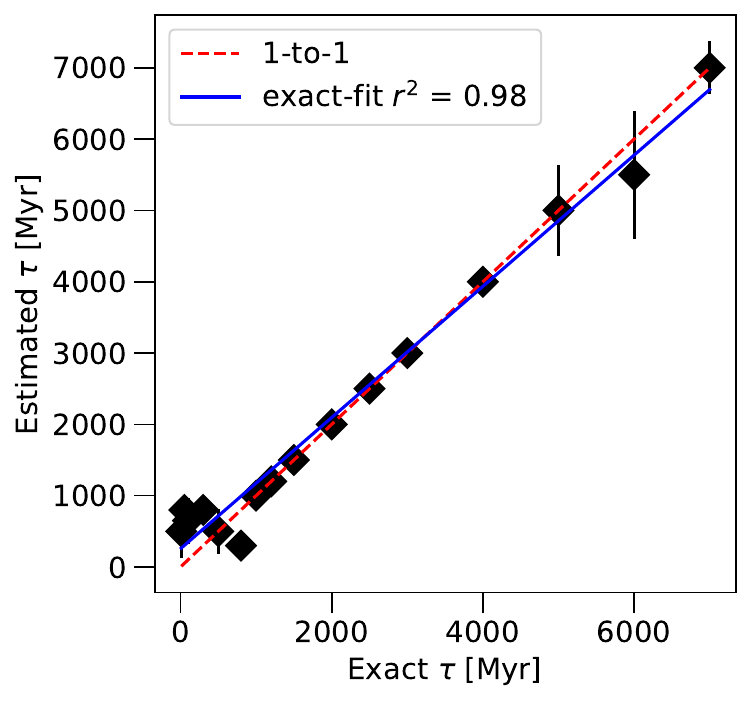}
\includegraphics[scale=0.48]{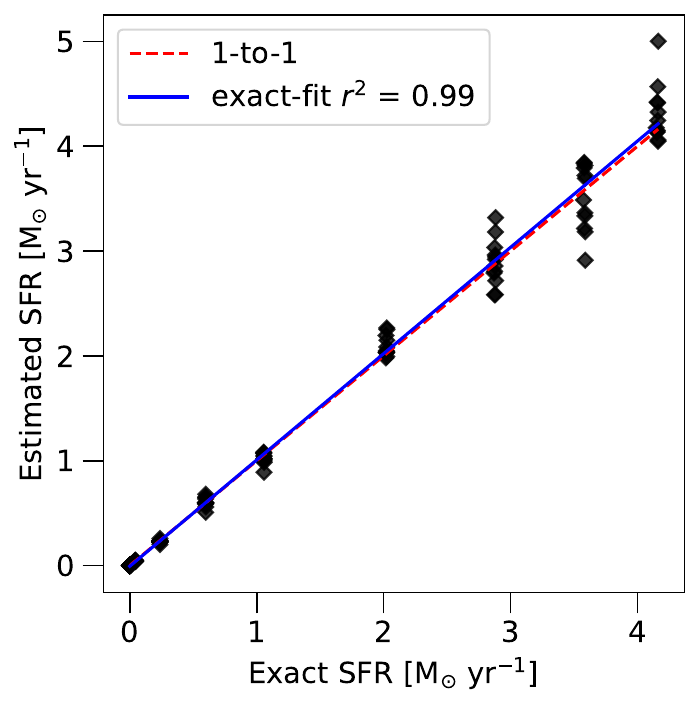}
\caption{Median estimated galaxy parameters as a function of the generated parameters for the mock galaxies. Errors are the 1$\sigma$ of the distribution. The red dashed line indicates the 1$-$to$-$1 relation, while the blue solid line represents a linear fit to the data. The squared Pearson correlation coefficient $r^2$ is reported.}
\label{fig:mock_1}
\end{figure*}

\begin{figure*}
\centering
\includegraphics[scale=0.45]{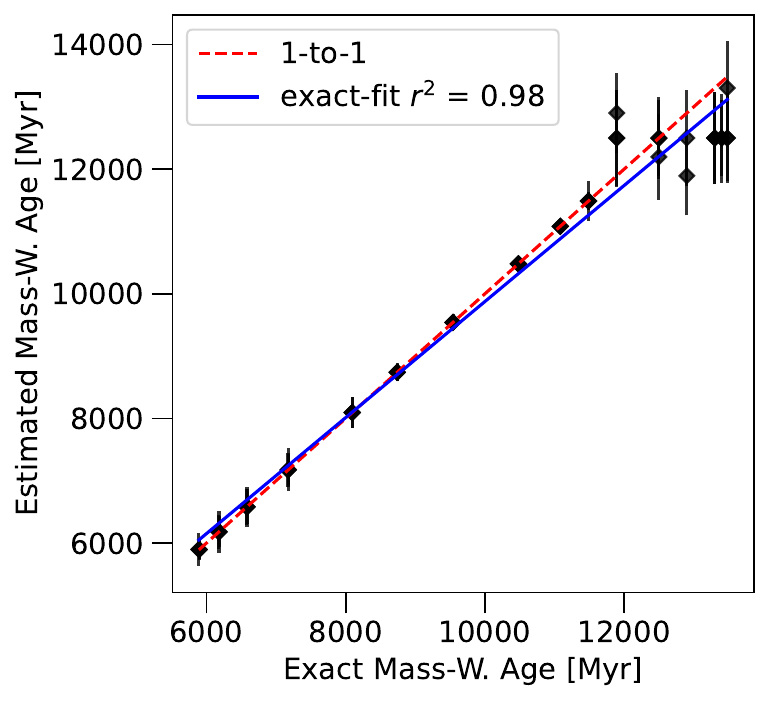}
\includegraphics[scale=0.45]{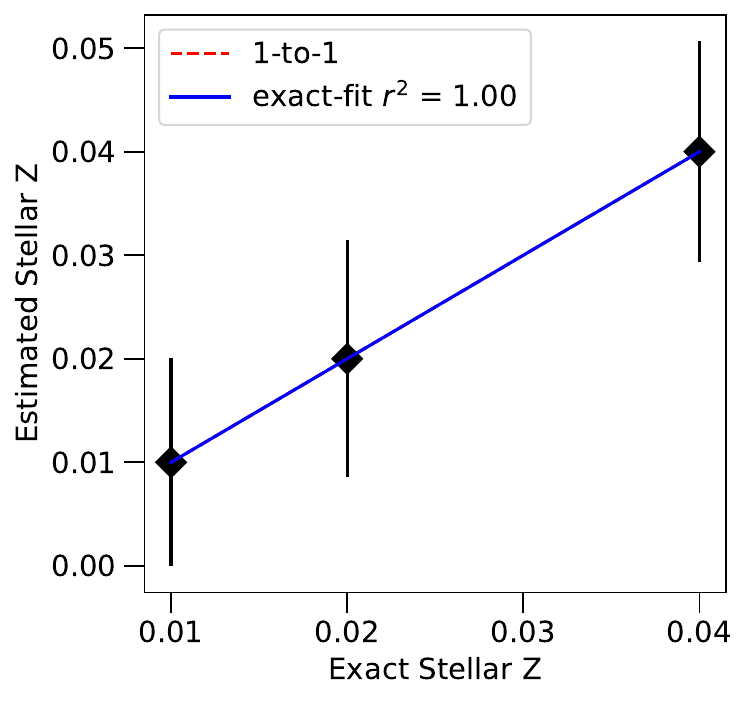}
\includegraphics[scale=0.45]{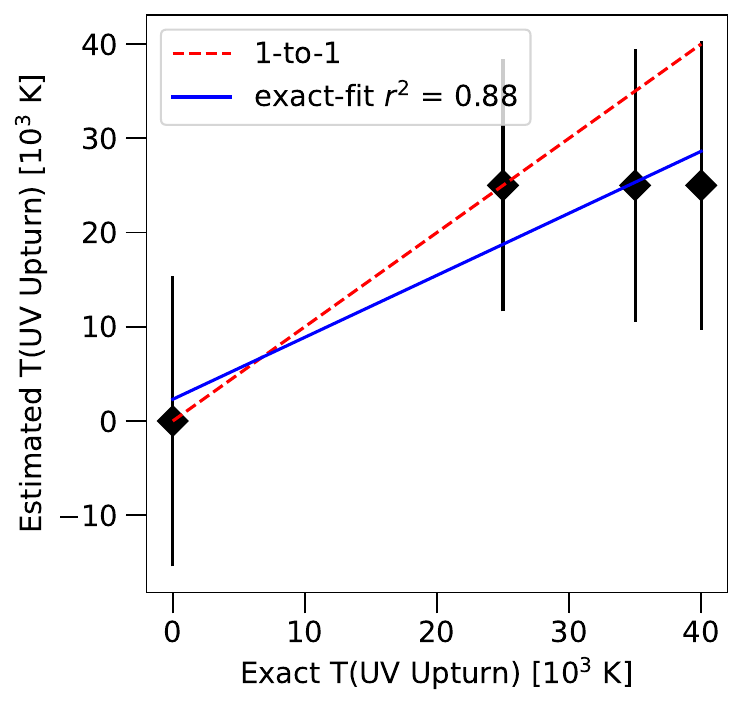}
\includegraphics[scale=0.45]{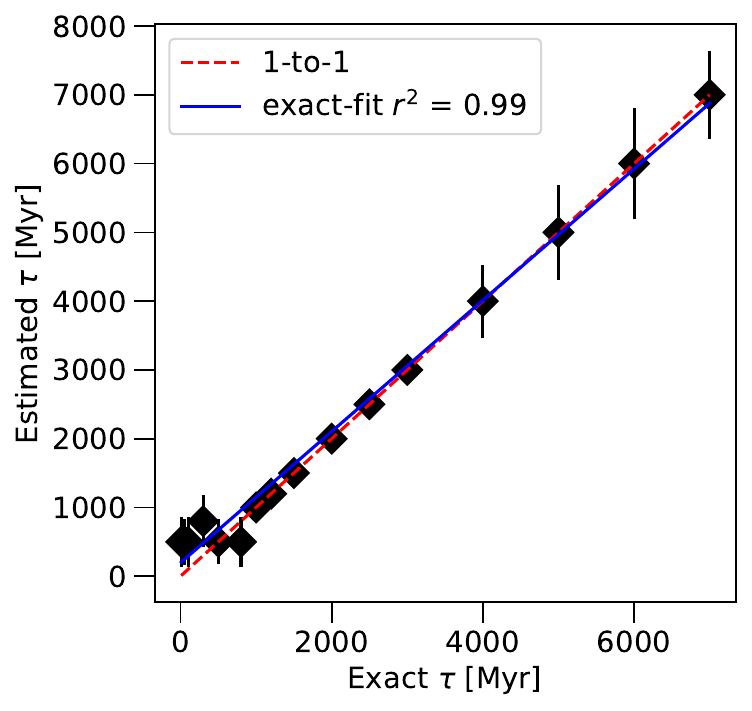}
\includegraphics[scale=0.45]{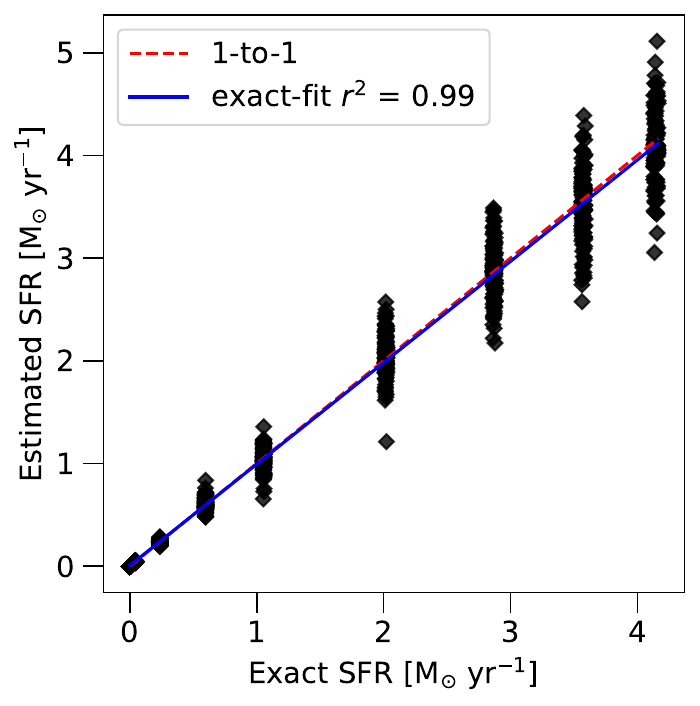}
\includegraphics[scale=0.45]{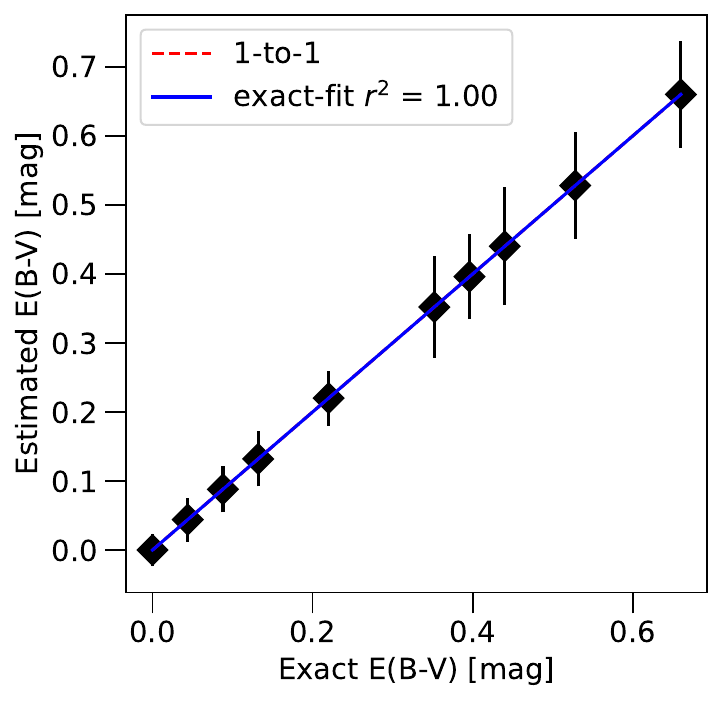}
\caption{Same as in Fig. \ref{fig:mock_1}, but for the mock catalogue considering dust extinction.}
\label{fig:mock_2}
\end{figure*}

Figure \ref{fig:mock_1} (\ref{fig:mock_2}) reports the estimated parameters of the generated mock galaxies as a function of the exact value given as input without (with) dust extinction module. From top left to bottom right: the mass-weighted age, the stellar metallicity, the temperature of the UV upturn, SF timescale, SFR and in Fig. \ref{fig:mock_2} the stellar colour excess.

\section{$E(B-V)$ SED fitting results for M84, M86 and M87}
\label{app:extinction}

\begin{figure*}
\centering
\includegraphics[scale=0.55]{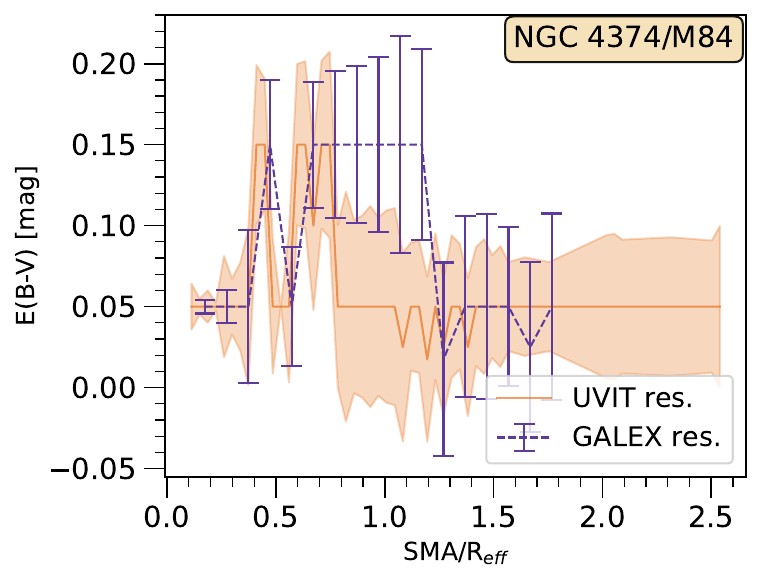}
\includegraphics[scale=0.55]{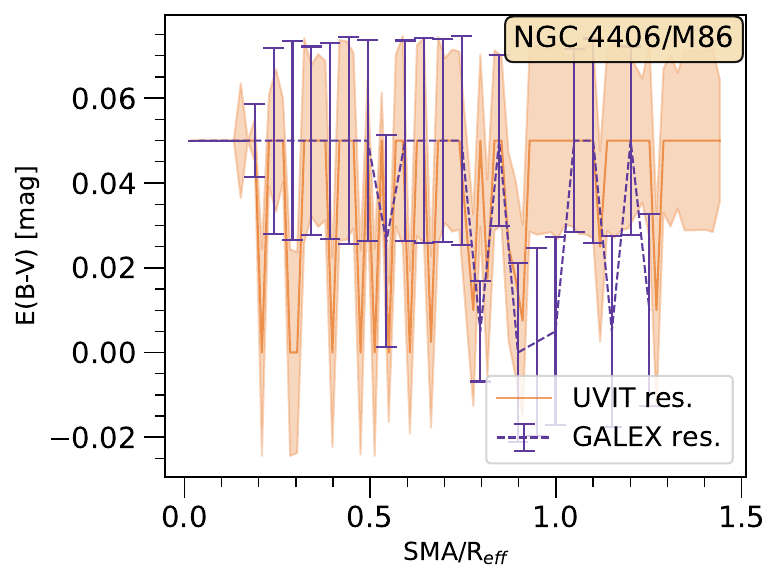}
\includegraphics[scale=0.55]{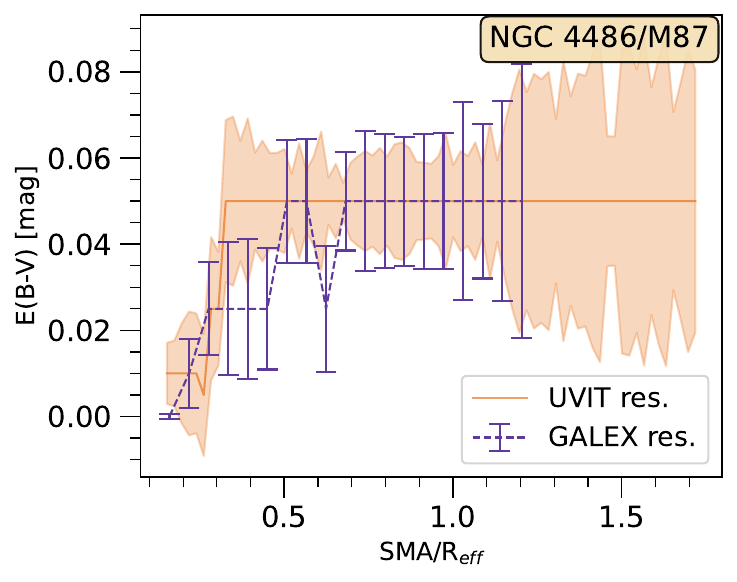}
\caption{Best fit stellar colour excess $E(B-V)$ as a function of SMA/R$_{\rm eff}$ for M84 (left), M86 (central) and M87 (right). Symbols as in Fig. \ref{fig:m49_m86_res}.} 
\label{fig:extinction}
\end{figure*}

Figure \ref{fig:extinction} show the best fit stellar colour excess $E(B-V)$ as a function of the SMA/R$_{\rm eff}$ for M84 (left), M86 (central) and M87 (right). Symbols as in Fig. \ref{fig:m49_m86_res}.

\end{document}